%% file: eeSlep.tex
\documentclass[a4paper,12pt]{article}
\pdfoutput=1
\usepackage{geometry}
\usepackage{amsmath,amssymb,amsfonts}
\usepackage{xcolor,graphicx,cite,soul}
\usepackage{array,booktabs,longtable}
\usepackage{caption,microtype}
\usepackage{hyperref}

\input{paperdef}

\graphicspath{{figs/}}
\allowdisplaybreaks
\begin{document}
\thispagestyle{empty}

\def\thefootnote{\fnsymbol{footnote}}

\begin{flushright}
\mbox{}
IFT--UAM/CSIC--18-007 \\
\end{flushright}

\vspace{0.5cm}

\begin{center}

{\large\sc 
{\bf Slepton Production at \boldmath{$e^+e^-$} Colliders}} 

\vspace{0.4cm}

{\large\sc {\bf in the Complex MSSM: A Full One-Loop Analysis}}

\vspace{1cm}

{\sc
S.~Heinemeyer$^{1,2,3}$%
\footnote{email: Sven.Heinemeyer@cern.ch}%
~and C.~Schappacher$^{4}$%
\footnote{email: schappacher@kabelbw.de}%
}

\vspace*{.7cm}

{\sl
$^1$Campus of International Excellence UAM+CSIC, 
Cantoblanco, 28049, Madrid, Spain 

\vspace*{0.1cm}

$^2$Instituto de F\'isica Te\'orica (UAM/CSIC), 
Universidad Aut\'onoma de Madrid, \\ 
Cantoblanco, 28049, Madrid, Spain

\vspace*{0.1cm}

$^3$Instituto de F\'isica de Cantabria (CSIC-UC), 
39005, Santander, Spain

\vspace*{0.1cm}

$^4$Institut f\"ur Theoretische Physik,
Karlsruhe Institute of Technology, \\
76128, Karlsruhe, Germany (former address)
}

\end{center}

\vspace*{0.1cm}

\begin{abstract}
\noindent
For the search for scalar leptons in the Minimal Supersymmetric 
Standard Model (MSSM) as well as for future precision analyses of these 
particles an accurate knowledge of their production and decay properties
is mandatory. 
We evaluate the cross sections for the slepton production at $e^+e^-$ 
colliders in the MSSM with complex parameters (cMSSM). 
The evaluation is based on a full one-loop calculation of the production 
mechanisms \eeSlSl\ including soft and hard photon radiation.  
The dependence of the slepton production cross sections on the relevant 
cMSSM parameters is analyzed numerically.  We find sizable contributions 
to many production cross sections.  They amount to roughly $15\,\%$ of 
the tree-level results but can go up to $40\,\%$ or higher in extreme 
cases. Also the complex phase dependence of the one-loop corrections
for charged slepton production was found non-negligible. 
The full one-loop contributions are thus crucial for physics analyses at 
a future linear $e^+e^-$ collider such as the ILC or CLIC.
\end{abstract}


\def\thefootnote{\arabic{footnote}}
\setcounter{page}{0}
\setcounter{footnote}{0}

\newpage


\section{Introduction}
\label{sec:intro}

One of the important tasks at the LHC is to search for physics beyond the 
Standard Model (SM), where the Minimal Supersymmetric Standard Model 
(MSSM)~\cite{Ni1984,Ba1988,HaK85,GuH86} is one of the leading candidates.
Two related important tasks are the investigation of the mechanism of
electroweak symmetry breaking, including the identification of the underlying 
physics of the Higgs boson discovered at 
$\sim 125\gev$~\cite{ATLASdiscovery,CMSdiscovery}, as well as the production 
and measurement of the properties of Cold Dark Matter (CDM).  Here the MSSM 
offers a natural candidate for CDM, the Lightest Supersymmetric Particle 
(LSP), the lightest neutralino,~$\neu{1}$~\cite{Go1983,ElHaNaOlSr1984} 
(see below).
These three (related) tasks will be the top priority in the future program of 
particle physics.  

Supersymmetry (SUSY) predicts two scalar partners for all SM fermions as well
as fermionic partners to all SM bosons. 
Contrary to the case of the SM, in the MSSM two Higgs doublets are required.
This results in five physical Higgs bosons instead of the single Higgs
boson in the SM.  These are the light and heavy $\cp$-even Higgs bosons, 
$h$ and $H$, the $\cp$-odd Higgs boson, $A$, and the charged Higgs bosons,
$H^\pm$.
In the MSSM with complex parameters (cMSSM) the three neutral Higgs
bosons mix~\cite{mhiggsCPXgen,Pi1998,Demir,mhiggsCPXRG1,mhiggsCPXFD1}, 
giving rise to the $\cp$-mixed states $\He, \Hz, \Hd$.
The neutral SUSY partners of the (neutral) Higgs and electroweak gauge
bosons are the four neutralinos, $\neu{1,2,3,4}$.  The corresponding
charged SUSY partners are the charginos, $\cha{1,2}$.

If SUSY is realized in nature and the scalar quarks and/or the gluino
are in the kinematic reach of the (HL-)LHC, it is expected that these
strongly interacting particles are eventually produced and studied.
On the other hand, SUSY particles that interact only via the electroweak
force, \eg, the scalar leptons, have a much smaller
production cross section at the LHC.  Correspondingly, the LHC
discovery potential as well as the current experimental bounds are
substantially weaker~\cite{ATLAS-SUSY,CMS-SUSY}.

At a (future) $e^+e^-$ collider sleptons, depending on their masses 
and the available center-of-mass energy, could be produced and
analyzed in detail.  Corresponding studies can be found for the ILC 
in \citeres{ILC-TDR,teslatdr,Ac2004,ilc1,ilc2,LCreport} and for CLIC 
in \citeres{CLIC1,CLIC2,LCreport}. 
(Results on the combination of LHC and ILC results can be found in 
\citeres{lhcilc1,lhcilc2,lhcilc3}.)  Such precision studies will be 
crucial to determine their nature and the underlying (SUSY) parameters.

In order to yield a sufficient accuracy, one-loop corrections to the 
various slepton production and decay modes have to be considered.
Full one-loop calculations in the cMSSM to (heavy) scalar tau decays was
evaluated in \citere{Stau2decay}, where the calculation can easily be taken
over to other slepton decays. 
Sleptons can also be produced in SUSY cascade decays. Full one-loop
evaluations in the cMSSM exist for the corresponding decays of Higgs
bosons~\cite{HiggsDecaySferm} as well as from charginos and
neutralinos~\cite{LHCxC1,LHCxC2,LHCxN,LHCxNprod}. 
In this paper we take the next step and concentrate on the slepton
production at $e^+e^-$ colliders, \ie we calculate 
\begin{alignat}{3}
\label{eq:eeSeSe}
&\sig(\eeSeSe) &\qquad & s,\spri = 1,2\,, \\
\label{eq:eeSnSn}
&\sig(\eeSnSn) &\qquad & g = 1,2,3\,,
\end{alignat}
with $\Se_{gs} = \{\tilde e_s, \tilde\mu_s, \tilde\tau_s\}$, 
$\Sn_g = \{\Sn_e, \Sn_\mu, \Sn_\tau\}$, generation index $g$ and slepton 
index $s$.
Our evaluation of the two channels (\ref{eq:eeSeSe}) and (\ref{eq:eeSnSn}) 
is based on a full one-loop calculation, \ie including electroweak (EW) 
corrections, as well as soft, hard and collinear QED radiation. 
The renormalization scheme employed is the same one as for the decay 
of sleptons~\cite{Stau2decay}.
Consequently, the predictions for the production and decay can be 
used together in a consistent manner.

Results for the cross sections (\ref{eq:eeSeSe}) and (\ref{eq:eeSnSn}) at 
various levels of sophistication have been obtained over the last three
decades. 
Tree-level results were published for \eeSeSe\ and \eeSnSn\ in the MSSM
with real parameters (rMSSM) in \citere{SchWa1985}, and later in a specific
supergravity (SUGRA) model in \citere{CaDi1995}. 
Tree-level results in the cMSSM were published only for the process
$\mu^+\mu^- \to \Stau{s}^{\pm}\Stau{s^{\prime}}^{\mp}$ in \citere{ChDrGaLe2001}. 
Several works dealt with the slepton production cross section at
threshold~\cite{FrMiZe2001,Fr2002}, also taking into account the 
electron/positron polarization~\cite{BlFrMoPo2002}, which is outside the 
scope of this article.
Full one-loop corrections in the rMSSM were presented for
$e^+e^- \to \Sele{s}^{\pm} \Sele{s^{\prime}}^{\mp}, 
            \Smue{s}^{\pm} \Smue{s^{\prime}}^{\mp}$~\cite{FrMaZe2004}
and for \eeSnSn~\cite{FrMaZe2005}. 
Using a renormalization scheme close to ours (see below), 
stop, sbottom, and stau production at $e^+e^-$ colliders was evaluated 
in \citere{ArHo2003} in the rMSSM at the full one-loop level.
Third generation sfermion production at the full one-loop level in the rMSSM,
including staus and tau sneutrinos were presented in
\citeres{KoWeEbMa2004,Ko2005,KoWeEbMa2005}. 

\medskip

In this paper we present for the first time a full and consistent 
one-loop calculation in the cMSSM for scalar lepton production at 
$e^+e^-$ colliders.  We take into account soft, hard and collinear QED 
radiation and the treatment of collinear divergences.  Again, here it 
is crucial to stress that the same renormalization scheme as for the 
decay of sleptons~\cite{Stau2decay}
(and for slepton production from Higgs boson decays~\cite{HiggsDecaySferm} 
as well as from chargino and neutralino 
decays~\cite{LHCxC1,LHCxC2,LHCxN,LHCxNprod}) has been used. 
Consequently, the predictions for the production and decay can be used 
together in a consistent manner (\eg, in a global phenomenological 
analysis of the slepton sector at the one-loop level). 
We analyze all processes \wrt the most relevant parameters, including the 
relevant complex phases.  In this way we go substantially beyond the 
existing analyses (see above).
In \refse{sec:calc} we briefly review the renormalization of the
relevant sectors of the cMSSM and give details as regards the calculation.
In \refse{sec:comparisons} various comparisons with results from other
groups are given.  The numerical results for the production channels 
(\ref{eq:eeSeSe}) and (\ref{eq:eeSnSn}) are presented in \refse{sec:numeval}.
The conclusions can be found in \refse{sec:conclusions}.


\subsection*{Prolegomena}

We use the following short-hands in this paper:
\begin{itemize}

\item \FT\ $\equiv$ \FA\ + \FC\ + \LT.

\item full = tree + loop.

\item $\SW \equiv \sin\theta_W$, $\CW \equiv \cos\theta_W$.

\item $\TB \equiv \tb$.

\end{itemize}
They will be further explained in the text below.
\vfill


\section{Calculation of diagrams}
\label{sec:calc}

In this section we give some details regarding the renormalization
procedure and the calculation of the tree-level and higher-order 
corrections to the production of sleptons in $e^+e^-$ collisions. 
The diagrams and corresponding amplitudes have been obtained with \FA\ 
(version 3.9) \cite{feynarts1,feynarts2,feynarts3}, using our MSSM model 
file (including the MSSM counterterms) of \citere{MSSMCT}. 
The further evaluation has been performed with \FC\ (version 9.5) and 
\LT\ (version 2.14)~\cite{formcalc1,formcalc2}.


\subsection{The complex MSSM}
\label{sec:renorm}

The cross sections (\ref{eq:eeSeSe}) and (\ref{eq:eeSnSn}) are calculated 
at the one-loop level, including soft, hard and collinear QED radiation; 
see the next section.  This requires the simultaneous renormalization of the 
gauge-boson sector, the lepton sector as well as the slepton sector of the 
cMSSM.  We give a few relevant details as regards these sectors and their 
renormalization.  More details and the application to Higgs-boson and SUSY 
particle decays can be found in 
\citeres{HiggsDecaySferm,HiggsDecayIno,MSSMCT,SbotRen1,SbotRen2,Stop2decay,%
Gluinodecay,Stau2decay,LHCxC1,LHCxC2,LHCxN,LHCxNprod}.  Similarly, the 
application to Higgs-boson and chargino/neutralino production cross sections 
at $e^+e^-$ colliders are given in \citeres{HiggsProd,HpProd,eeIno}.

The renormalization of the fermion and gauge-boson sector follows strictly 
\citere{MSSMCT} and the references therein (see especially \citere{mhcMSSMlong}). 
This defines in particular the counterterm $\dTB$, as well as the counterterms 
for the $Z$~boson mass, $\de\MZ^2$, and for the sine of the weak mixing angle, 
$\de\SW$ (with $\SW = \sqrt{1 - \CW^2} = \sqrt{1 - \MW^2/\MZ^2}$, where $\MW$ 
and $\MZ$ denote the $W$~and $Z$~boson masses, respectively).
For the fermion sector we use the default values as given in \citere{MSSMCT}. 

The renormalization of the slepton sector is implemented just as the
``$m_b, A_b$ \DRbar'' (RS2) scheme of \citeres{SbotRen1,SbotRen2,Stop2decay}, 
but extended to sleptons and all generations, also including the 
corresponding sfermion shifts.%
\footnote{
  The main difference between the renormalization in \citere{MSSMCT} 
  and the one used in this paper is that we impose a further on-shell 
  renormalization condition for the $\Fd_g$- and $\Fe_g$-type sfermion 
  masses, including an explicit restoration of the $SU(2)_L$ relation;
  see below.
}

The up-type squarks
($\Su_g = \{\tilde u, \tilde c, \tilde t\}$),
the neutrino-type sleptons
($\Sn_g = \{\tilde \nu_e, \tilde \nu_\mu, \tilde\nu_\tau\}$),
the down-type squarks 
($\Sd_g = \{\tilde d, \tilde s, \tilde b\}$),
and the electron-type sleptons
($\Se_g = \{\tilde e, \tilde\mu, \tilde\tau\}$)
are renormalized on-shell.  
The latter two via option $\mathcal{O}2$ of \citeres{SbotRen1,SbotRen2}.

The ``$m_b, A_b$ \DRbar'' scheme of \citeres{SbotRen1,SbotRen2,Stop2decay}, 
extended to all generations is herein after referred to as 
\textbf{mixed scheme}, of course not to be confused with the mixed 
scheme of \citere{MSSMCT}.

The schemes affecting $\Se_g$ and $\Sd_g$ are chosen with the 
variable \Code{\$SfScheme[t,\,\Vg]}:
\begin{subequations}
\label{eq:sqscheme}
\begin{alignat}{2}
\Code{\$SfScheme[2,\,\Vg]}~ &\Code{= DR[s]} &\qquad
	& \text{mixed scheme with $A_{\Fe_g}$ \DRbar} \notag \\
\Code{\$SfScheme[4,\,\Vg]}~ &\Code{= DR[s]} &\qquad
	& \text{mixed scheme with $A_{\Fd_g}$ \DRbar} \notag
\end{alignat}
\end{subequations}

The sfermions are on-shell, \ie the sfermion index \Vs\ 
runs over both values 1, 2.
\begin{subequations}
\label{eq:dMSf}
\begin{alignat}{2}
\Code{dMSfsq1[1,\,1,\,1,\,\Vg]} &\equiv{} & \delta\msn{g1}^2 
	&= \ReTilde\mati{\SE{\Sn_g}(\msn{g1}^2)}_{11}\,, \\
\Code{dMSfsq1[\Vs,\,\Vs,\,2,\,\Vg]} &\equiv{} & \delta\mse{gs}^2 
	&= \ReTilde\mati{\SE{\Se_g}(\mse{gs}^2)}_{ss}\,, \\
\Code{dMSfsq1[\Vs,\,\Vs,\,3,\,\Vg]} &\equiv {} & \delta\msu{gs}^2
	&= \ReTilde\mati{\SE{\Su_g}(\msu{gs}^2)}_{ss}\,, \\
\Code{dMSfsq1[\Vs,\,\Vs,\,4,\,\Vg]} &\equiv {} & \delta\msd{gs}^2
	&= \ReTilde\mati{\SE{\Sd_g}(\msd{gs}^2)}_{ss}\,.
\end{alignat}
The non-diagonal entries of the up-type mass matrix are determined 
by \cite{SbotRen1,SbotRen2}
\begin{alignat}{2}
\Code{dMSfsq1[1,\,2,\,3,\,\Vg]} &\equiv \delta Y_{\Fu_g}
	= \frac 12\ReTilde\mati{\SE{\Su_g}(\msu{g1}^2) +
	                         \SE{\Su_g}(\msu{g2}^2)}_{12}\,, \\
\Code{dMSfsq1[2,\,1,\,3,\,\Vg]}	&\equiv \delta Y_{\Fu_g}^*
	= \frac 12\ReTilde\mati{\SE{\Su_g}(\msu{g1}^2) +
	                         \SE{\Su_g}(\msu{g2}^2)}_{21}\,.
\end{alignat}
\end{subequations}
For clarity of notation we furthermore define the auxiliary constants
\begin{subequations}
\begin{align}
\label{eq:SlepAux}
\Code{dMsq12Sf1[2,\,\Vg]} \equiv \delta M_{\Se_g,12}^2
	&= \mfe{g} (\delta A_{\Fe_g}^* - \mu\,\dTB - \TB\,\delta\mu) +
	   (A_{\Fe_g}^* - \mu\,\TB)\,\delta\mfe{g}\,, \\
\Code{dMsq12Sf1[4,\,\Vg]} \equiv \delta M_{\Sd_g,12}^2
	&= \mfd{g} (\delta A_{\Fd_g}^* - \mu\,\dTB - \TB\,\delta\mu) +
	   (A_{\Fd_g}^* - \mu\,\TB)\,\delta\mfd{g}\,.
\end{align}
\end{subequations}
For the bottom quark we choose: $\delta\mb = \delta\mb^{\DRbar}$.

In the \textbf{mixed scheme} the down-type off-diagonal mass 
counterterms are related as
\begin{subequations}
\begin{align}
\label{eq:SlepY}
\Code{dMSfsq1[1,\,2,\,2,\,\Vg]} &\equiv \delta Y_{\Fe_g}
	= \frac 1{|U^{\Se_g}_{11}|^2 - |U^{\Se_g}_{12}|^2}
	\Bigl\{
	  U^{\Se_g}_{11} U^{\Se_g*}_{21}
	  \bigl(\delta\mse{g1}^2 - \delta\mse{g2}^2\bigr) + {} \\[-1ex]
&\kern 12em
	  U^{\Se_g}_{11} U^{\Se_g*}_{22} \delta M_{\Se_g,12}^2 -
	  U^{\Se_g}_{12} U^{\Se_g*}_{21} \delta M_{\Se_g,12}^{2*}
	\Bigr\}\,, \notag \\
\Code{dMSfsq1[2,\,1,\,2,\,\Vg]} &= \delta Y_{\Fe_g}^*\,, \\
\Code{dMSfsq1[1,\,2,\,4,\,\Vg]} &\equiv \delta Y_{\Fd_g}
	= \frac 1{|U^{\Sd_g}_{11}|^2 - |U^{\Sd_g}_{12}|^2}
	\Bigl\{
	  U^{\Sd_g}_{11} U^{\Sd_g*}_{21}
	  \bigl(\delta\msd{g1}^2 - \delta\msd{g2}^2\bigr) + {} \\[-1ex]
&\kern 12em
	  U^{\Sd_g}_{11} U^{\Sd_g*}_{22} \delta M_{\Sd_g,12}^2 -
	  U^{\Sd_g}_{12} U^{\Sd_g*}_{21} \delta M_{\Sd_g,12}^{2*}
	\Bigr\}\,, \notag \\
\Code{dMSfsq1[2,\,1,\,4,\,\Vg]} &= \delta Y_{\Fd_g}^*\,.
\end{align}
\end{subequations}
The trilinear couplings $A_{f_{tg}}\equiv\mati{\matr{A}_{f_t}}_{gg}$ are 
renormalized by
\begin{subequations}
\begin{alignat}{2}
\Code{dAf1[2,\,\Vg,\,\Vg]} &\equiv \delta A_{\Fe_g} &
	&= \biggl\{\frac 1{\mfe{g}}\Bigl[
	\begin{aligned}[t]
	& U^{\Se_g}_{11} U^{\Se_g*}_{12}
	  (\delta\mse{g1}^2 - \delta\mse{g2}^2) + {} \\
	& U^{\Se_g}_{11} U^{\Se_g*}_{22} \delta Y_{\Fe_g}^* +
	  U^{\Se_g*}_{12} U^{\Se_g}_{21} \delta Y_{\Fe_g} -
	  (A_{\Fe_g} - \mu^*\TB)\,\delta\mfe{g}
	\rlap{$\Bigr] + {}$}
	\end{aligned} \\
&&	&\qquad \TB\,\delta\mu^* + \mu^*\dTB
        \biggr\}_{\mathrm{[div]}}\,, \notag \\[1ex] 
\Code{dAf1[3,\,\Vg,\,\Vg]} &\equiv \delta A_{\Fu_g} &
	&= \frac 1{\mfu{g}} \Bigl[
	\rlap{$\begin{aligned}[t]
	& U^{\Su_g}_{11} U^{\Su_g*}_{12}
	    (\delta\msu{g1}^2 - \delta\msu{g2}^2) + {} \\
	& U^{\Su_g}_{11} U^{\Su_g*}_{22} \delta Y_{\Fu_g}^* +
	  U^{\Su_g*}_{12} U^{\Su_g}_{21} \delta Y_{\Fu_g} -
	  \bigl(A_{\Fu_g} - \mu^*/\TB\bigr)\,\delta\mfu{g}
	\rlap{$\Bigr] + {}$}
	\end{aligned}$} \\
&&	&\qquad \delta\mu^*/\TB - \mu^* \dTB/\TB^2\,, \notag \\[1ex]
\Code{dAf1[4,\,\Vg,\,\Vg]} &\equiv \delta A_{\Fd_g} &
	&= \biggl\{\frac 1{\mfd{g}}\Bigl[
	\rlap{$\begin{aligned}[t]
	& U^{\Sd_g}_{11} U^{\Sd_g*}_{12}
	  (\delta\msd{g1}^2 - \delta\msd{g2}^2) + {} \\
	& U^{\Sd_g}_{11} U^{\Sd_g*}_{22} \delta Y_{\Fd_g}^* +
	  U^{\Sd_g*}_{12} U^{\Sd_g}_{21} \delta Y_{\Fd_g} -
	  (A_{\Fd_g} - \mu^*\TB)\,\delta\mfd{g}
	\rlap{$\Bigr] + {}$}
	\end{aligned}$} \\
&&	&\qquad \TB\,\delta\mu^* + \mu^*\dTB
        \biggr\}_{\mathrm{[div]}}\,, \notag
\end{alignat}
\end{subequations}
where the subscripted [div] means to take the divergent part in the 
mixed scheme only, to effect \DRbar\ renormalization of $A_{\Fe_g}$ 
and $A_{\Fd_g}$ \cite{SbotRen1,SbotRen2}.

The squark and slepton $Z$-factors are derived in the OS scheme and can 
be found in Section 3.6.1 and 3.6.2 of \citere{MSSMCT}.

\medskip

As now all the sfermion masses are renormalized as on-shell an explicit 
restoration of the $SU(2)_L$~relation is needed.  Requiring the $SU(2)_L$
relation to be valid at the one-loop level induces the following
shifts in the soft SUSY-breaking parameters:
\begin{align}
\label{eq:SfShiftL}
M_{\tilde L}^2(\Se_g) &= M_{\tilde L}^2(\Sn_g) + 
                      \de M_{\tilde L}^2(\Sn_g) - \de M_{\tilde L}^2(\Se_g)\,, \\
M_{\tilde Q}^2(\Sd_g) &= M_{\tilde Q}^2(\Su_g) +
                      \de M_{\tilde Q}^2(\Su_g) - \de M_{\tilde Q}^2(\Sd_g)
\label{eq:SfShiftQ}
\end{align}
with
\begin{align}
\label{eq:dSfShift}
\de M_{\tilde L, \tilde Q}^2(\Sf) &= |U^{\Sf}_{11}|^2 \de\msf1^2 + 
                                |U^{\Sf}_{12}|^2 \de\msf2^2 - 
                                 U^{\Sf}_{22} U^{\Sf *}_{12} \de Y_f - 
                                 U^{\Sf}_{12} U^{\Sf *}_{22} \de Y_f^* - 
                                 2 \mf{} \de \mf{} \notag \\
&\quad  + \MZ^2\, \CBB\, Q_f\, \de \SW^2 
        - (I_f^3 - Q_f \SW^2) (\CBB\, \de \MZ^2 + \MZ^2\, \de \CBB)\,.
\end{align}
Now $M_{\tilde L}^2(\Se_g)$ and $M_{\tilde Q}^2(\Sd_g)$ are used in the scalar 
electron- and down-type mass matrix instead of the parameters 
$M_{\tilde L, \tilde Q}^2$ in the sfermion mass matrix when calculating the 
values of $\mse{gs}$ and $\msd{gs}$. 
However, with this procedure, both ($s = 1,2$) sfermion masses are shifted, 
which contradicts our choice of independent parameters. 
To keep this choice, also the right-handed soft SUSY-breaking mass parameters
$M_{\tilde E, \tilde D}$ receive a shift:
\begin{align}
\label{eq:BackShiftE}
M_{\tilde E}^2 &= \frac{\mfe{g}^2\, |A_{\Fe_g}^* - \mu\, \TB|^2}
               {M_{\tilde L}^2(\Se_{g}) + \mfe{g}^2 + 
               \MZ^2 \CBB (I_{\Fe_g}^3 - Q_{\Fe_g} \SW^2) - \mse{gs}^2} - 
               \mfe{g}^2 - \MZ^2 \CBB Q_{\Fe_g} \SW^2 + \mse{gs}^2 \,, \\
M_{\tilde D}^2 &= \frac{\mfd{g}^2\, |A_{\Fd_g}^* - \mu\, \TB|^2}
               {M_{\tilde Q}^2(\Sd_{g}) + \mfd{g}^2 + 
               \MZ^2 \CBB (I_{\Fd_g}^3 - Q_{\Fd_g} \SW^2) - \msd{gs}^2} - 
               \mfd{g}^2 - \MZ^2 \CBB Q_{\Fd_g} \SW^2 + \msd{gs}^2
\label{eq:BackShiftD}
\end{align}
with our choice of mass ordering, $\msf1 < \msf2$, we have
\begin{align}
s &= \left\{
\begin{array}{rl}
1 & \text{for } M_{\tilde L}^2 > M_{\tilde E}^2 \text{ and/or }
                M_{\tilde Q}^2 > M_{\tilde D}^2\,, \\
2 & \text{for } M_{\tilde E}^2 > M_{\tilde L}^2 \text{ and/or } 
                M_{\tilde D}^2 > M_{\tilde Q}^2\,.
\end{array} \right.
\end{align}
Taking into account the shift \refeq{eq:BackShiftE} in $M_{\tilde E}$
and \refeq{eq:BackShiftD} in $M_{\tilde D}$, up to one-loop order%
\footnote{
  In the case of a pure OS scheme for the rMSSM the shifts 
  \refeqs{eq:SfShiftL}, (\ref{eq:SfShiftQ}), (\ref{eq:BackShiftE}), 
  and (\ref{eq:BackShiftD}) result in mass parameters $\msd{gs}$ and 
  $\msd{gs}$, which are exactly the same as in \refeqs{eq:mseOS} and 
  (\ref{eq:msdOS}).  This constitutes an important consistency check 
  of these two different methods.
}%
, the new resulting mass parameters $\mse{gs}$ and $\msd{gs}$ are the 
same as the on-shell masses:
\begin{align}
\label{eq:mseOS}
\big(\mse{gs}^{\OS}\big)^2 &= \big(\mse{gs}^{}\big)^2 +
                             \big(\de\mse{gs}^{\text{dep.}}\big)^2 -
                             \ReTilde\mati{\SE{\Se_g}(\mse{gs}^2)}_{ss}\,, \\
\big(\msd{gs}^{\OS}\big)^2 &= \big(\msd{gs}^{}\big)^2 +
                             \big(\de\msd{gs}^{\text{dep.}}\big)^2 -
                             \ReTilde\mati{\SE{\Sd_g}(\msd{gs}^2)}_{ss}
\label{eq:msdOS}
\end{align}
where $\de\mse{gs}^{\text{dep.}}$ and $\de\msd{gs}^{\text{dep.}}$ are the
dependent mass counterterms.

A slightly different slepton sector renormalization is also described in
detail in \citere{MSSMCT} and references therein.


\subsection{Contributing diagrams}
\label{sec:diagrams}

\begin{figure}
\begin{center}
\framebox[14cm]{\includegraphics[width=0.24\textwidth]{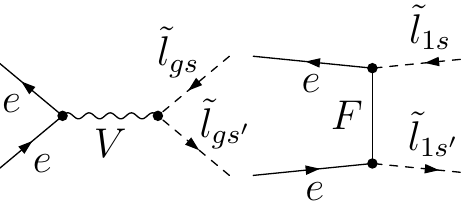}}
\framebox[14cm]{\includegraphics[width=0.75\textwidth]{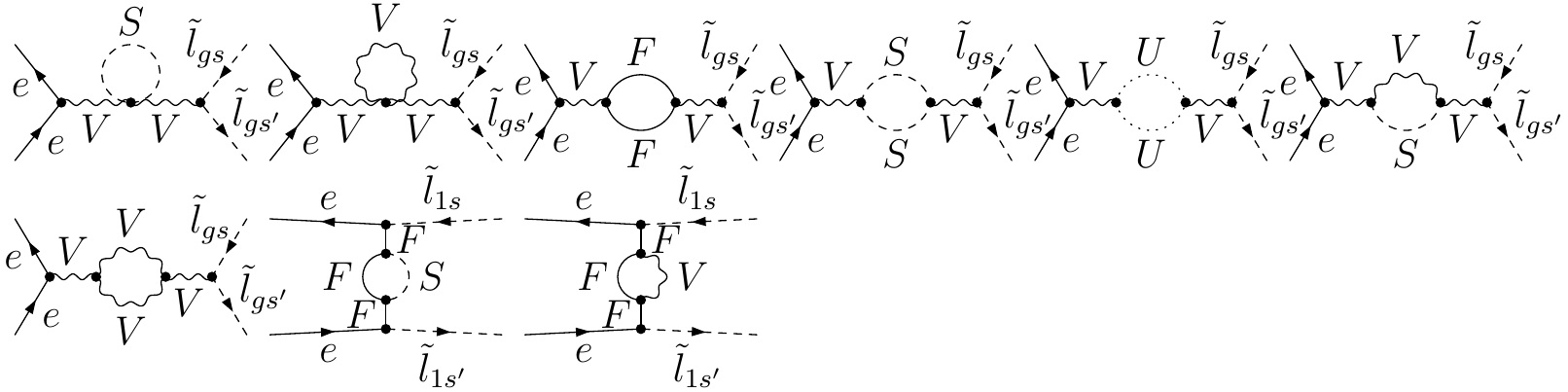}}
\framebox[14cm]{\includegraphics[width=0.75\textwidth]{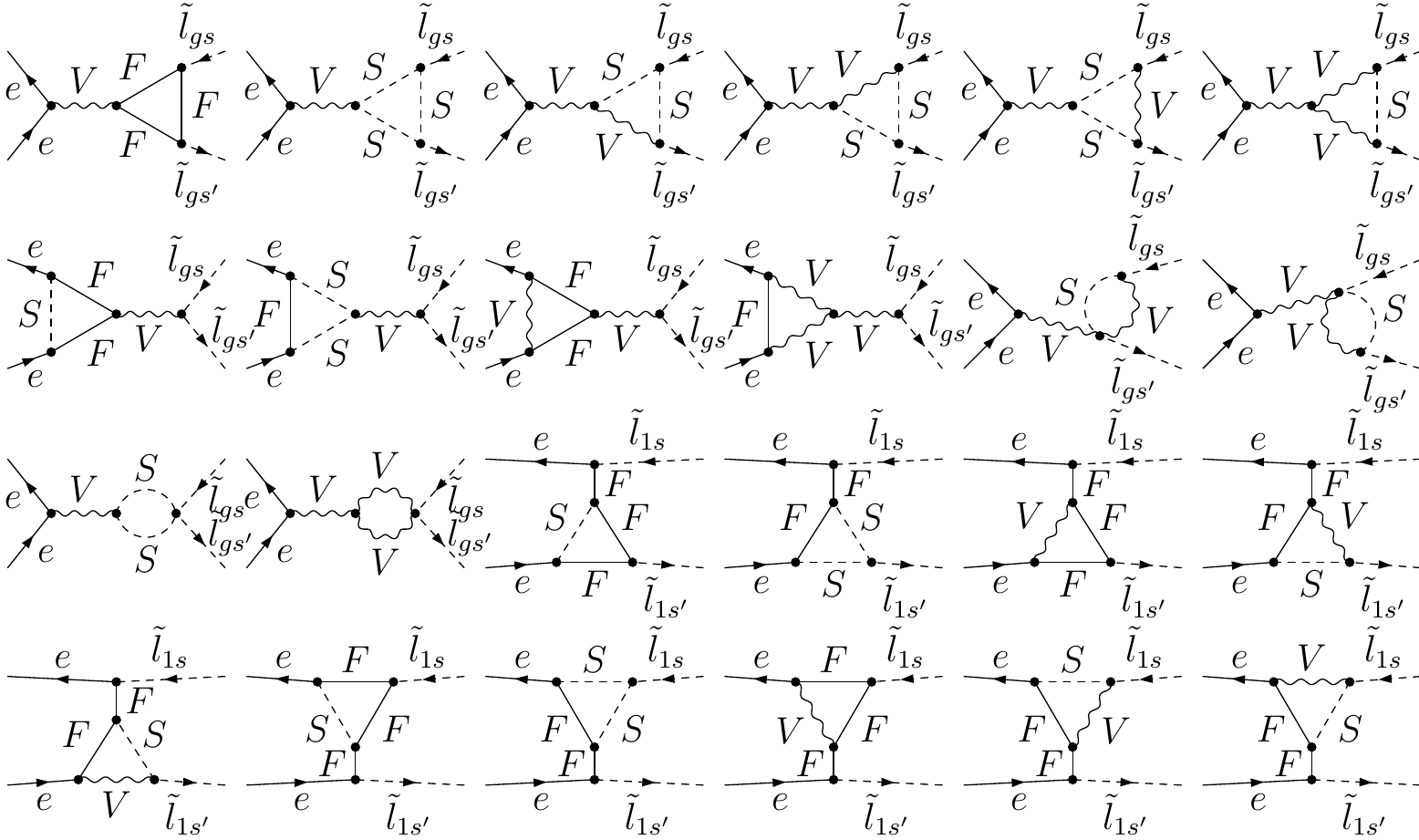}}
\framebox[14cm]{\includegraphics[width=0.75\textwidth]{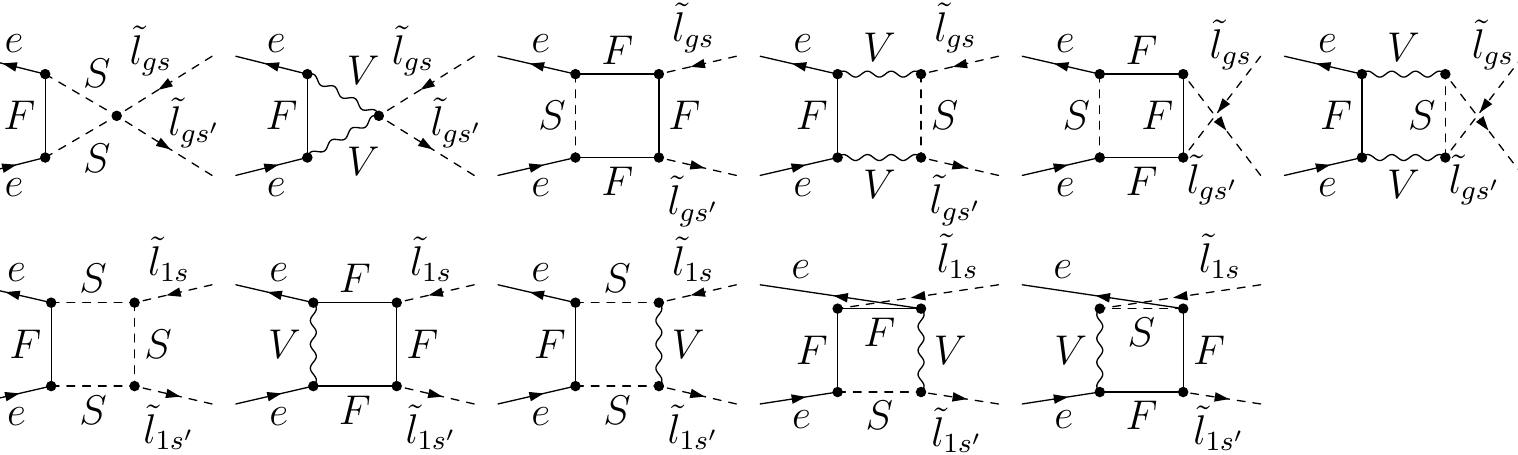}}
\framebox[14cm]{\includegraphics[width=0.75\textwidth]{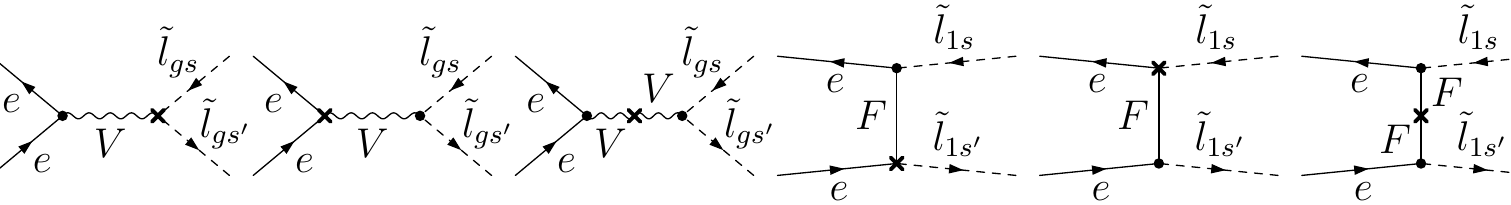}}
\caption{
  Generic tree, self-energy, vertex, box, and counterterm diagrams for the 
  process \eeSlSl\ ($\Sl_{gs} = \{\Se_{gs},\Sn_g\};\; g = 1,2,3;\; s,\spri = 1,2$). 
  The additional diagrams, which occur only in the case of first generation 
  slepton production, are denoted with $\Sl_{1s}$.
  $F$ can be a SM fermion, chargino or neutralino; 
  $S$ can be a sfermion or a Higgs/Goldstone boson; 
  $V$ can be a $\ga$, $Z$ or $W^\pm$. 
  It should be noted that electron--Higgs couplings are neglected. 
}
\label{fig:eeSlSl}
\end{center}
\end{figure}

Sample diagrams for the process \eeSeSe\ and \eeSnSn\ are shown in 
\reffi{fig:eeSlSl}.
Not shown are the diagrams for real (hard and soft) photon radiation. 
They are obtained from the corresponding tree-level diagrams by attaching a 
photon to the (incoming/outgoing) electron or slepton.
The internal particles in the generically depicted diagrams in 
\reffi{fig:eeSlSl} are labeled as follows: 
$F$ can be a SM fermion $f$, chargino $\cha{c}$ or neutralino 
$\neu{n}$; $S$ can be a sfermion $\Sf_s$ or a Higgs (Goldstone) boson 
$h^0, H^0, A^0, H^\pm$ ($G, G^\pm$); $U$ denotes the ghosts $u_V$;
$V$ can be a photon $\ga$ or a massive SM gauge boson, $Z$ or $W^\pm$. 
We have neglected all electron--Higgs couplings and terms proportional 
to the electron mass whenever this is safe, \ie except when the electron 
mass appears in negative powers or in loop integrals.
We have verified numerically that these contributions are indeed totally 
negligible.  For internally appearing Higgs bosons no higher-order
corrections to their masses or couplings are taken into account; 
these corrections would correspond to effects beyond one-loop order.%
\footnote{
  We found that using loop corrected Higgs-boson masses 
  in the loops leads to a UV divergent result.
}

Moreover, in general, in \reffi{fig:eeSlSl} we have omitted diagrams 
with self-energy type corrections of external (on-shell) particles. 
While the contributions from the real parts of the loop functions are 
taken into account via the renormalization constants defined by OS 
renormalization conditions, the contributions coming from the imaginary 
part of the loop functions can result in an additional (real) correction 
if multiplied by complex parameters.  In the analytical and numerical 
evaluation, these diagrams have been taken into account via the 
prescription described in \citere{MSSMCT}.

Within our one-loop calculation we neglect finite width effects that 
can help to cure threshold singularities.  Consequently, in the close 
vicinity of those thresholds our calculation does not give a reliable
result.  Switching to a complex mass scheme \cite{complexmassscheme} 
would be another possibility to cure this problem, but its application 
is beyond the scope of our paper.

The tree-level formulas $\sigtree(\eeSeSe)$ and $\sigtree(\eeSnSn)$ 
are given in \citeres{CaDi1995,FrMiZe2001} and \citere{FrMaZe2005}, 
respectively. Concerning our evaluation of $\sig(\eeSeSe)$ we define:
\begin{align}
\label{eq:eeSeSesum}
\sig(\eeSeSe) \equiv \sig(e^+e^- \to \Se^+_{gs} \Se^-_{gs^{\prime}}) + 
                     \sig(e^+e^- \to \Se^-_{gs} \Se^+_{gs^{\prime}})
                     \qquad \forall\; s \neq s^{\prime}\,, 
\end{align}
if not indicated otherwise.  Differences between the two charge conjugated 
processes can appear at the loop level when complex parameters are taken 
into account, as will be discussed in \refse{sec:eeSeSe}.


\subsection{Ultraviolet, infrared and collinear divergences}

As regularization scheme for the UV divergences we have used constrained 
differential renormalization~\cite{cdr}, which has been shown to be 
equivalent to dimensional reduction~\cite{dred1,dred2} at the \onel\ 
level~\cite{formcalc1,formcalc2}. 
Thus the employed regularization scheme preserves SUSY~\cite{dredDS,dredDS2}
and guarantees that the SUSY relations are kept intact, \eg, that the gauge 
couplings of the SM vertices and the Yukawa couplings of the corresponding 
SUSY vertices also coincide to \onel\ order in the SUSY limit. 
Therefore no additional shifts, which might occur when using a different 
regularization scheme, arise.  All UV divergences cancel in the final result.

\begin{figure}
\centering
\includegraphics[width=0.49\textwidth,height=7.5cm]{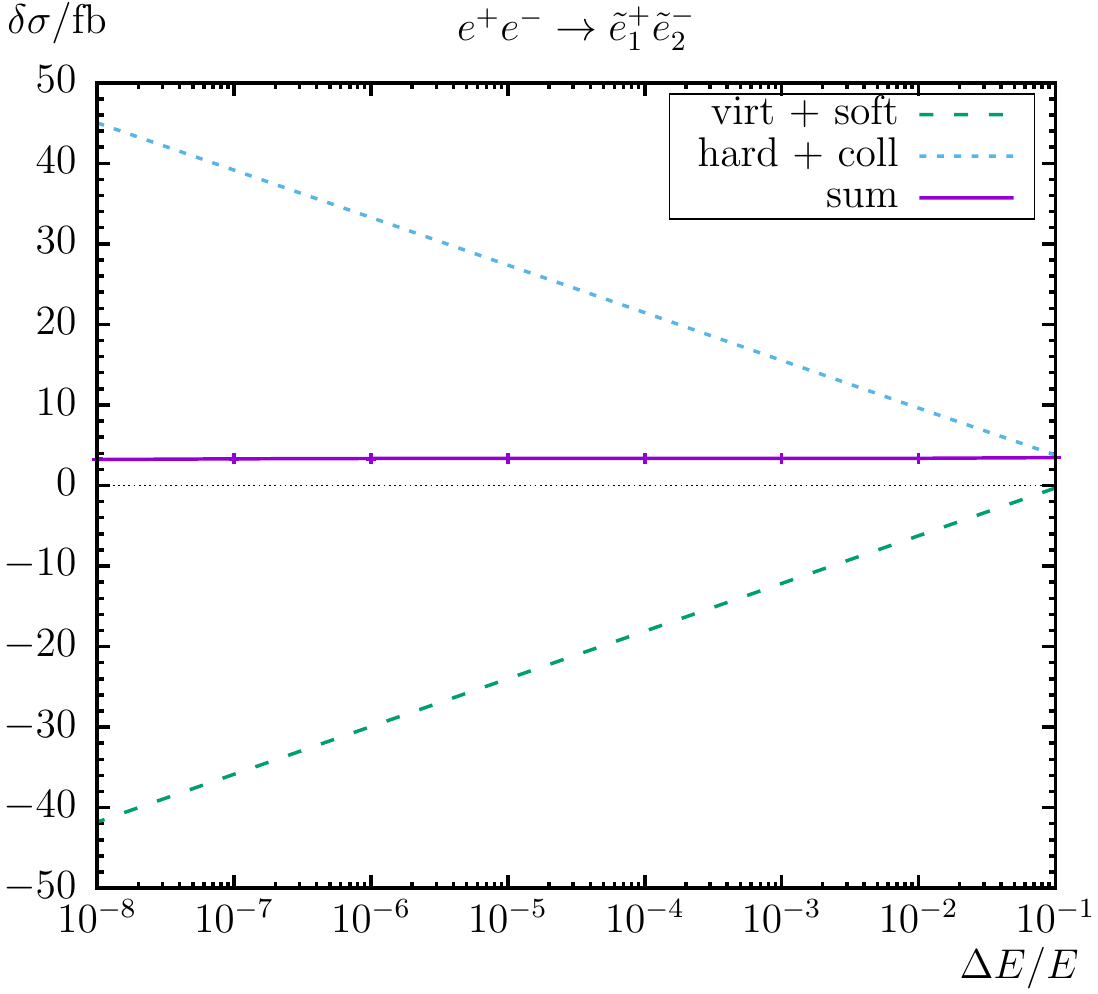}
\vspace{1em}
\begin{minipage}[b]{0.4\textwidth}
\centering
\begin{tabular}[b]{lr}
\toprule
$\Delta E/E$ & $\delta\sig$/fbarn \\
\midrule
$10^{-1}$ & $ 3.460 \pm 0.003$ \\
$10^{-2}$ & $ 3.365 \pm 0.007$ \\
$10^{-3}$ & $ 3.358 \pm 0.011$ \\
$10^{-4}$ & $ 3.359 \pm 0.015$ \\
$10^{-5}$ & $ 3.361 \pm 0.020$ \\
$10^{-6}$ & $ 3.342 \pm 0.023$ \\
$10^{-7}$ & $ 3.299 \pm 0.028$ \\
$10^{-8}$ & $ 3.231 \pm 0.028$ \\
\bottomrule
\end{tabular}
\vspace{2em}
\end{minipage}
\caption{\label{fig:collE}
  Phase space slicing method.  The different contributions to the 
  one-loop corrections $\delta\sig(e^+e^- \to \tilde{e}^+_1 \tilde{e}^-_2)$ 
  for our input parameter scenario \Scs\ (see \refta{tab:para} below) as a 
  function of $\Delta E/E$ with fixed $\Delta \theta/\text{rad} = 10^{-2}$.
}
\end{figure}

Soft photon emission implies numerical problems in the phase space 
integration of radiative processes.  The phase space integral diverges 
in the soft energy region where the photon momentum becomes very small,
leading to infrared (IR) singularities.  Therefore the IR divergences from 
diagrams with an internal photon have to cancel with the ones from the 
corresponding real soft radiation.  We have included the soft photon contribution 
via the code already implemented in \FC\ following the description given 
in \citere{denner}.  The IR divergences arising from the diagrams involving 
a photon are regularized by introducing a photon mass parameter, $\la$. 
All IR divergences, \ie all divergences in the limit $\la \to 0$, cancel 
once virtual and real diagrams for one process are added. 
We have numerically checked that our results do not depend on $\la$ or 
on $\Delta E = \delta_s E = \delta_s \sqrt{s}/2$ defining the energy 
cut that separates the soft from the hard radiation.  As one can see
from the example in \reffi{fig:collE} this holds for several orders of 
magnitude.  Our numerical results below have been obtained for fixed 
$\delta_s = 10^{-3}$.

Numerical problems in the phase space integration of the radiative 
process arise also through collinear photon emission.  Mass singularities 
emerge as a consequence of the collinear photon emission off massless
particles.  But already very light particles (such as electrons) can 
produce numerical instabilities.
For the treatment of collinear singularities in the photon radiation off 
initial state electrons and positrons we used the 
\textit{phase space slicing method}~\cite{slicing1,slicing2,slicing3,slicing4}, 
which is not (yet) implemented in \FC\ and therefore we have developed and 
implemented the code necessary for the evaluation of collinear contributions; 
see also \citeres{HiggsProd,HpProd}.

\begin{figure}
\centering
\includegraphics[width=0.49\textwidth,height=7.5cm]{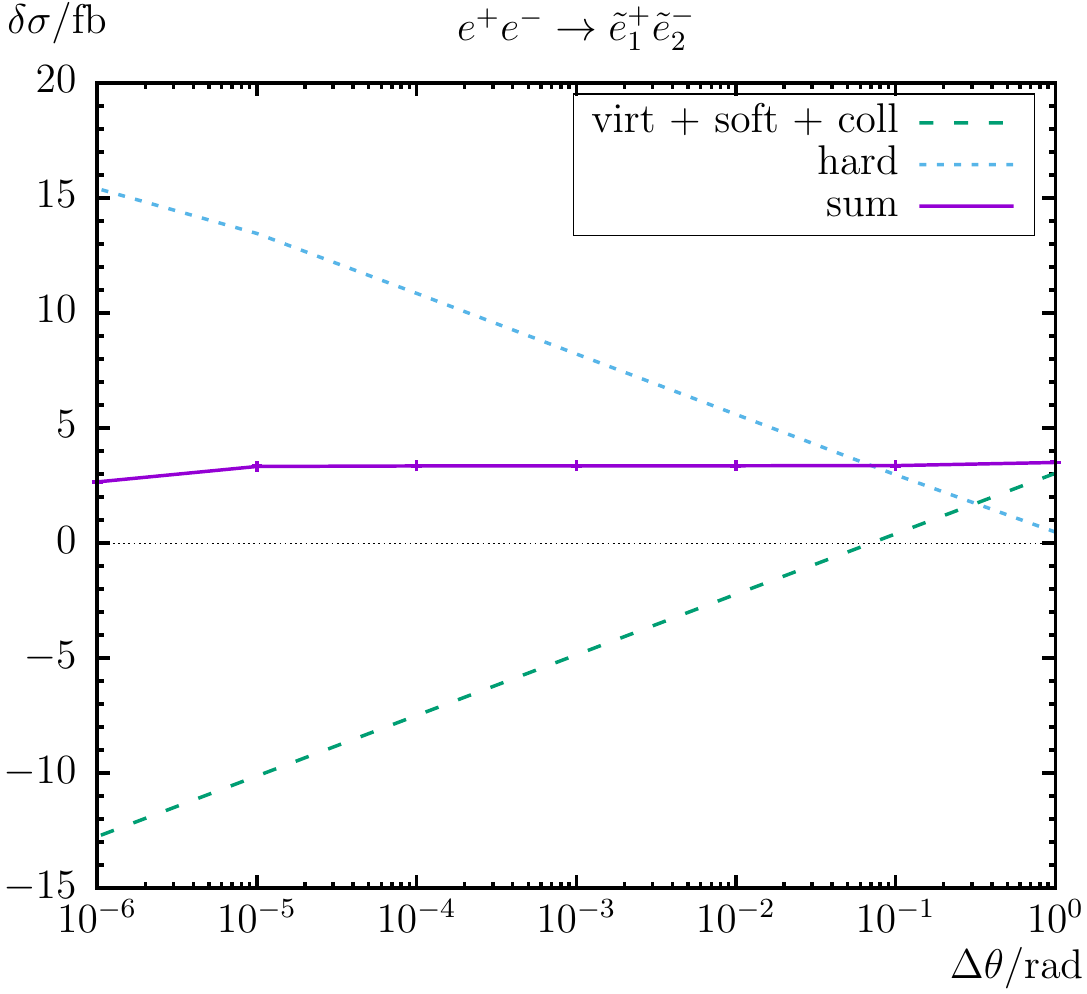}
\begin{minipage}[b]{0.4\textwidth}
\centering
\begin{tabular}[b]{lr}
\toprule
$\Delta \theta$/rad & $\delta\sig$/fbarn \\
\midrule
$10^{ 0}$ & $ 3.503 \pm 0.014$ \\
$10^{-1}$ & $ 3.362 \pm 0.012$ \\
$10^{-2}$ & $ 3.358 \pm 0.010$ \\
$10^{-3}$ & $ 3.356 \pm 0.011$ \\
$10^{-4}$ & $ 3.352 \pm 0.012$ \\
$10^{-5}$ & $ 3.329 \pm 0.013$ \\
$10^{-6}$ & $ 2.653 \pm 0.015$ \\
\bottomrule
\end{tabular}
\vspace{2em}
\end{minipage}
\caption{\label{fig:collT}
  Phase space slicing method.  The different contributions to the 
  one-loop corrections $\delta\sig(e^+e^- \to \tilde{e}^+_1 \tilde{e}^-_2)$ 
  for our input parameter scenario \Scs\ (see \refta{tab:para} below) as a 
  function of $\Delta\theta/\text{rad}$ with fixed $\Delta E/E = 10^{-3}$.
}
\end{figure}

In the phase space slicing method, the phase space is divided into
regions where the integrand is finite (numerically stable) and 
regions where it is divergent (or numerically unstable).
In the stable regions the integration is performed numerically, whereas
in the unstable regions it is carried out (semi-) analytically using 
approximations for the collinear photon emission.

The collinear part is constrained by the angular cut-off parameter 
$\Delta\theta$, imposed on the angle between the photon and the
(in our case initial state) electron/positron.

The differential cross section for the collinear photon radiation 
off the initial state $e^+e^-$ pair corresponds to a convolution
\begin{align}
\text{d}\sigcoll(s) = \frac{\alpha}{\pi} \int_0^{1-\delta_s} \text{d}z\,
  \text{d}\sigtree(\sqrt{z s}) \LV \LB 2\, \ln \LP 
  \frac{\Delta \theta \sqrt{s}}{2\, m_e} \RP - 1 \RB P_{ee}(z) + 1 - z \RV\,,
\end{align}
with $P_{ee}(z) = (1 + z^2)/(1 - z)$
denoting the splitting function of a photon from the initial $e^+e^-$ pair.
The electron momentum is reduced (because of the radiated photon) by 
the fraction $z$ such that the center-of-mass frame of the hard process 
receives a boost.  The integration over all possible factors $z$ is 
constrained by the soft cut-off $\delta_s = \Delta E/E$, to prevent 
over-counting in the soft energy region.

We have numerically checked that our results do not depend on the angular 
cut-off parameter $\Delta\theta$ over several orders of magnitude; 
see the example in \reffi{fig:collT}.  Our numerical results below have 
been obtained for fixed $\Delta \theta/\text{rad} = 10^{-2}$.

The one-loop corrections of the differential cross section are decomposed 
into the virtual, soft, hard, and collinear parts as follows:
\begin{align}
\text{d}\sigloop = \text{d}\sigvirt(\la) + 
                   \text{d}\sigsoft(\la, \Delta E) + 
                   \text{d}\sighard(\Delta E, \Delta\theta) + 
                   \text{d}\sigcoll(\Delta E, \Delta\theta)\,.
\end{align}
The hard and collinear parts have been calculated via Monte Carlo 
integration algorithms of the \texttt{CUBA} library \cite{cuba1,cuba2} 
as implemented in \FC~\cite{formcalc1,formcalc2}.


\section{Comparisons}
\label{sec:comparisons}

In this section we present the comparisons with results from other groups 
in the literature for slepton production in $e^+e^-$ collisions.
These comparisons were restricted to the MSSM with real parameters, with
one exception for tree-level tau slepton pair production.
The level of agreement of such comparisons (at one-loop order) depends on 
the correct transformation of the input parameters from our renormalization 
scheme into the schemes used in the respective literature, as well as on the 
differences in the employed renormalization schemes as such.
In view of the non-trivial conversions and the large number of comparisons 
such transformations and/or change of our renormalization prescription are
beyond the scope of our paper.  In the following we list all relevant papers 
in the literature and explain either our comparison, or why no (meaningful) 
comparison could be performed.

\begin{itemize}

\item
In \citere{SchWa1985} the production of slepton and squark pairs in 
$e^+e^-$ annihilation and $Z$ decay have been calculated in the rMSSM 
at tree level (including arbitrarily polarized beams). Unfortunately, 
in \citere{SchWa1985} not sufficient information as regards their 
input parameters where given, rendering a comparison impossible.

\item
Selectron pair production at $e^+e^-$ colliders in SUGRA models were 
presented in \citere{CaDi1995} at tree level.  We omitted a comparison with 
\citere{CaDi1995}, since implementing the SUGRA spectrum is beyond the 
scope of our paper. 

\item
Tree-level tau slepton pair production at muon colliders in the cMSSM were 
analyzed in \citere{ChDrGaLe2001} including $\CP$ violation. 
The center-of-mass energy was assumed to be around the resonances of the 
heavy neutral Higgs bosons, at $\sqrt{s} \simeq 500\gev$.
We used their input parameters as far as possible, but we differ from their 
results, especially in the case of complex input parameters. This can most 
likely be attributed to the differences in the Higgs-boson mass calculations 
employed in \citere{ChDrGaLe2001} and the current version 2.13.0 of 
\FH~\cite{mhcMSSMlong,FeynHiggs1,FeynHiggs2,FeynHiggs3,FeynHiggs5,%
FeynHiggs6,FeynHiggs7,FeynHiggs8}, which we use.  The cross section 
(close to the heavy Higgs boson thresholds) depends strongly on 
\textit{tiny} mass differences of the two heavy neutral Higgs bosons, which 
deviate clearly between the two employed calculations, rendering this 
comparison not significant.

\item
In \citeres{FrMiZe2001,Fr2002} pair production of smuons and selectrons 
near threshold in $e^+e^-$ and $e^-e^-$ collisions have been computed in 
the rMSSM, including Coulomb rescattering effects.%
\footnote{
  It should be noted that \citere{FrMiZe2001} 
  is mainly an extraction of \citere{Fr2002}.
}
Because these near production threshold effects are beyond the scope of 
our paper we have omitted a comparison with \citeres{FrMiZe2001,Fr2002}.

\item
\citere{BlFrMoPo2002} deals with selectron pair production at $e^-e^-$ 
and $e^+e^-$ colliders with polarized beams. The calculations have been 
computed in the rMSSM near the production threshold. Again, these effects 
are beyond the scope of our paper and we have omitted a comparison with 
\citere{BlFrMoPo2002}.

\begin{figure}[t]
\begin{center}
\begin{tabular}{c}
\includegraphics[width=0.48\textwidth,height=6cm]{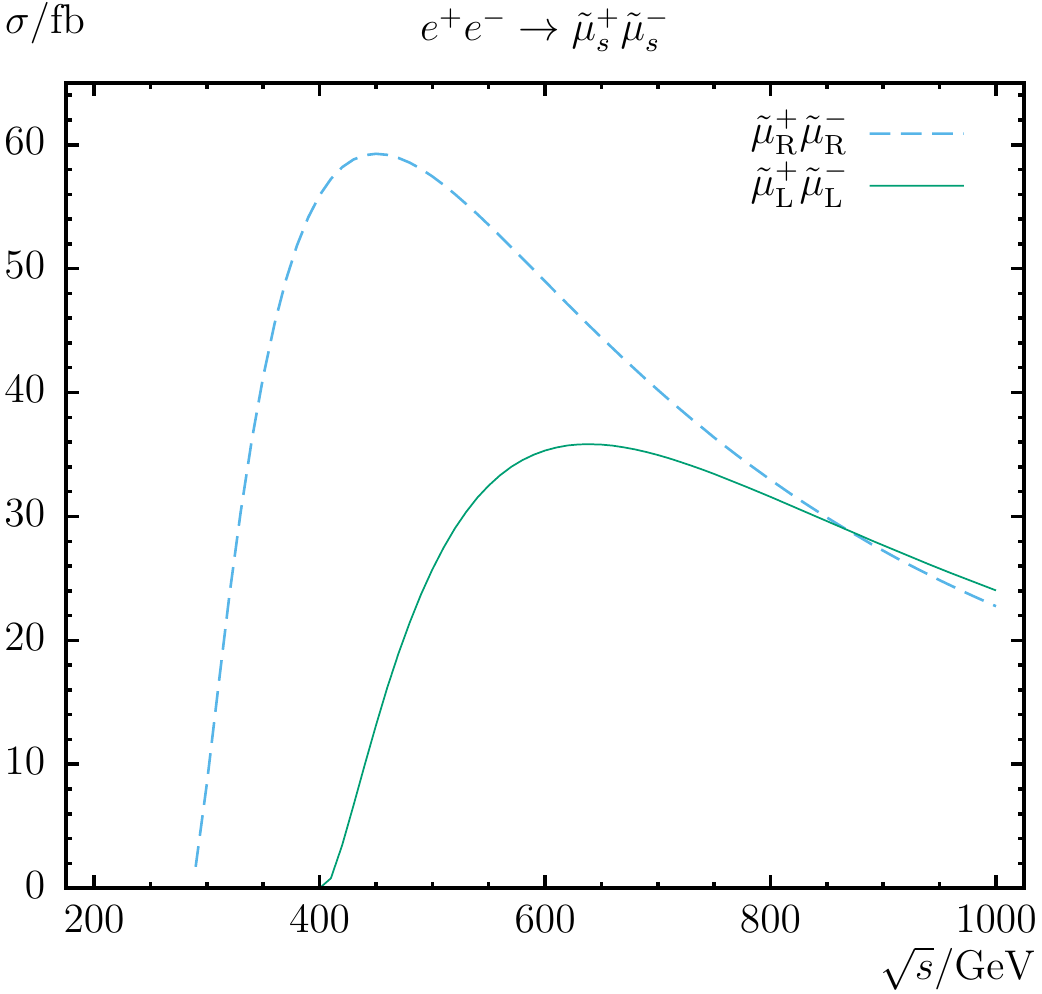}
\includegraphics[width=0.48\textwidth,height=6cm]{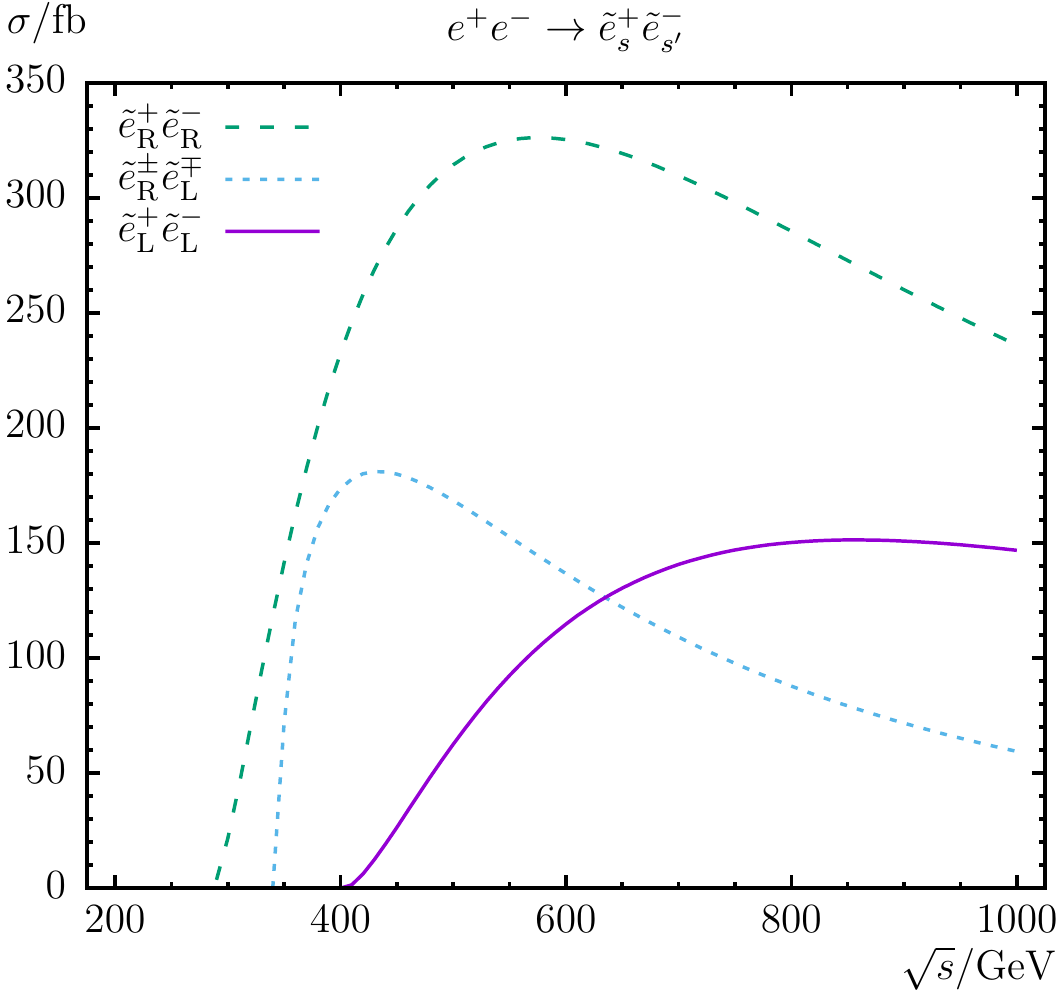}
\\[1em]
\includegraphics[width=0.48\textwidth,height=6cm]{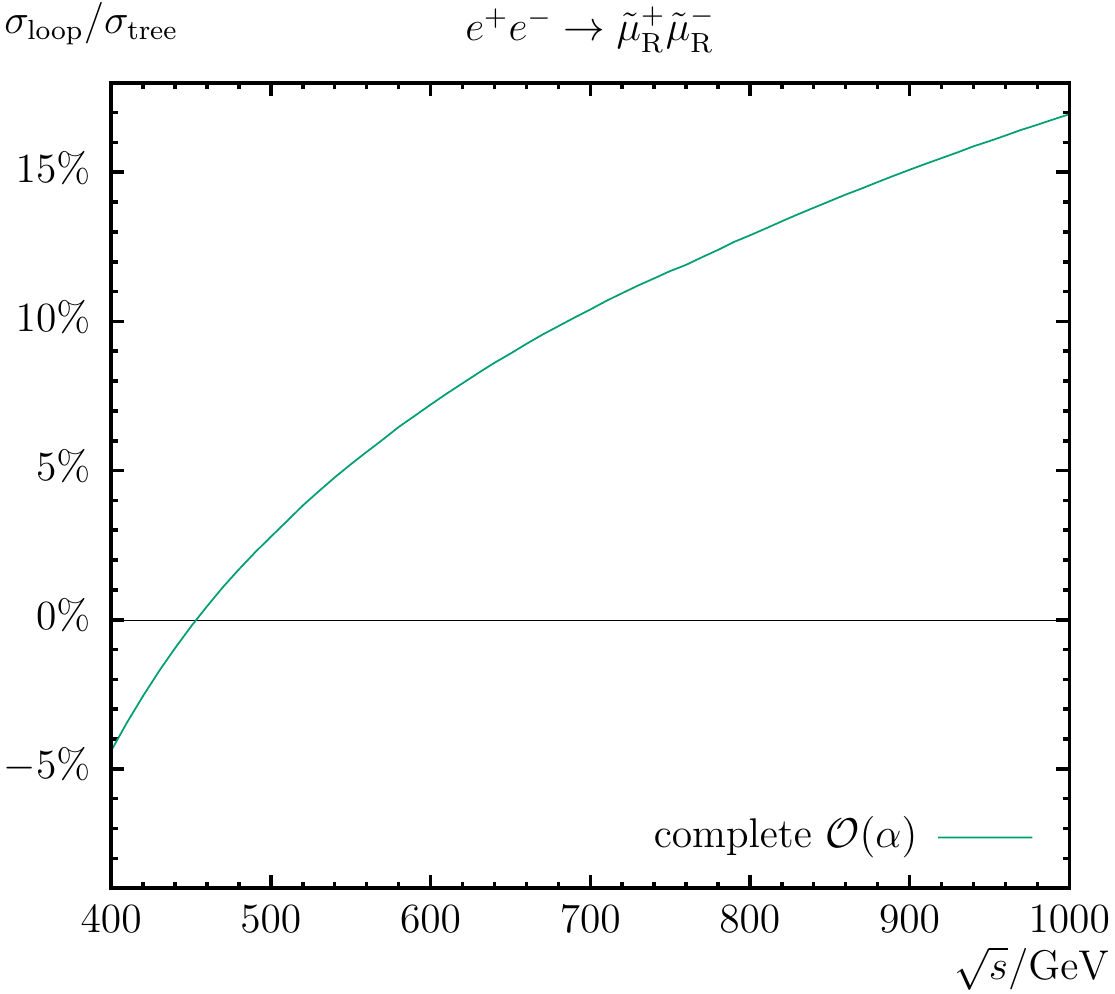}
\includegraphics[width=0.48\textwidth,height=6cm]{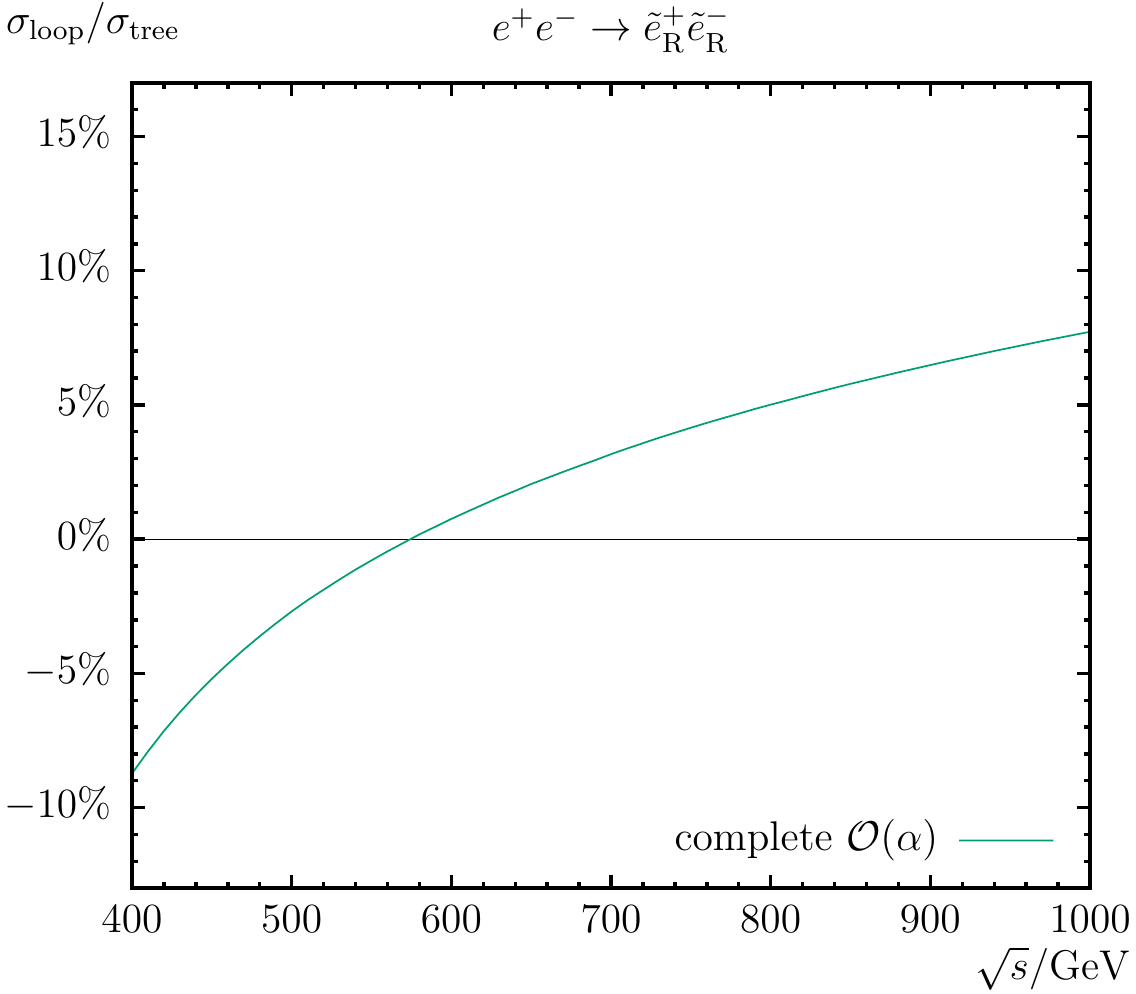}
\end{tabular}
\caption{\label{fig:FrMaZe2004}
  Comparison with \citere{FrMaZe2004} for 
  $e^+e^- \to \Sele{s} \Sele{s^{\prime}}, \Smue{s} \Smue{s^{\prime}}$.
  Born cross sections (upper row) and complete relative one-loop 
  results (lower row) are shown for the SPS$1a$ parameter point as 
  given according to \citere{FrMaZe2004} as a function of $\sqrt{s}$.
}
\end{center}
\end{figure}

\item
Also \citere{FrMaZe2004} is (mainly) based on \citere{Fr2002}. 
The general production of scalar leptons at linear $e^+e^-$ colliders has 
been computed in \citere{FrMaZe2004}, including multi-photon initial state 
radiation and polarized beams.  
The authors used (older versions of) \FA, \texttt{FeynCalc}~\cite{feyncalc}
and \LT\ for their calculations. As input parameters they used the mSUGRA 
parameter point SPS$1a$ \cite{SPS1a}, translated from the \DRbar\ to 
on-shell values; see the appendix of \citere{FrMaZe2004}.  
We also used this parameter point (as far as possible) and reproduced 
successfully their tree-level results in their Figs.~2 and 3a 
(see \citere{FrMaZe2004}) in the upper row of our \reffi{fig:FrMaZe2004}. 
Our (relative) one-loop results are in qualitative agreement with the ones 
in their Figs.~17 and 18a of \citere{FrMaZe2004}; see the lower row in our 
\reffi{fig:FrMaZe2004}. 
The quantitative numerical differences can be explained with the different 
renormalization schemes, slightly different input parameters, and the 
different treatment of the photon bremsstrahlung, where they have included 
multi-photon emission while we kept our calculation at $\order\al$.
It should also be kept in mind that the \textit{relative} one-loop 
corrections are sensitive to every kind of difference.

\begin{figure}[t]
\begin{center}
\begin{tabular}{c}
\includegraphics[width=0.48\textwidth,height=6cm]{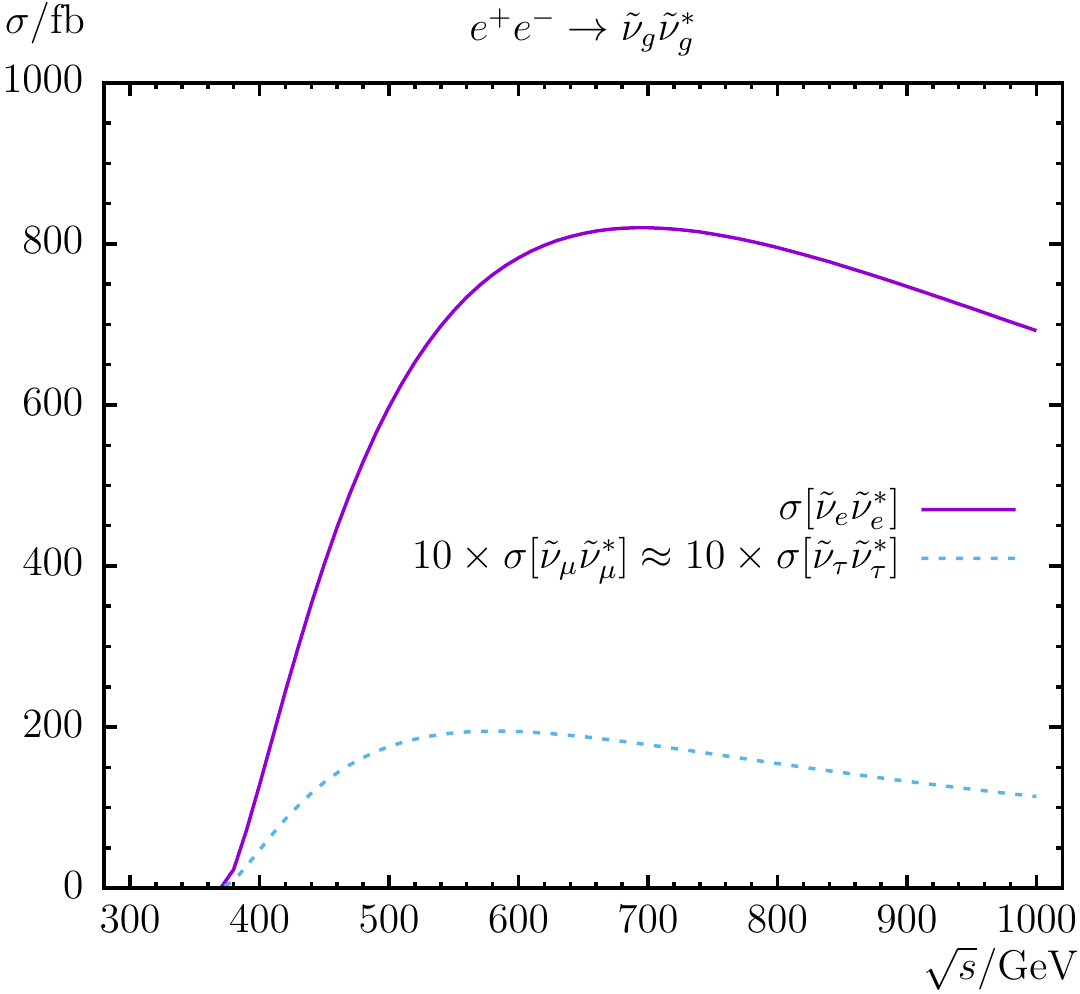}
\\[1em]
\includegraphics[width=0.48\textwidth,height=6cm]{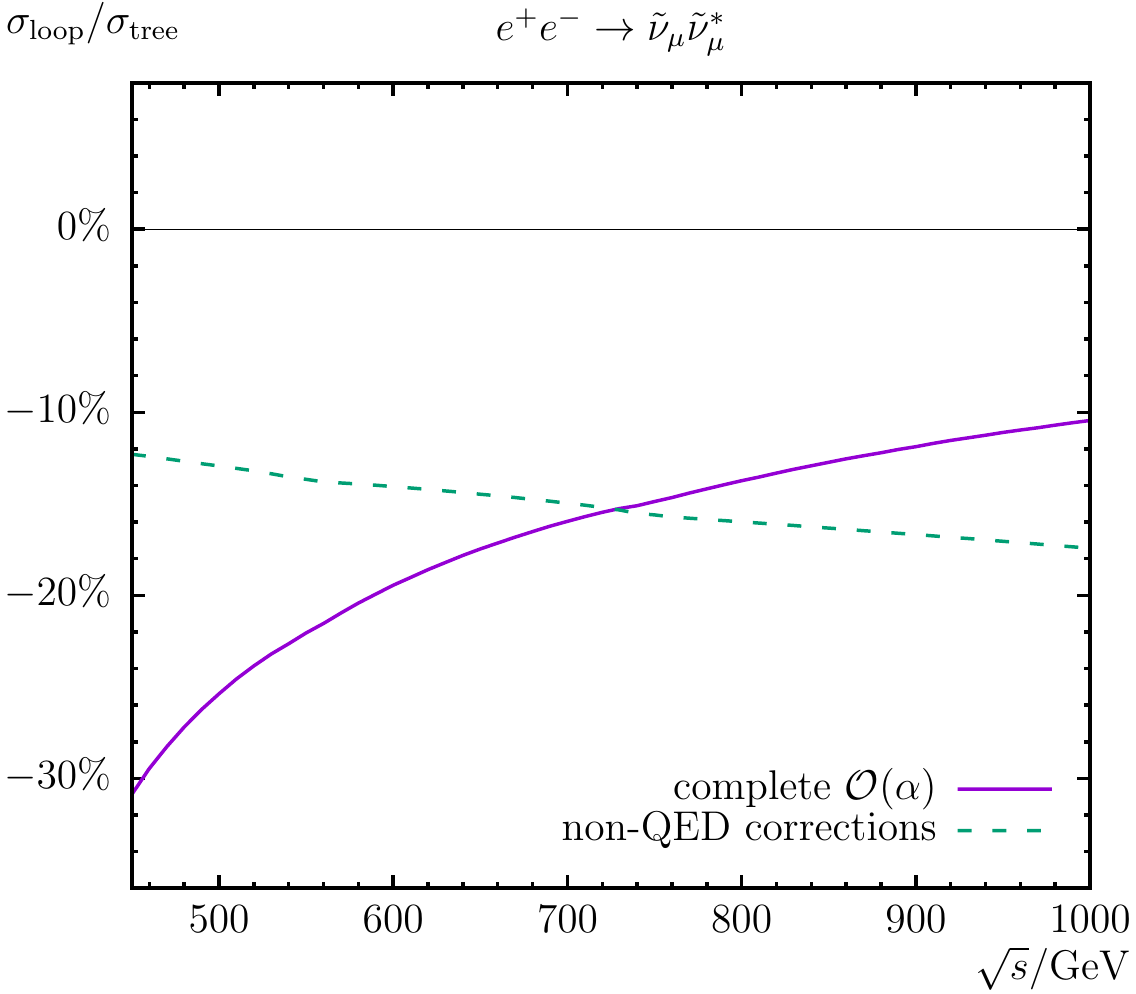}
\includegraphics[width=0.48\textwidth,height=6cm]{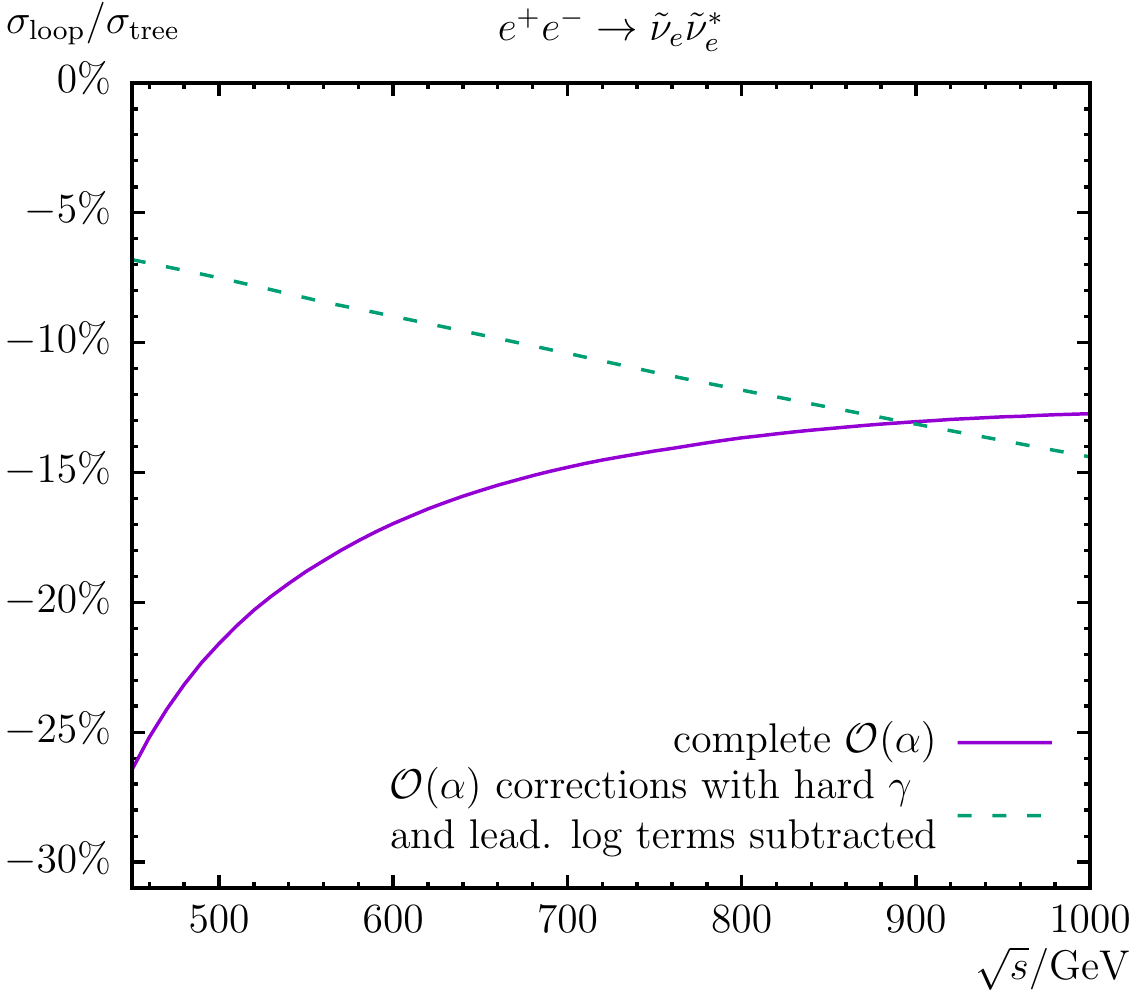}
\end{tabular}
\caption{\label{fig:FrMaZe2005}
  Comparison with \citere{FrMaZe2005} for $\eeSnSn$.
  Born cross sections (upper plot) and complete relative one-loop 
  results (lower row) are shown for the SPS$1a$ parameter point as 
  given according to \citere{FrMaZe2005} as a function of $\sqrt{s}$.
}
\end{center}
\end{figure}

\item
\citere{FrMaZe2005} is an addendum to \citere{FrMaZe2004} dealing with
$\eeSnSn$.  As input parameters they used the SUSY parameter point SPS$1a$
translated from the \DRbar\ to on-shell values; see \citere{SPS1a}.  We 
reproduced successfully $\eeSnmSnm$ and $\eeSnaSna$ of \citere{FrMaZe2005} 
in our \reffi{fig:FrMaZe2005}. 
The (quantitative) difference in the relative loop corrections can be 
explained with the different renormalization schemes, slightly different 
input parameters, and the different treatment of the photon radiation, 
where they have included multi-photon emission while we kept our calculation 
at $\order\al$.  The process $\eeSneSne$, on the other hand, while in rather 
good qualitative agreement, differ quantitative significantly already at the 
tree level.  Unfortunately, we were not able to trace back the source of the 
difference. However, since (in our automated approach) we agree with other 
tree-level calculations, we are confident that our results are correct.

\begin{figure}
\begin{center}
\begin{tabular}{c}
\includegraphics[width=0.48\textwidth,height=6cm]{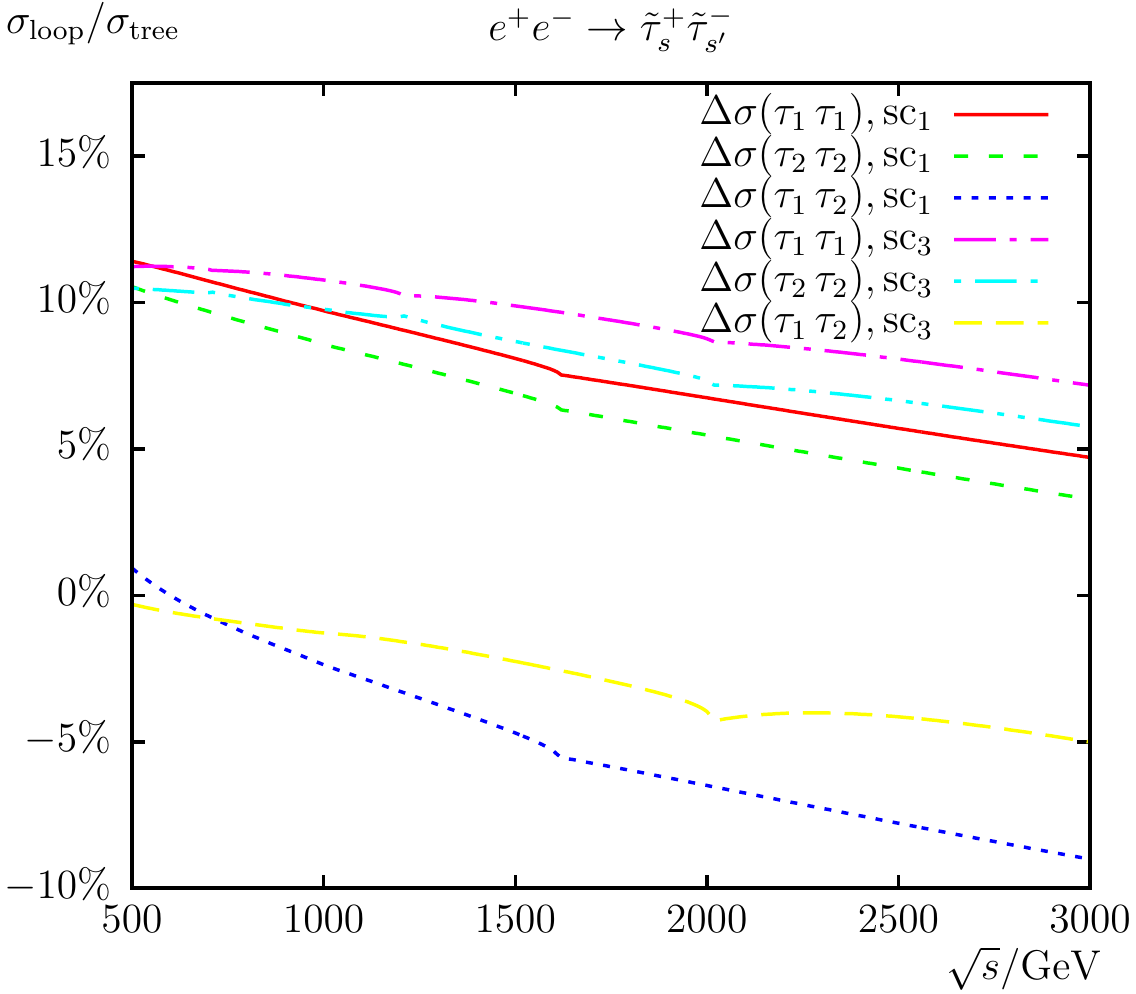}
\includegraphics[width=0.48\textwidth,height=6cm]{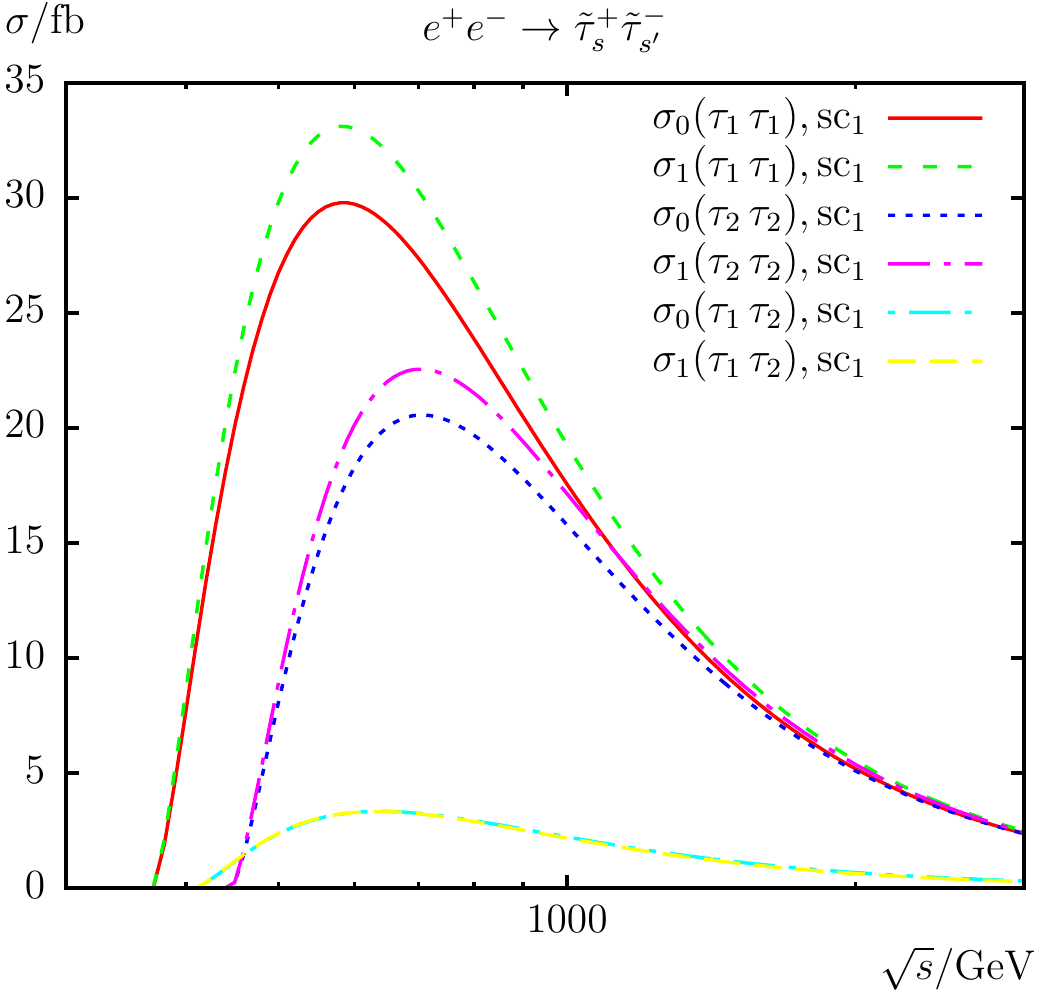}
\end{tabular}
\caption{\label{fig:ArHo2003}
  Comparison with \citere{ArHo2003} for $\sig(\eeSasSas)$.  
  Total relative corrections (left) and tree-level and one-loop cross 
  sections (right) as as functions of $\sqrt{s}$ are shown with 
  scenarios sc$_1$ and sc$_3$ chosen according to \citere{ArHo2003}. 
}
\end{center}
\end{figure}

\item
We performed a comparison with \citere{ArHo2003} using their input 
parameters (as far as possible).  They have calculated (third generation)
scalar fermion production in the rMSSM within an on-shell scheme close to 
ours at the ``full'' one-loop level (but without explicit QED radiation). 
They also used (older versions of) \FT\ for their calculations.  We found 
very good agreement with their Fig.~13, as can be seen in our 
\reffi{fig:ArHo2003}.  The \textit{tiny} differences can easily be explained 
with the slightly different SM input parameters and the slightly different 
renormalization scheme.

\begin{figure}[t]
\begin{center}
\begin{tabular}{c}
\includegraphics[width=0.48\textwidth,height=6cm]{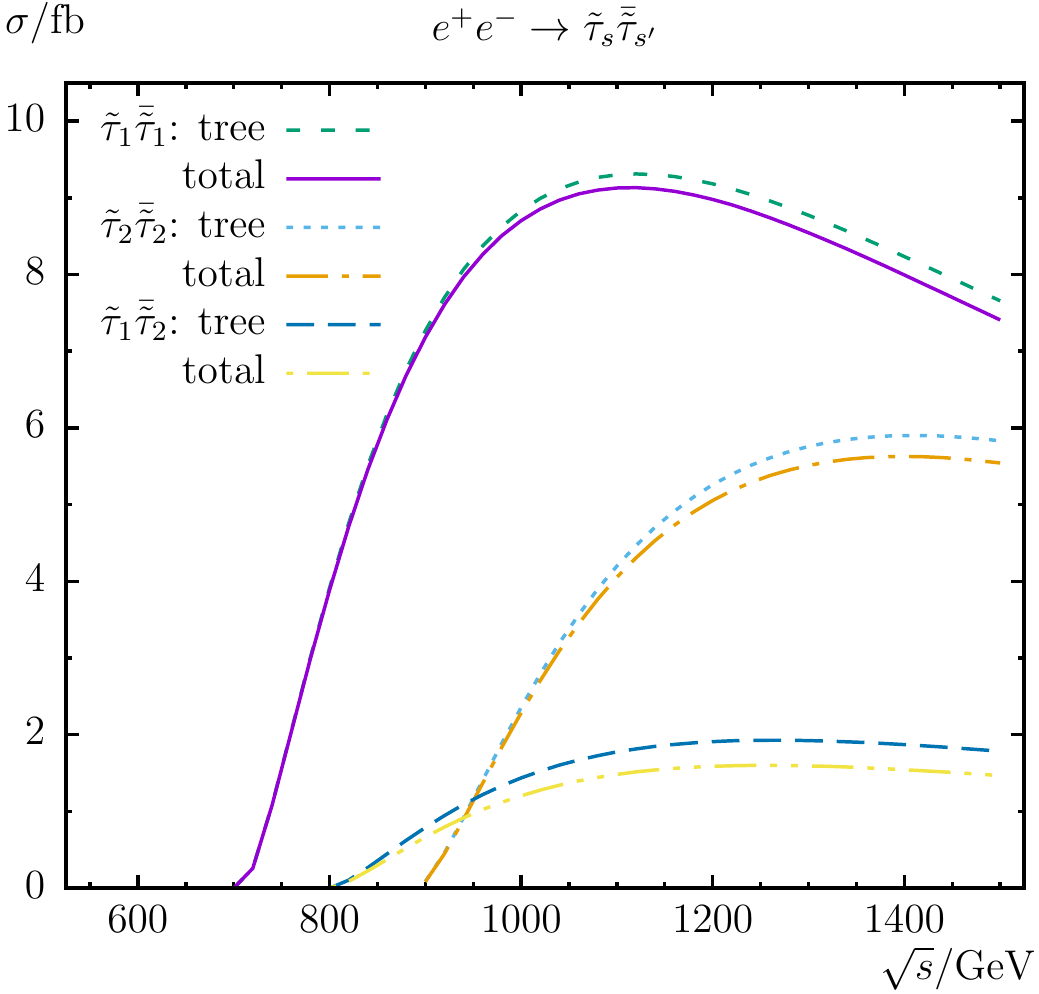}
\includegraphics[width=0.48\textwidth,height=6cm]{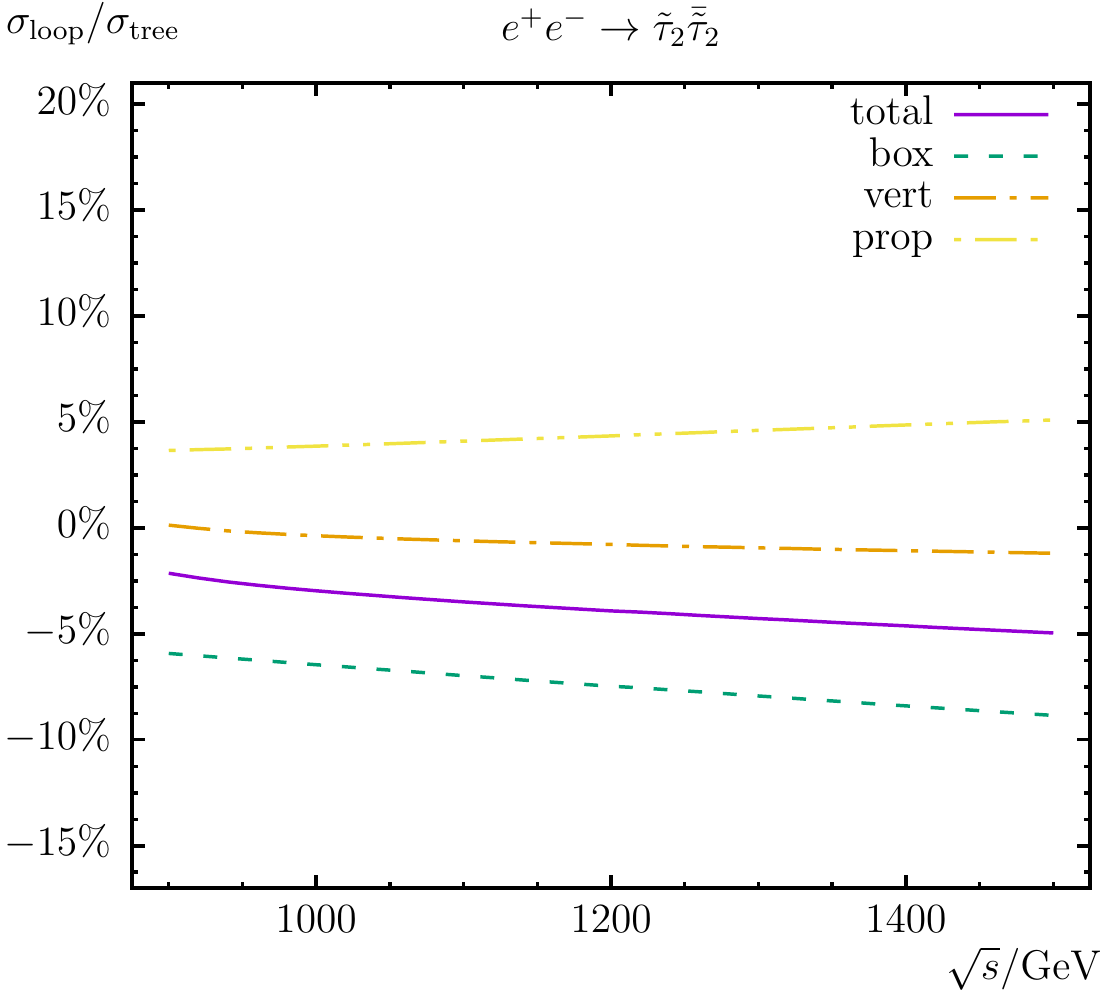}
\\[1em]
\includegraphics[width=0.48\textwidth,height=6cm]{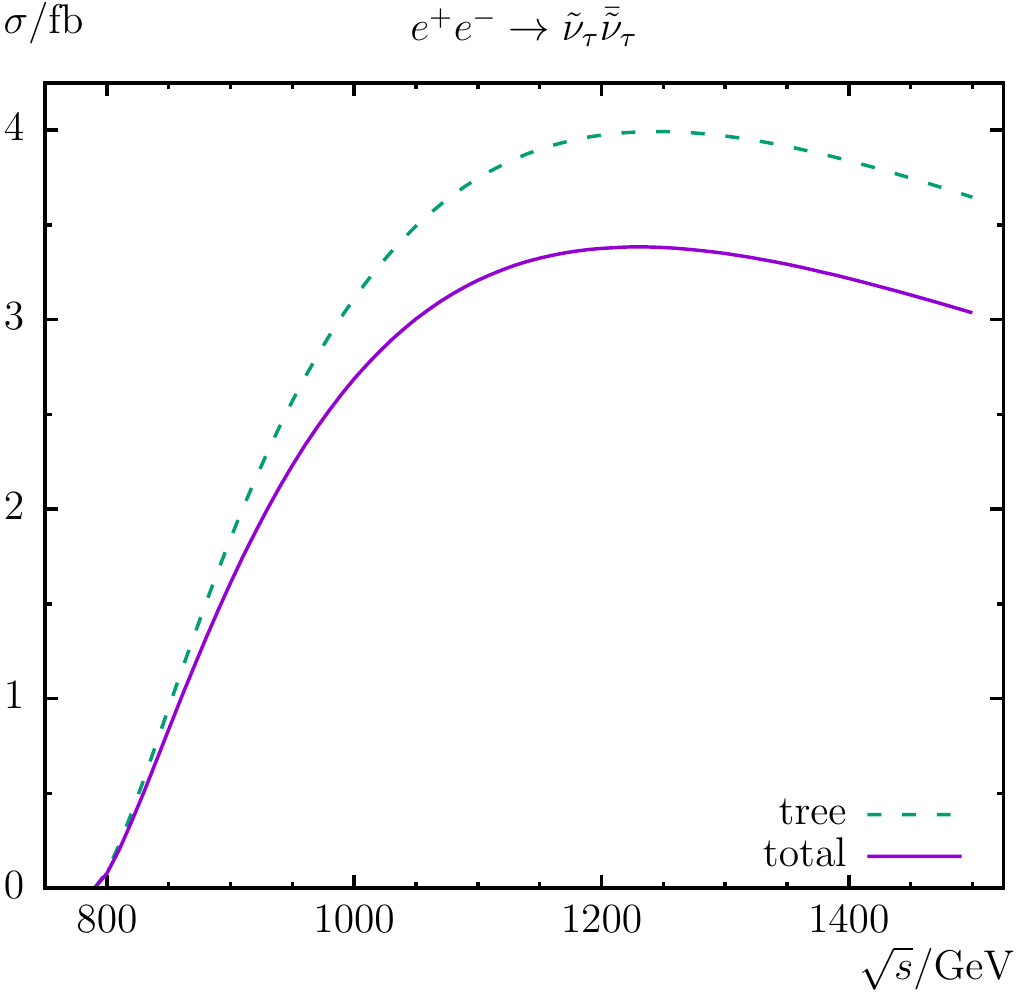}
\includegraphics[width=0.48\textwidth,height=6cm]{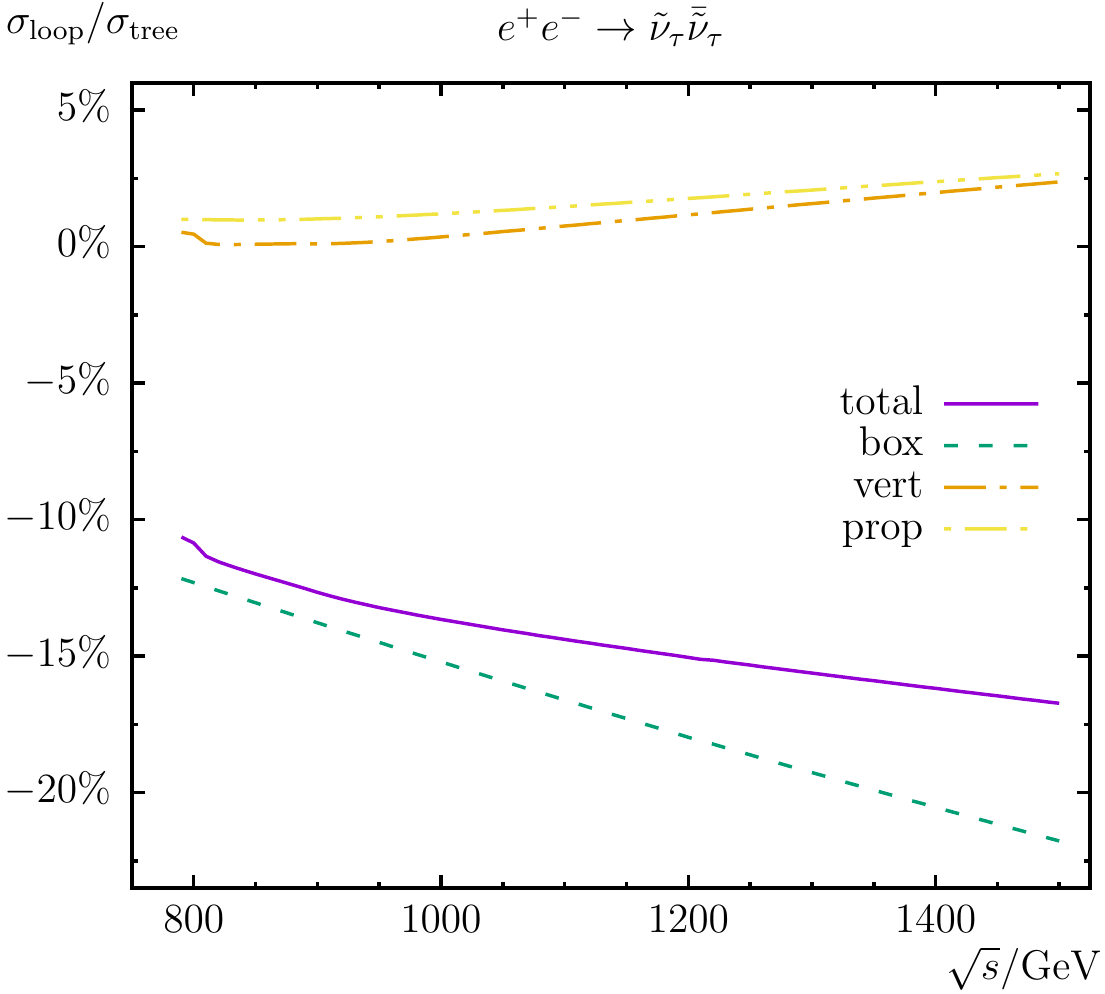}
\end{tabular}
\caption{\label{fig:KoWeEbMa2004}
  Comparison with \citere{KoWeEbMa2004} for $\sig(\eeSasSas)$ 
  (upper row) and $\sig(\eeSnaSna)$ (lower row).  The left (right) 
  plot shows cross sections (relative corrections) with $\sqrt{s}$ 
  varied within the Scenario~1 according to \citere{KoWeEbMa2004}.
}
\end{center}
\end{figure}

\item
In \citeres{KoWeEbMa2004,Ko2005} the ``complete'' one-loop corrections 
to $\eeSfSf$ (third generation) in the rMSSM were analyzed numerically 
including photon corrections.%
\footnote{
  It should be noted that \citere{Ko2005} is the ``source'' 
  of \citere{KoWeEbMa2004} and \citere{KoWeEbMa2005}.
}
However, their numerical results have been presented taking into account 
only weak corrections.  They used (an older version of) \LT\ for their 
calculations and their own on-shell renormalization procedure together 
with a $\alpha(\MZ)|_{\MSbar} = 1/127.934$ scheme. 
We used their ``Scenario 1- gaugino'' (as far as possible) for our 
comparison.  We are in rather good agreement with their Figs.~4c, 4f, 
7a, and 7d which can be seen in our \reffi{fig:KoWeEbMa2004}. 
The minor differences can be explained (as usual) with the slightly 
different input parameters and the different renormalization scheme.%
\footnote{
  It should be noted that the sum, but not the 
  individual contributions, of vertex (vert) and 
  propagator (prop) contributions (in our upper 
  right plot of \reffi{fig:KoWeEbMa2004}) are 
  (nearly) the same as the corresponding sum in 
  Fig.~4f of \citere{KoWeEbMa2004}.  This is 
  because of the different renormalization of 
  the (charged) slepton sector.
}

\item
Finally, in \citeres{Ko2005,KoWeEbMa2005} full $\order{\alpha}$ corrections 
to $\eeSfSf$ have been calculated in the MSSM with real parameters.
Polarized electrons and multi-photon bremsstrahlung were included in the 
phenomenological analysis.  As input parameters they used the mSUGRA 
parameter point SPS$1a'$ \cite{SPS1a}, translated from the \DRbar\ to 
on-shell values.  We also used this parameter point (as far as possible) 
and reproduced parts of their results in their Figs.~7 and 8 
(see \citere{KoWeEbMa2005}) in our \reffi{fig:KoWeEbMa2005}.
We are in good agreement for the processes $\eeSazSaz$ and $\eeSaeSaz$,
while we disagree in $\eeSaeSae$ and $\eeSnaSna$ already at tree-level.
Unfortunately, we were not able to track down the source of the 
differences.  However, since we agree (in our automated approach) with 
other tree-level calculations, we are confident that our results are 
correct.

\begin{figure}
\begin{center}
\begin{tabular}{c}
\includegraphics[width=0.48\textwidth,height=6cm]{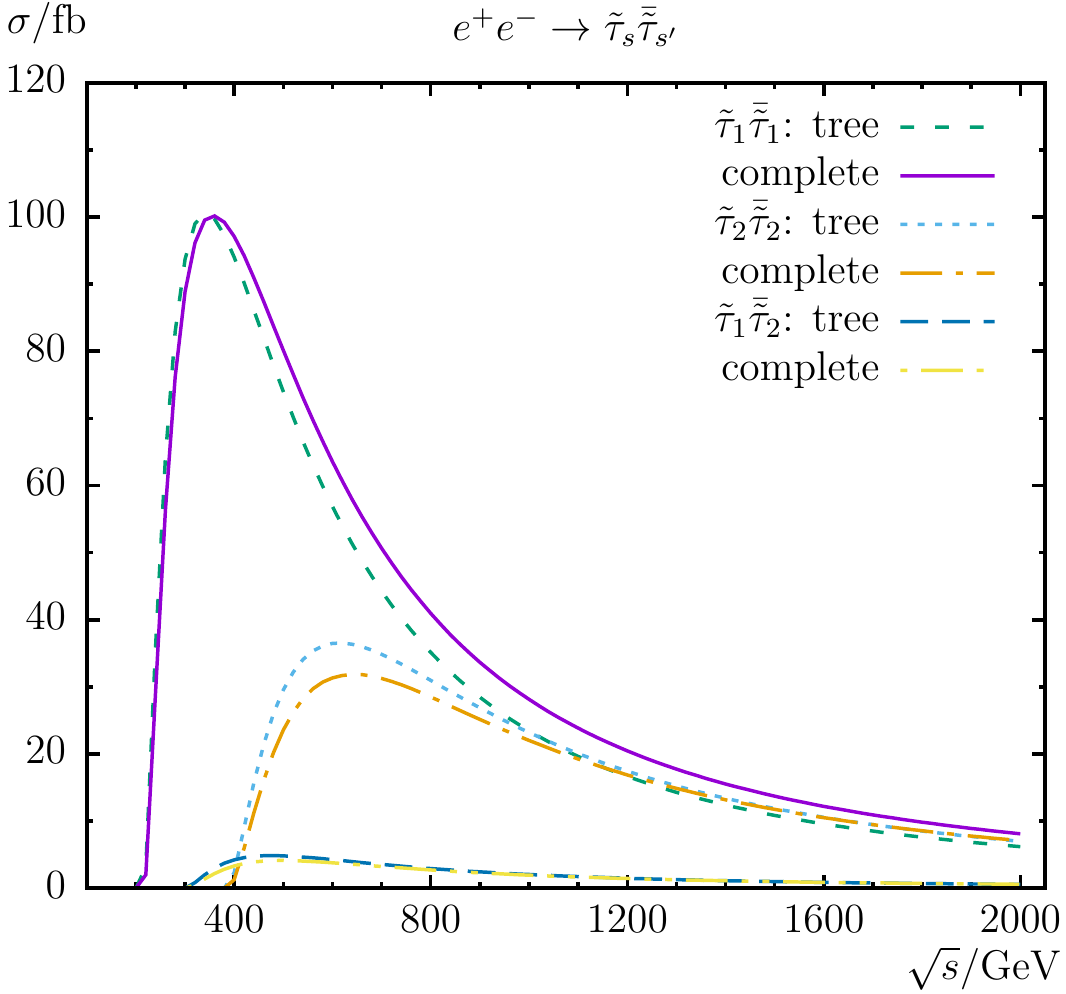}
\includegraphics[width=0.48\textwidth,height=6cm]{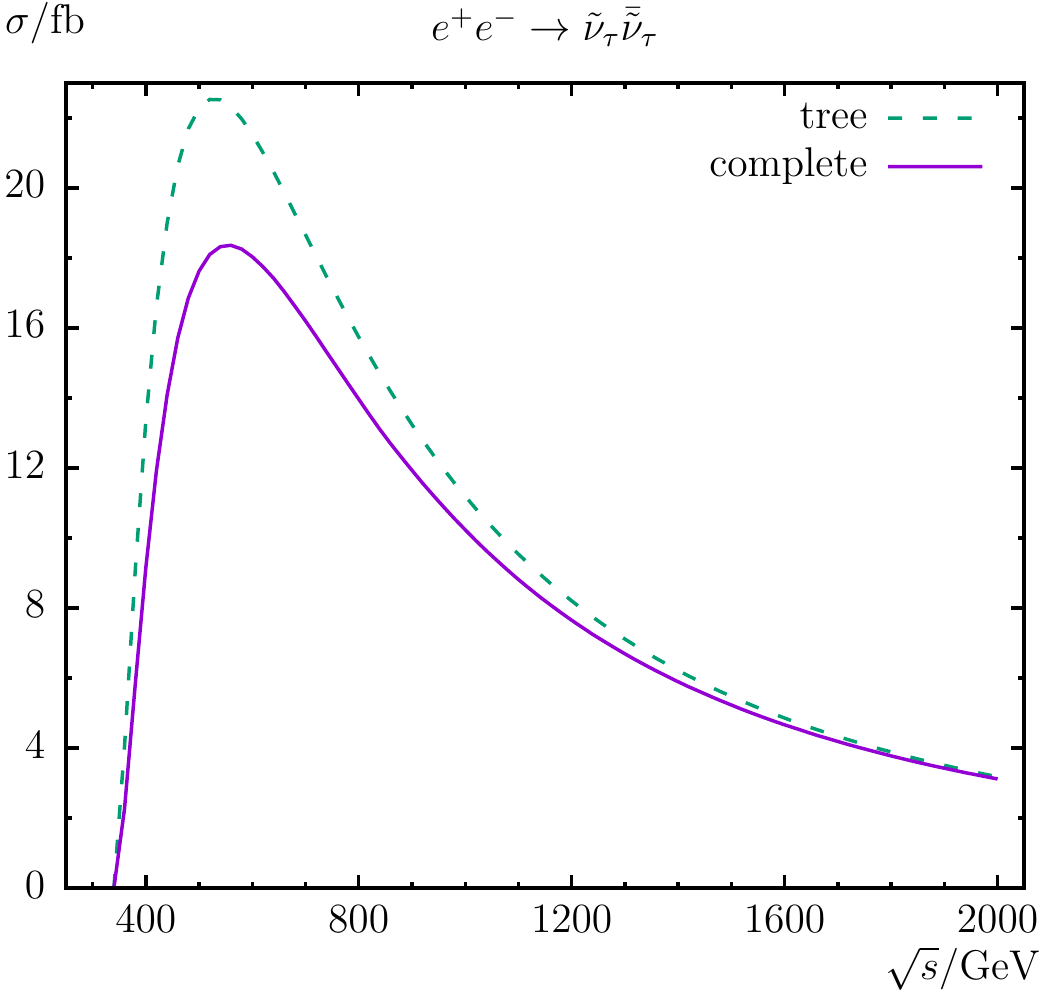}
\end{tabular}
\caption{\label{fig:KoWeEbMa2005}
  Comparison with \citere{KoWeEbMa2005} for $\sig(\eeSasSas)$ 
  left plot and $\sig(\eeSnaSna)$ right plot.  Tree and complete 
  cross sections are shown within the parameter set SPS$1a^{\prime}$ 
  chosen according to \citere{KoWeEbMa2005}, varied with $\sqrt{s}$. 
}
\end{center}
\end{figure}

\end{itemize}

To conclude, in most cases where a meaningful comparison could be 
performed, we found good agreement with the literature where expected, 
and the encountered differences can be traced back to different 
renormalization schemes, corresponding mismatches in the input
parameters and small differences in the SM parameters. Nevertheless,
in some cases we disagree already significantly at tree-level but we were 
not able to track down the source of these differences.  This does not 
disprove the reliability of our calculation because our computational 
method/code has already been successfully tested and compared with quite 
a few other programs; see \citeres{HiggsDecaySferm,HiggsDecayIno,MSSMCT,%
SbotRen1,Stop2decay,Gluinodecay,Stau2decay,LHCxC1,LHCxN}. After comparing
with the existing literature we would like to stress again that here we
present for the first time a full one-loop calculation of $\sig(\eeSeSe)$
and $\sig(\eeSnSn)$ in the cMSSM, using the scheme that was employed
successfully already for the full one-loop decays of the (produced)
sleptons.  The various calculations can readily be used together
for the full production and decay chain.


\section{Numerical analysis}
\label{sec:numeval}

In this section we present our numerical analysis of slepton production 
at $e^+e^-$ colliders in the cMSSM. 
In the figures below we show the cross sections at the tree level 
(``tree'') and at the full one-loop level (``full''), which is the 
cross section including \textit{all} one-loop corrections as described 
in \refse{sec:calc}.  The \Code{CCN[1]} renormalization scheme 
(\ie OS conditions for the two charginos and the lightest neutralino) 
has been used for most evaluations. 
In cases where the \Code{CCN[1]} scheme is divergent (\ie $M_2 = \mu$)
and/or unreliable, also some \Code{CNN[$c,n,n'$]} schemes 
(OS conditions for one chargino and two neutralinos) have been used, 
as indicated below; see \citere{MSSMCT} for further informations to 
these renormalization schemes.
It should be noted that within the \Code{CCN[1]} scheme at $\mu = M_2$ 
a divergence already in the tree-level result can be induced by 
$\delta \mu$ through the shifted scalar lepton masses
\footnote{
  We use the same (shifted) sfermion masses for 
  the tree-level and the full one-loop corrected 
  cross section.
}%
: $\delta \mu$ in \refeq{eq:SlepAux} enters via \refeq{eq:SlepY} into 
\refeq{eq:dSfShift} from which the slepton shifts are calculated. 
The divergence is suppressed with $\mfe{g}$ (see \refeq{eq:SlepAux}). 
In several analyses in \refse{sec:numeval} in order to overcome the 
problem with a divergent tree-level result, we switched to the tree-level 
result of the \Code{CNN[1,2,3]} scheme, which is free of such a divergence.
When several schemes are shown in one plot, a full comparison would
require the transition of the relevant input parameters (which are 
varied). However, we do \textit{not} intend to perform an analysis of 
the advantages and disadvantages of the various renormalization schemes. 
We want to demonstrate, however, that it is always possible to choose a 
``good'' renormalization scheme, \ie a scheme that leads to stable and 
not excessively large higher-order corrections.
Consequently, the above mentioned parameter conversion is not (yet)
included in our calculation.

We first define the numerical scenario for the cross section evaluation. 
Then we start the numerical analysis with the cross sections of \eeSeSe\
($g = 1,2,3;\, s,\spri = 1,2$) in \refse{sec:eeSeSe}, evaluated as a 
function of $\sqrt{s}$, $\MSE$, $\mu$, $|M_1|$ and/or $M_2$, $\phiAeg$ 
or $\phiMe$, the phase of $M_1$.
In some cases also the $\TB$ dependence is shown.
Then we turn to the processes \eeSnSn\ in \refse{sec:eeSnSn}. 
All these processes are of particular interest for ILC and CLIC 
analyses~\cite{ILC-TDR,teslatdr,Ac2004,ilc1,ilc2,CLIC1,CLIC2} 
(as emphasized in \refse{sec:intro}).


\subsection{Parameter settings}
\label{sec:paraset}

The renormalization scale $\mu_R$ has been set to the center-of-mass energy, 
$\sqrt{s}$.  The SM parameters are chosen as follows; see also \cite{pdg}:
\begin{itemize}

\item Fermion masses (on-shell masses, if not indicated differently):
\begin{align}
m_e    &= 0.5109989461\mev\,, & m_{\nu_e}    &= 0\,, \notag \\
m_\mu  &= 105.6583745\mev\,,  & m_{\nu_{\mu}} &= 0\,, \notag \\
m_\tau &= 1776.86\mev\,,      & m_{\nu_{\tau}} &= 0\,, \notag \\
m_u &= 70.59\mev\,,          & m_d         &= 70.59\mev\,, \notag \\ 
m_c &= 1.280\gev\,,          & m_s         &= 96.00\mev\,, \notag \\
m_t &= 173.1\gev\,,          & m_b         &= 2.954\gev\,.
\end{align}
According to \citere{pdg}, $m_s$ is an estimate of a so-called 
"current quark mass" in the \MSbar\ scheme at the scale 
$\mu \approx 2\gev$.  $m_c \equiv m_c(m_c)$ is the "running" mass in 
the \MSbar\ scheme and $m_b \equiv m^{\DRbar}_b$ is the \DRbar\ bottom quark 
mass as calculated in \citere{Stop2decay}. 
$m_u$ and $m_d$ are effective parameters, calculated through the 
hadronic contributions to
\begin{align}
\Delta\alpha_{\text{had}}^{(5)}(M_Z) &= 
      \frac{\alpha}{\pi}\sum_{f = u,c,d,s,b}
      Q_f^2 \Bigl(\ln\frac{M_Z^2}{m_f^2} - \frac 53\Bigr) \approx 0.02764\,.
\end{align}

\item Gauge-boson masses\index{gaugebosonmasses}:
\begin{align}
M_Z = 91.1876\gev\,, \qquad M_W = 80.385\gev\,.
\end{align}

\item Coupling constant\index{couplingconstants}:
\begin{align}
\alpha(0) = 1/137.035999139\,.
\end{align}
\end{itemize}

The SUSY parameters are chosen according to the scenario \Scs, shown 
in \refta{tab:para}.  This scenario is viable for the various cMSSM 
slepton production modes, \ie not picking specific parameters for 
each cross section.  They are in particular in agreement with 
the relevant SUSY searches of ATLAS and CMS:
Our electroweak spectrum is not covered by the latest ATLAS/CMS exclusion 
bounds.  Two limits have to be distinguished. The limits not taking into
account a possible intermediate slepton exclude a lightest neutralino only 
well below $300\gev$~\cite{ATLAS1,CMS1}, whereas in \Scs\ we have 
$\mneu1 \approx 323\gev$.  Limits with intermediary sleptons often assume 
a chargino decay to lepton and sneutrino, while in our scenario 
$\mcha1 < \msneu$. Furthermore, the exclusion bounds given in the 
$\mneu1$-$\mneu2$ mass plane (with $\mneu2 \approx \mcha1$ assumed) above 
$\mneu2 \sim 300 \gev$ do not cover a compressed spectrum~\cite{ATLAS1,CMS2} 
for $\neu1$, $\neu2$, and $\cha1$.  In particular our scenario \Scs\ assumed 
masses of $\mneu1 \approx 323\gev$ and $\mneu2 \approx 354\gev$, which are 
not excluded.

\begin{table}
\caption{\label{tab:para}
  MSSM default parameters for the numerical investigation; all 
  parameters (except of $\TB$) are in GeV.  The values for the trilinear 
  sfermion Higgs couplings, $A_f$ are chosen to be real (except for 
  $A_{\Fe_g}$ which can be complex) and such that charge- and/or 
  color-breaking minima are avoided \cite{ccb1,ccb2,ccb3,ccb4,ccb5,ccb6,ccb7}.  
  It should be noted that we chose common values 
  $M_{\tilde Q, \tilde U, \tilde D} = 2000\gev$ for all squark generations,
  and $\MSL = \MSE + 50\gev$ for all slepton generations. 
  For the sleptons we show the tree-level values as well as their OS masses 
  in our \Code{DR[2]} renormalization scheme; see \refse{sec:renorm}. 
}
\centering
\begin{tabular}{lrrrrrrrrrrrr}
\toprule
Scen. & $\sqrt{s}$ & $\TB$ & $\mu$ & $\MHp$ & $M_{\tilde Q, \tilde U, \tilde D}$ & 
$\MSE$ & $A_{\Fu_g}$ & $A_{\Fd_g}$ & $|A_{\Fe_g}|$ & $|M_1|$ & $M_2$ & $M_3$ \\ 
\midrule
\Scs & 1000 & 10 & 350 & 1200 & 2000 & 300 & 2600 & 2000 & 2000 & 400 & 
600 & 2000 \\
\bottomrule
\end{tabular}

\vspace{0.5em}

\begin{tabular}{lrrrrrrrrrr}
\toprule
& $\msneu$ & $\msele1$ & $\msele2$ & $\msmue1$ & $\msmue2$ & $\mstau1$ & $\mstau2$ \\
\midrule
tree & 344.129 & 303.013 & 353.212 & 303.012 & 353.213 & 302.664 & 353.519 \\
\midrule
OS   & 344.129 & 303.013 & 352.973 & 303.012 & 352.974 & 302.664 & 353.264 \\
\bottomrule
\end{tabular}
\end{table}

It should be noted that higher-order corrected Higgs-boson masses do not 
enter our calculation.%
\footnote{
  Since we work in the MSSM with complex parameters, 
  $\MHp$ is chosen as input parameter, and higher-order 
  corrections affect only the neutral Higgs-boson spectrum; 
  see \citere{chargedmhiggs2L} for the most recent evaluation.
}
However, we ensured that over larger parts of the parameter space the 
lightest Higgs-boson mass is around $\sim 125 \pm 3\gev$ to indicate the
phenomenological validity of our scenarios. 
In our numerical evaluation we will show the variation with $\sqrt{s}$ 
(up to $3\tev$, shown in the upper left plot of the respective figures), 
$M_{\tilde E}$ (from 100 to 500 GeV, upper right plot), $\mu$ 
(starting at $\mu = 100\gev$ up to $\mu = 1000\gev$, shown in the 
middle/lower left plots), $|M_1|$ or $M_2$ 
(from 100 to 1000 GeV, middle/lower right plots), and $\phiAeg$ or 
$\phiMe$ (between $0^{\circ}$ and $360^{\circ}$, lower right plots).
The dependence of $\TB$ turned out to be rather small, therefore we 
show it only in a few cases, where it is of special interest.

Concerning the complex parameters, some more comments are in order.
Potentially complex parameters that enter the selectron and
electron sneutrino production cross sections at tree level 
(via the $t$-channel exchange of a neutralino or chargino) are the 
soft SUSY-breaking parameters $M_1$ and $M_2$ as well as the Higgs 
mixing parameter~$\mu$.  Also trilinear slepton couplings $A_{\Fe_g}$
enter the tree-level production cross sections.
However, when performing an analysis involving complex parameters it 
should be noted that the results for physical observables are affected 
only by certain combinations of the complex phases of the parameters 
$\mu$, the trilinear couplings $A_f$ and the gaugino mass parameters 
$M_{1,2,3}$~\cite{MSSMphases,SUSYphases}.
It is possible, for instance, to rotate the phase $\phiMz$ away.
Experimental constraints on the (combinations of) complex phases arise, 
in particular, from their contributions to electric dipole moments of 
the electron and the neutron (see \citeres{EDMrev2,EDMPilaftsis,Pi1999} 
and the references therein), of the deuteron~\cite{EDMRitz} and of heavy 
quarks~\cite{EDMDoink1,EDMDoink2}.
While SM contributions enter only at the three-loop level, due to its
complex phases the MSSM can contribute already at one-loop order.
Large phases in the first two generations of sfermions can only be 
accommodated if these generations are assumed to be very heavy 
\cite{EDMheavy1,EDMheavy2} or large cancellations 
occur~\cite{EDMmiracle1,EDMmiracle2,EDMmiracle3};
see, however, the discussion in \citere{EDMrev1}. 
A review can be found in \citere{EDMrev3}.
Recently additional constraints at the two-loop level on some $\CP$ 
phases of SUSY models have been investigated in \citere{Ya2013}.
Accordingly (using the convention that $\phiMz = 0$, as done in this paper), 
in particular, the phase $\phimu$ is tightly constrained~\cite{plehnix}, 
and we set it to zero.  On the other hand, the bounds on the phases of the
third-generation trilinear couplings are much weaker.  Consequently, 
the largest effects on the slepton production cross sections at the 
tree level are expected from the complex gaugino mass parameter $M_1$, 
\ie from $\phiMe$.  
As mentioned above, the only other phase entering at the tree level, is 
$\phiAeg$.  This motivates our choice of $\phiMe$ and $\phiAeg$ as 
parameters to be varied.

Since now complex parameters can appear in the couplings, contributions 
from absorptive parts of self-energy type corrections on external legs 
can arise.  The corresponding formulas for an inclusion of these 
absorptive contributions via finite wave function correction factors can 
be found in \citeres{MSSMCT,Stop2decay}.

The numerical results shown in the next subsections are of course 
dependent on the choice of the SUSY parameters.  Nevertheless, they 
give an idea of the relevance of the full one-loop corrections.


\subsection{\texorpdfstring{The process \boldmath{\eeSeSe}}
                           {The process e+e- -> Slepton Slepton}}
\label{sec:eeSeSe}

\begin{figure}
\begin{center}
\begin{tabular}{c}
\includegraphics[width=0.48\textwidth,height=6cm]{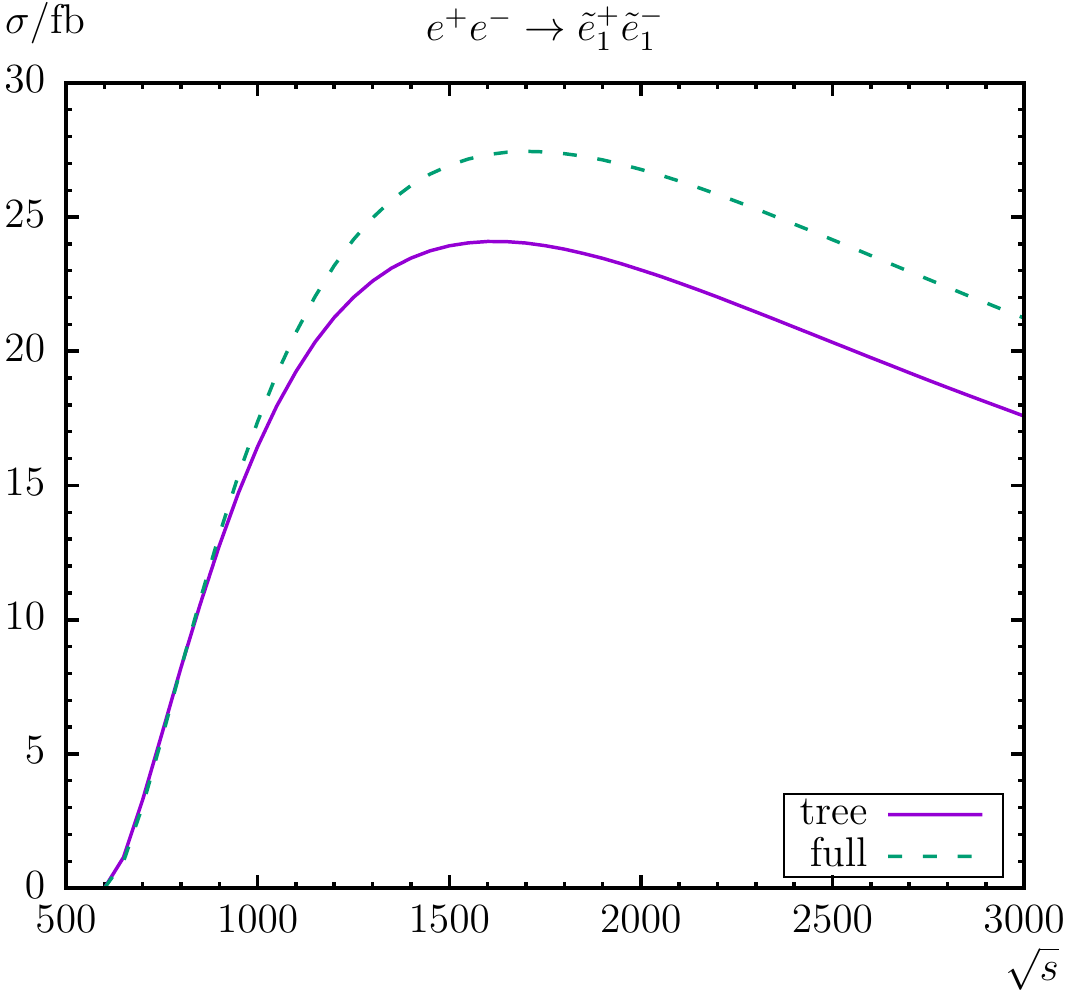}
\includegraphics[width=0.48\textwidth,height=6cm]{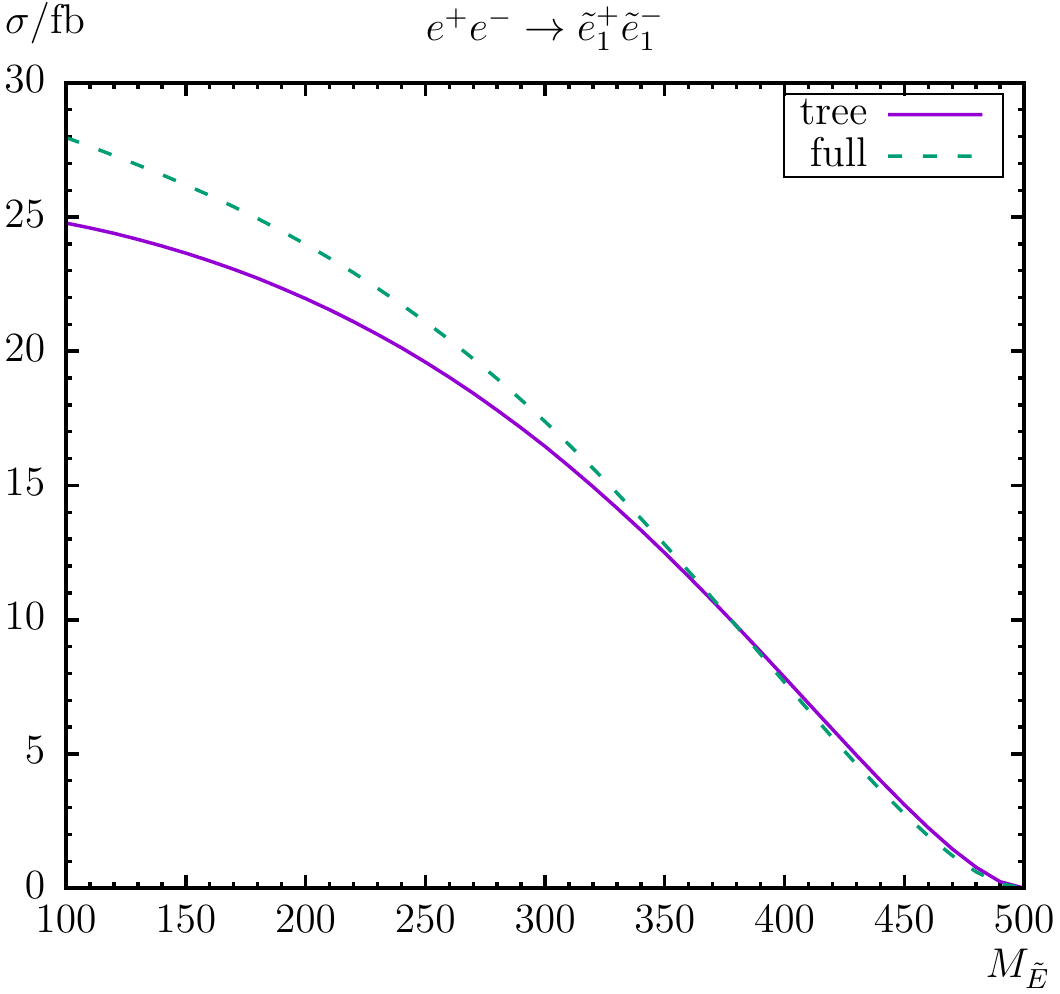}
\\[1em]
\includegraphics[width=0.48\textwidth,height=6cm]{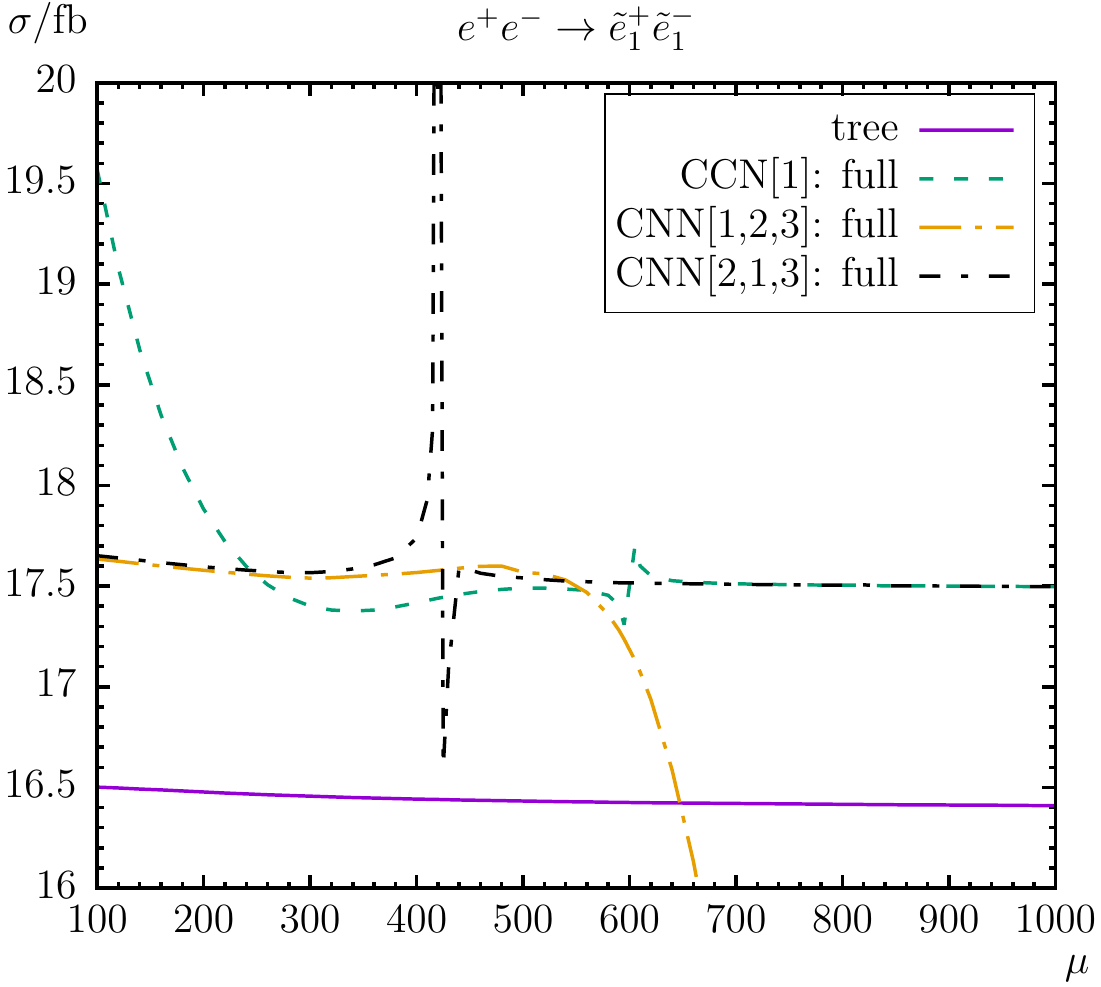}
\includegraphics[width=0.48\textwidth,height=6cm]{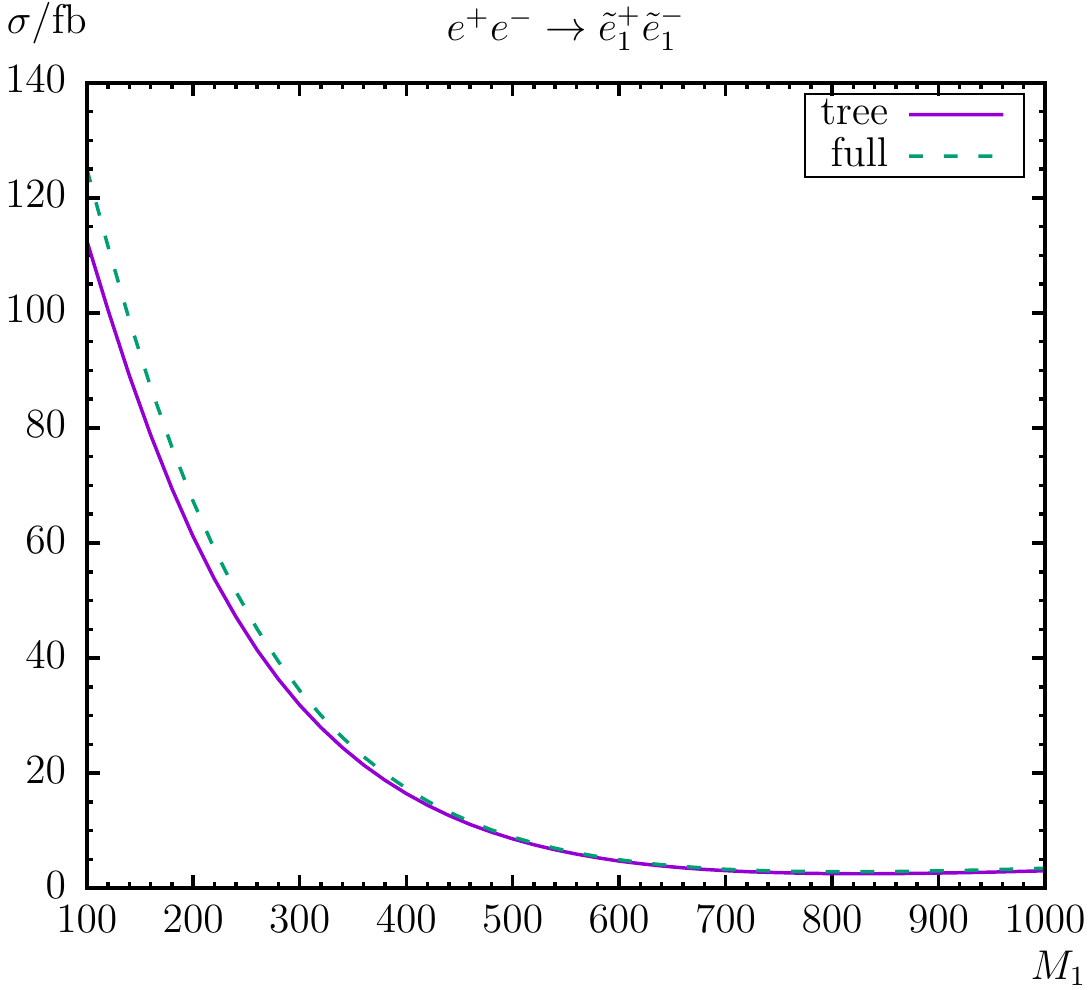}
\\[1em]
\includegraphics[width=0.48\textwidth,height=6cm]{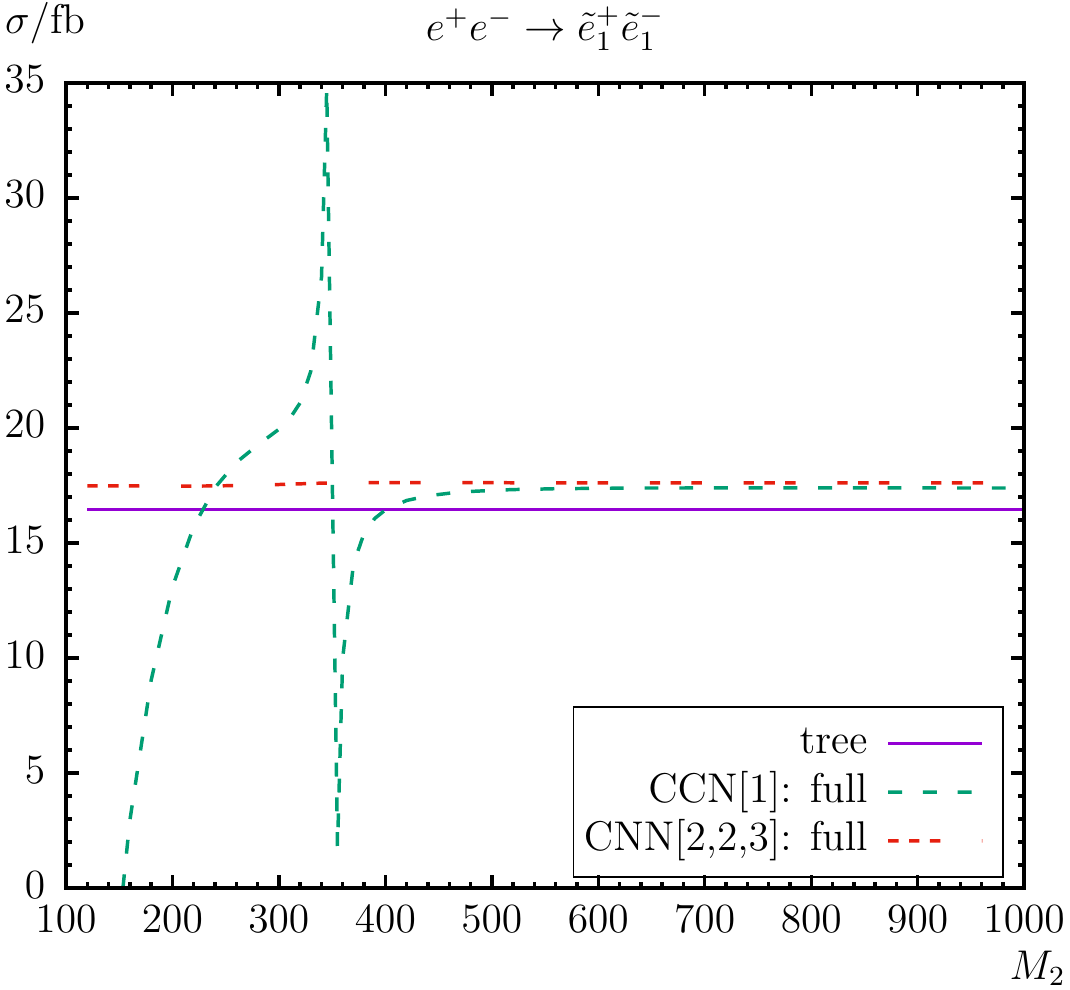}
\includegraphics[width=0.48\textwidth,height=6cm]{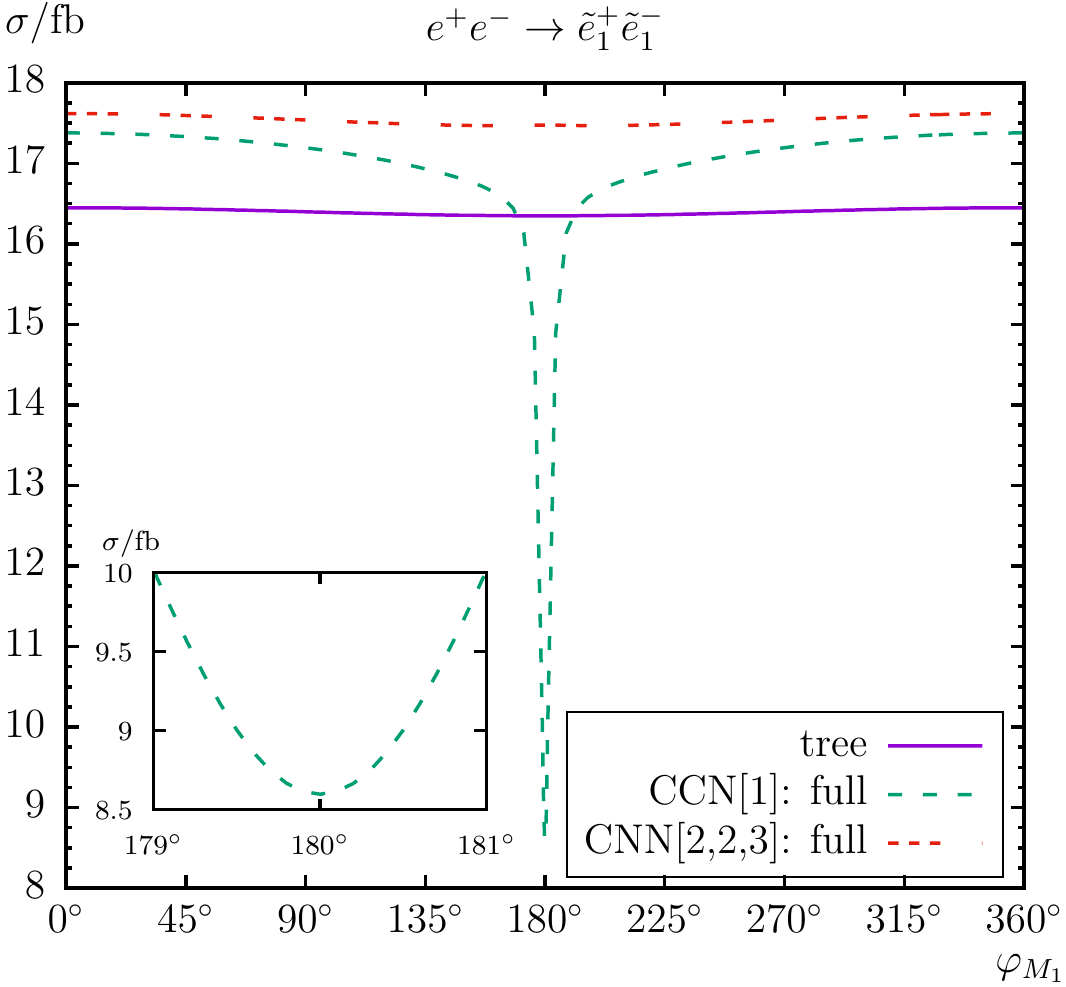}
\end{tabular}
\caption{\label{fig:eeSe1Se1}
  $\sig(\eeSeeSee)$.
  Tree-level and full one-loop corrected cross sections are shown 
  with parameters chosen according to \Scs; see \refta{tab:para}.
  The upper plots show the cross sections with $\sqrt{s}$ (left) 
  and $\MSE$ (right) varied; the middle plots show $\mu$ (left) 
  and $|M_1|$ (right) varied; the lower plots show $M_2$ (left) 
  and $\phiMe$ (right) varied. All masses and energies are in GeV.
}
\end{center}
\end{figure}

\begin{figure}
\begin{center}
\begin{tabular}{c}
\includegraphics[width=0.48\textwidth,height=6cm]{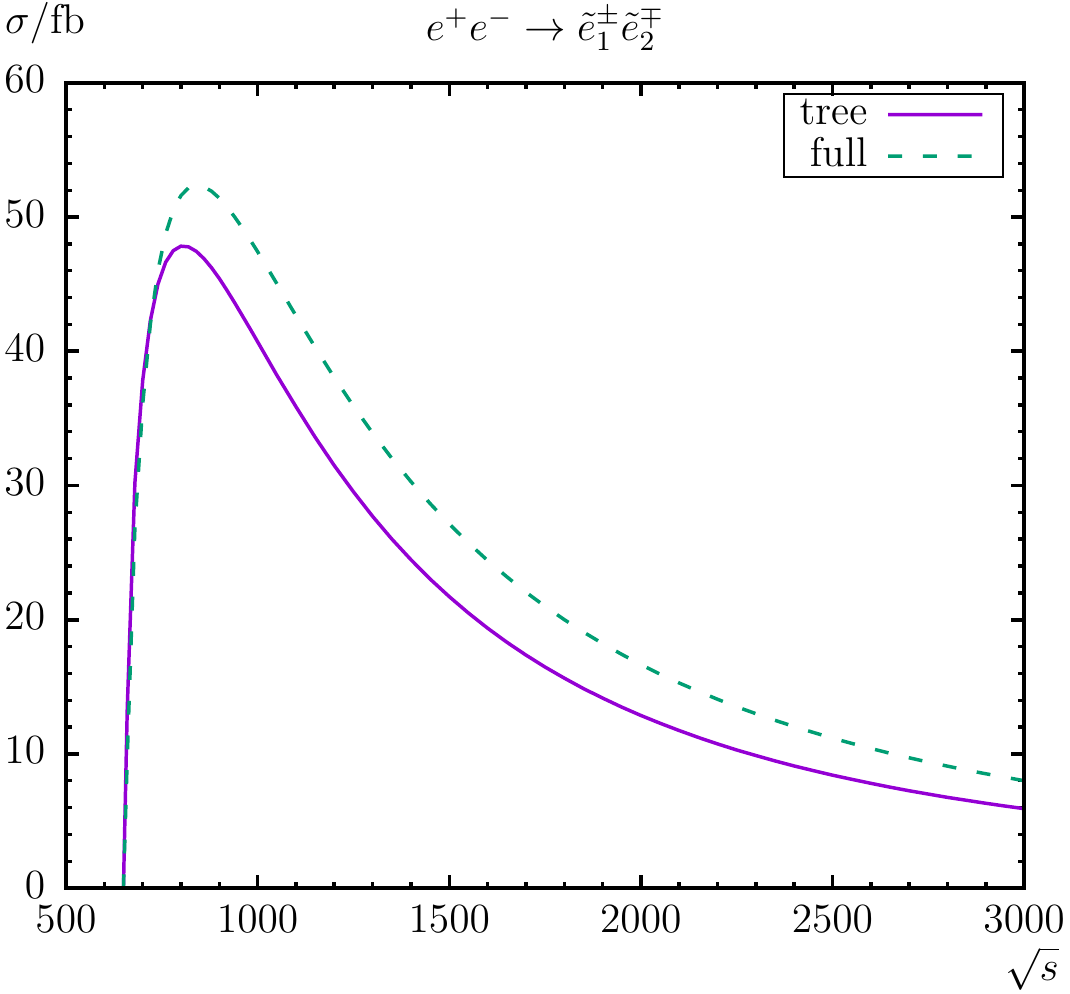}
\includegraphics[width=0.48\textwidth,height=6cm]{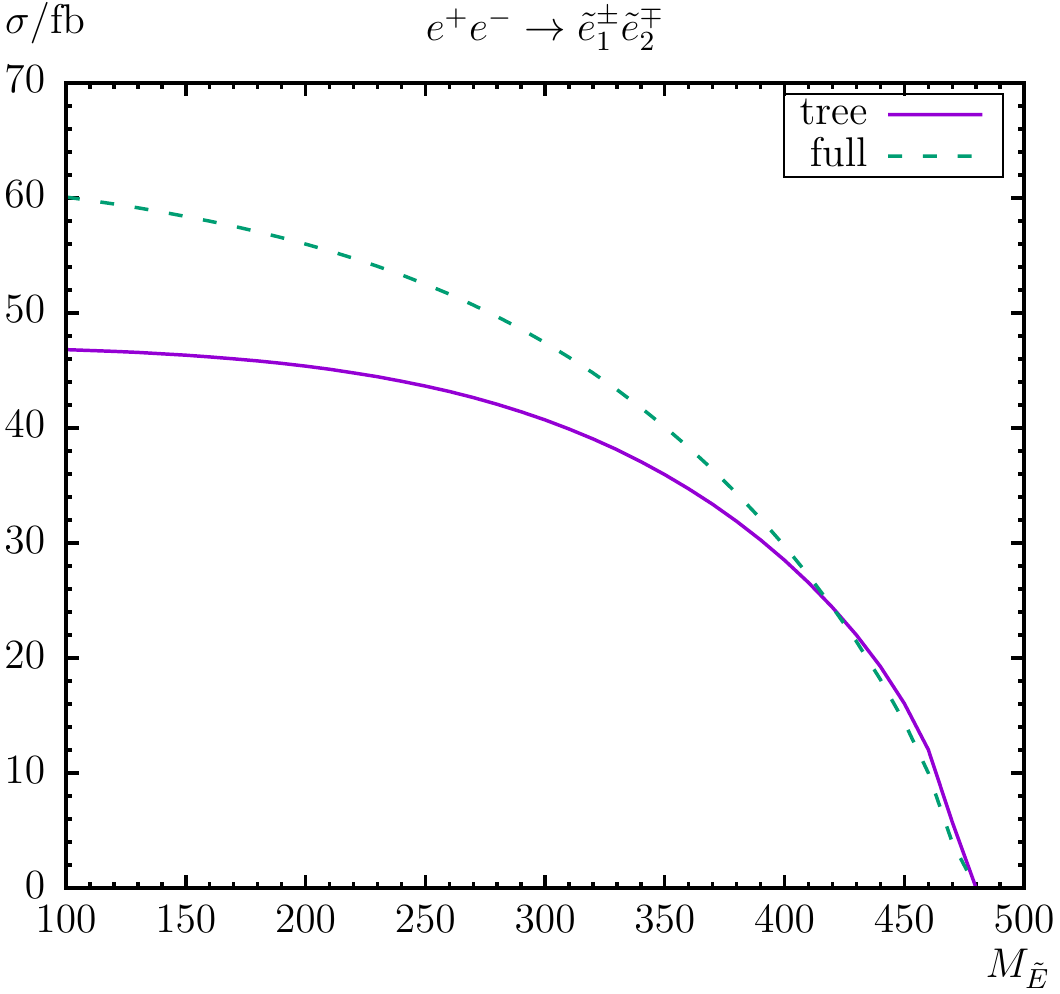}
\\[1em]
\includegraphics[width=0.48\textwidth,height=6cm]{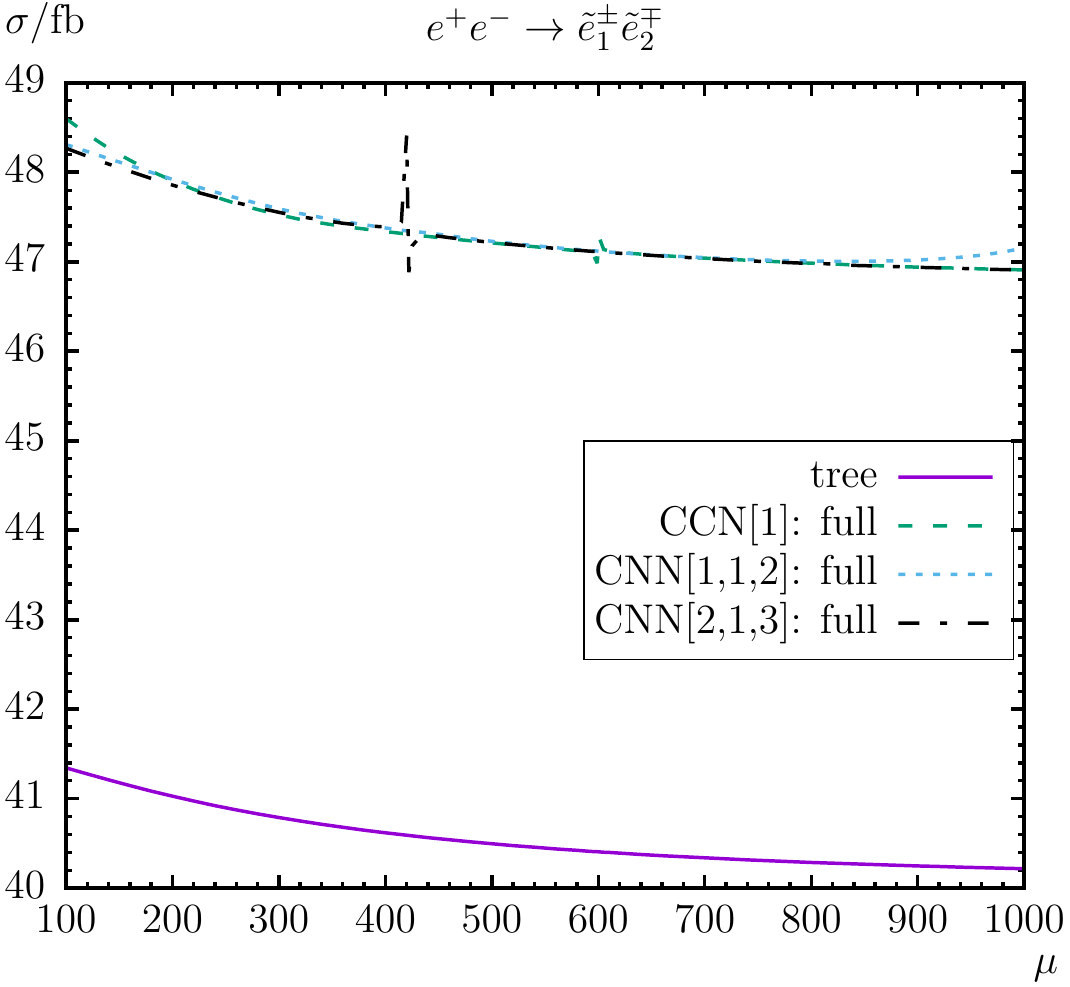}
\includegraphics[width=0.48\textwidth,height=6cm]{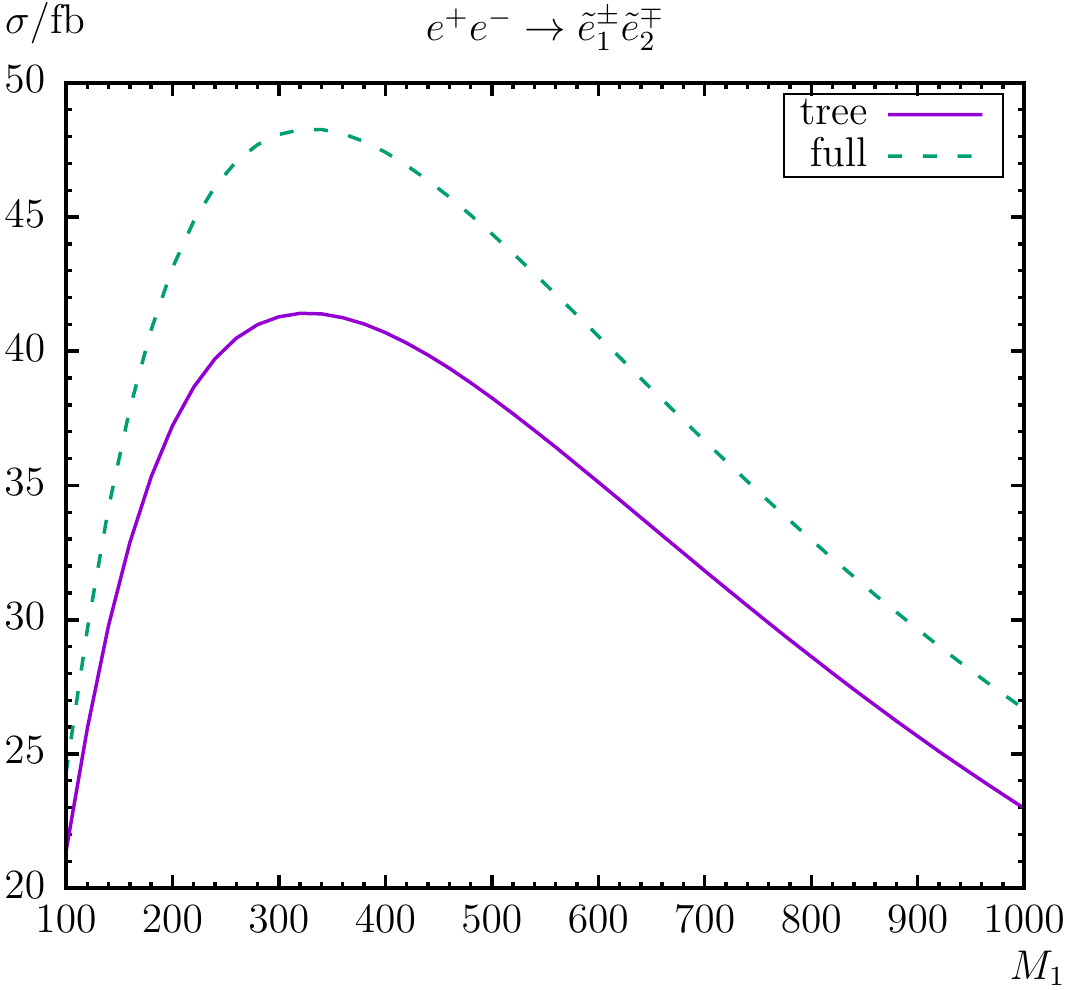}
\\[1em]
\includegraphics[width=0.48\textwidth,height=6cm]{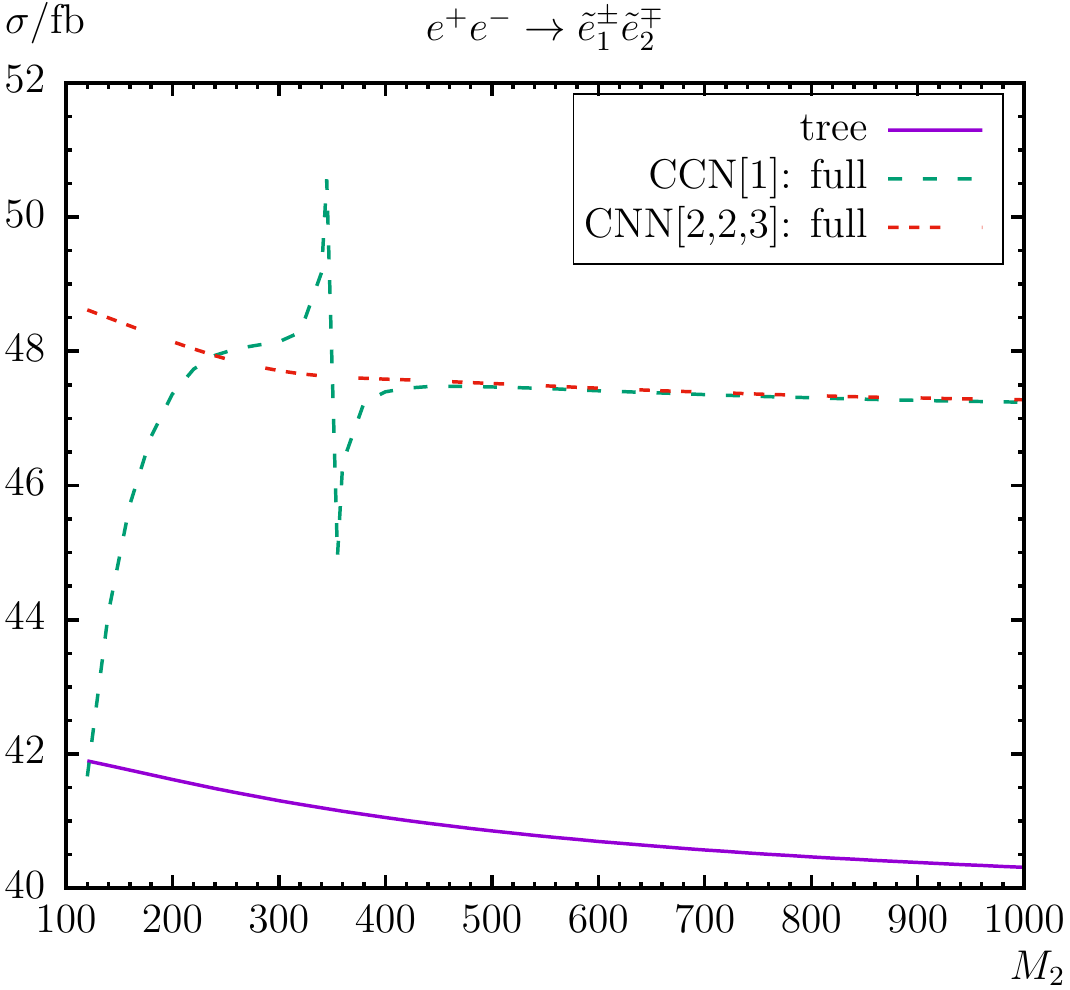}
\includegraphics[width=0.48\textwidth,height=6cm]{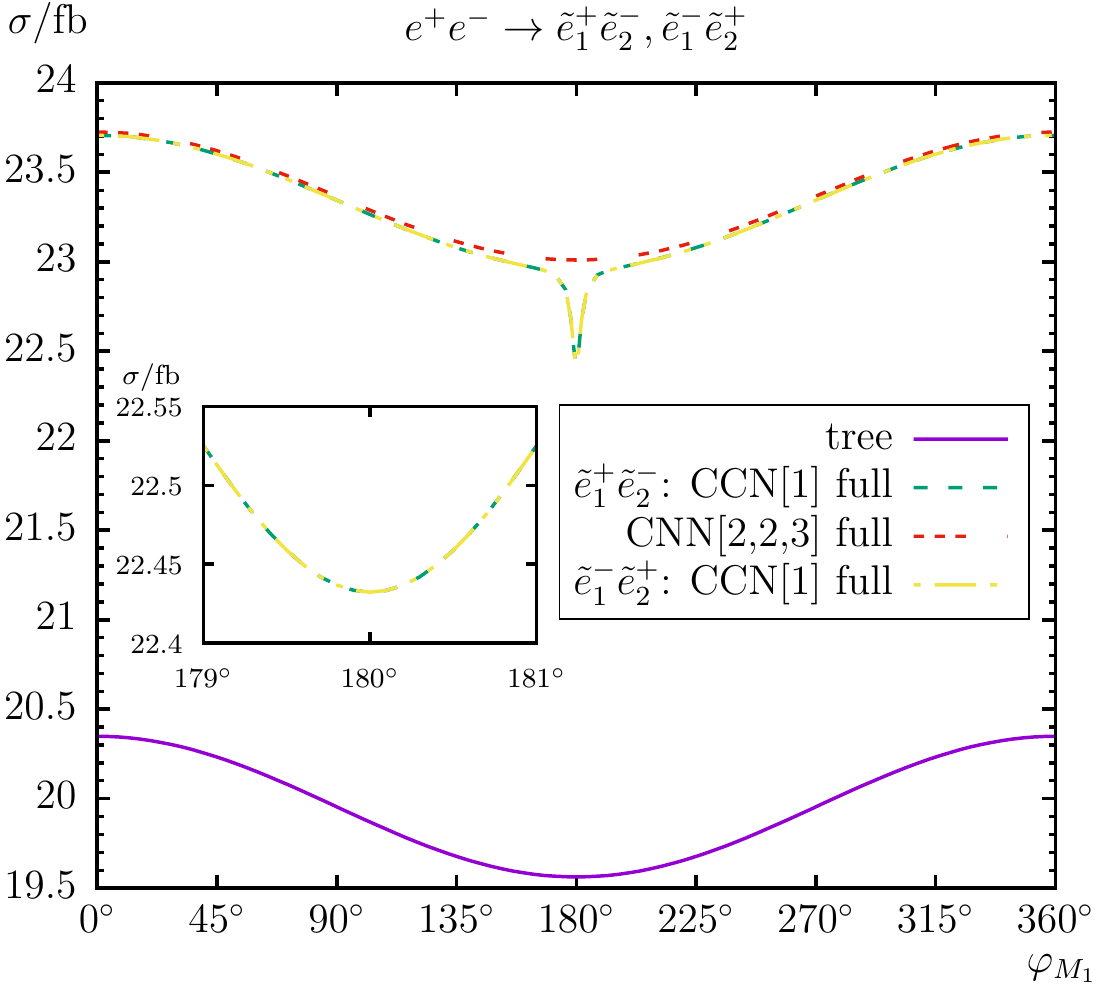}
\end{tabular}
\caption{\label{fig:eeSe1Se2}
  $\sig(\eeSeeSez)$.
  Tree-level and full one-loop corrected cross sections are shown 
  with parameters chosen according to \Scs; see \refta{tab:para}.
  The upper plots show the cross sections with $\sqrt{s}$ (left) 
  and $\MSE$ (right) varied; the middle plots show $\mu$ (left) 
  and $|M_1|$ (right) varied; the lower plots show $M_2$ (left) 
  and $\phiMe$ (right) varied. All masses and energies are in GeV.
}
\end{center}
\end{figure}

\begin{figure}
\begin{center}
\begin{tabular}{c}
\includegraphics[width=0.48\textwidth,height=6cm]{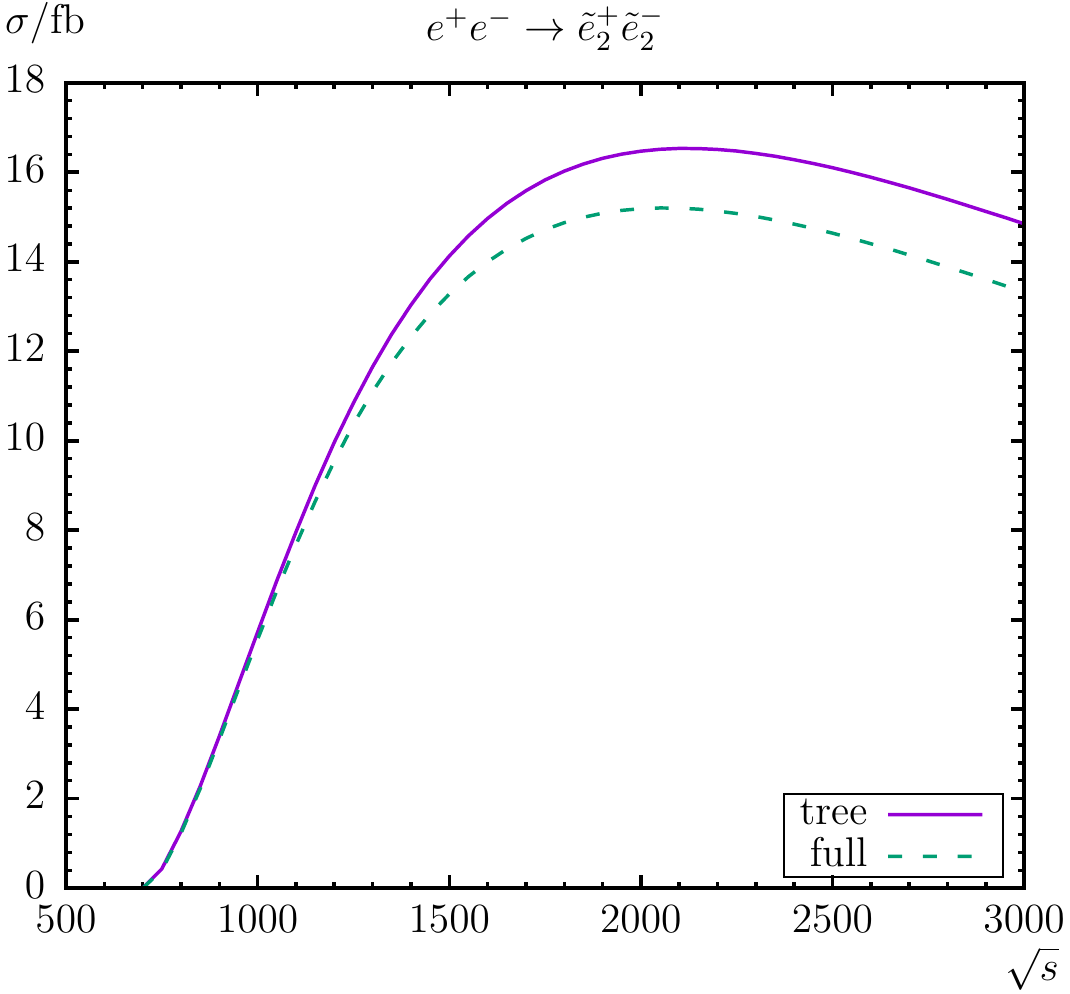}
\includegraphics[width=0.48\textwidth,height=6cm]{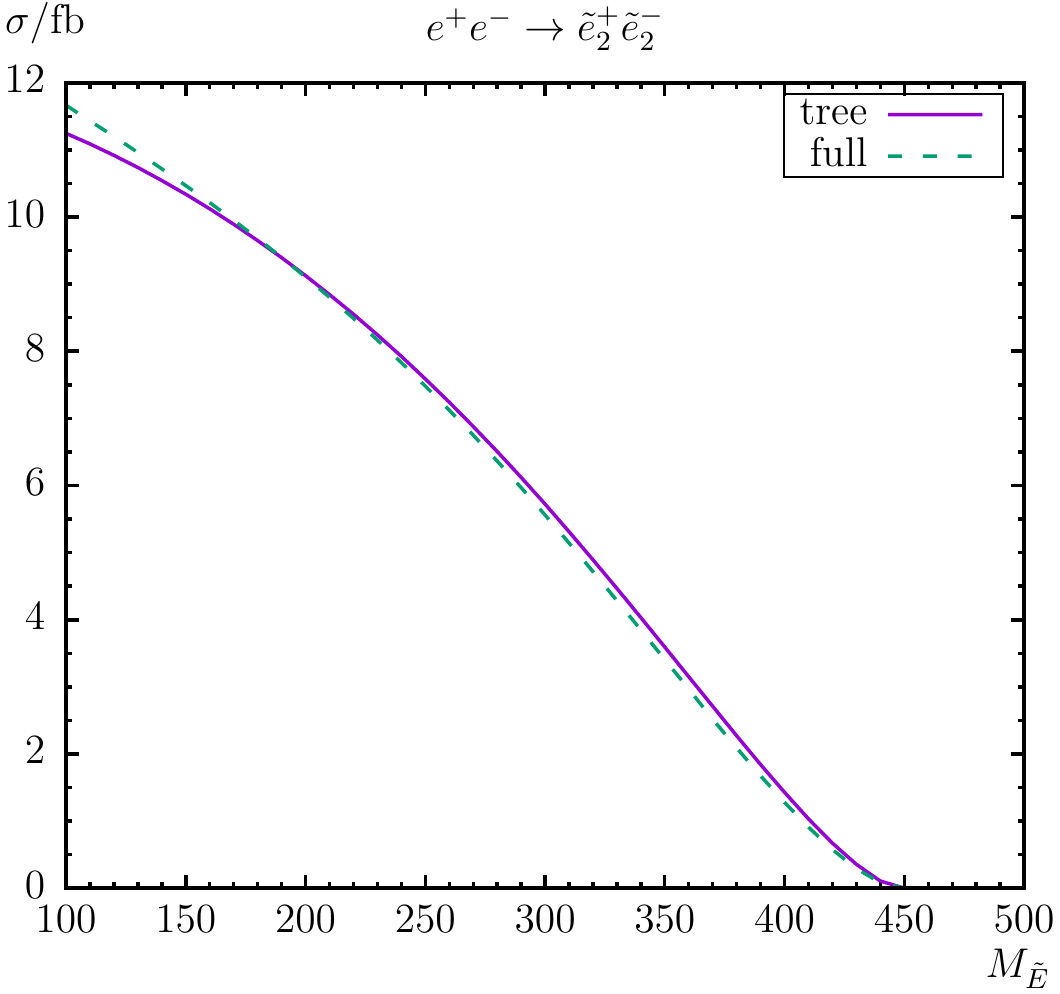}
\\[1em]
\includegraphics[width=0.48\textwidth,height=6cm]{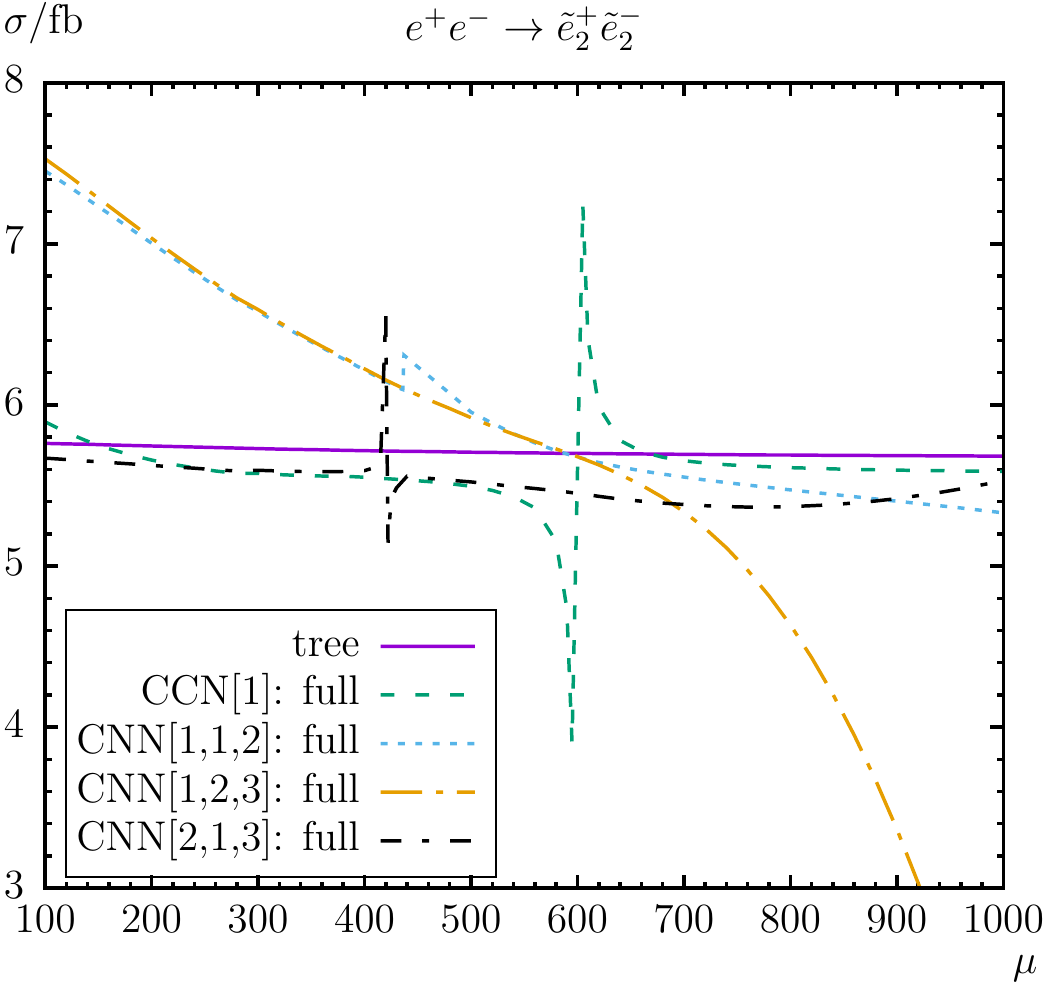}
\includegraphics[width=0.48\textwidth,height=6cm]{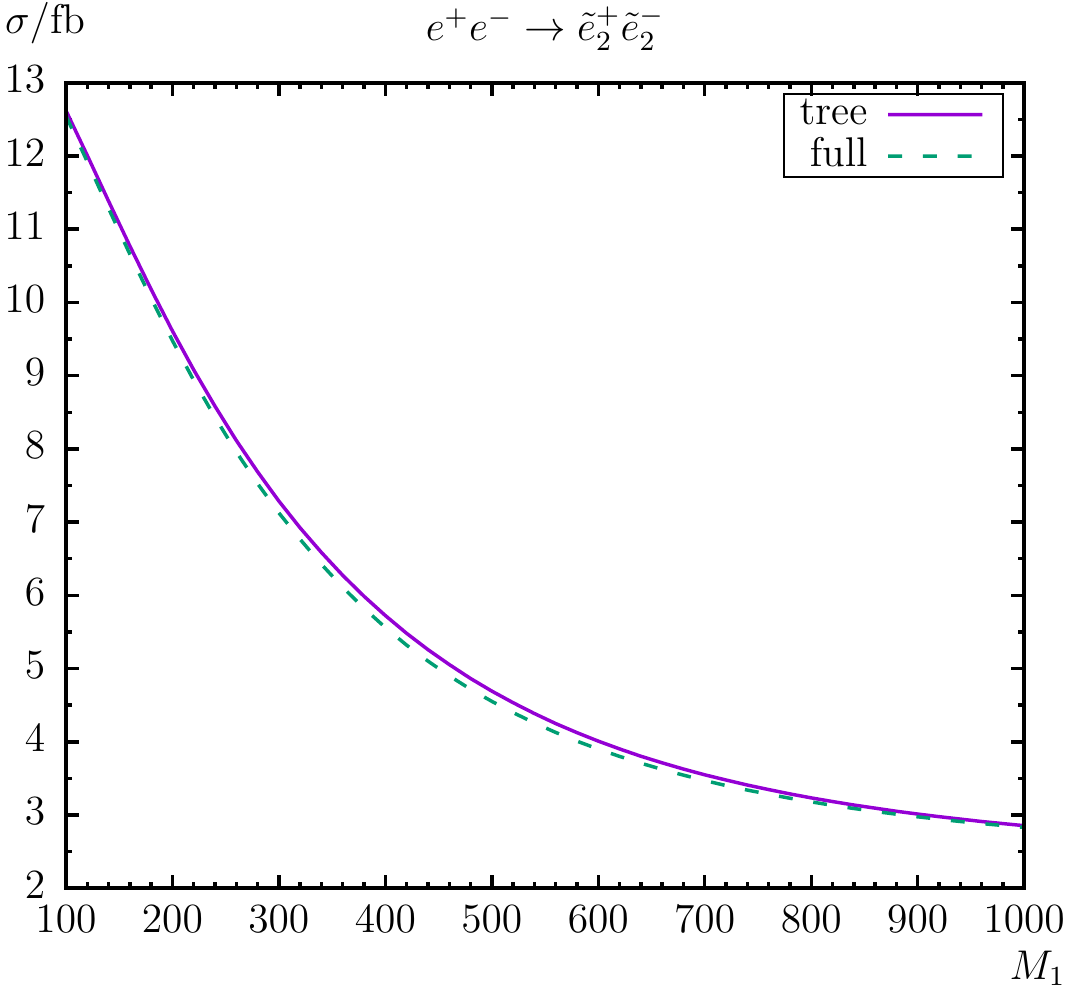}
\\[1em]
\includegraphics[width=0.48\textwidth,height=6cm]{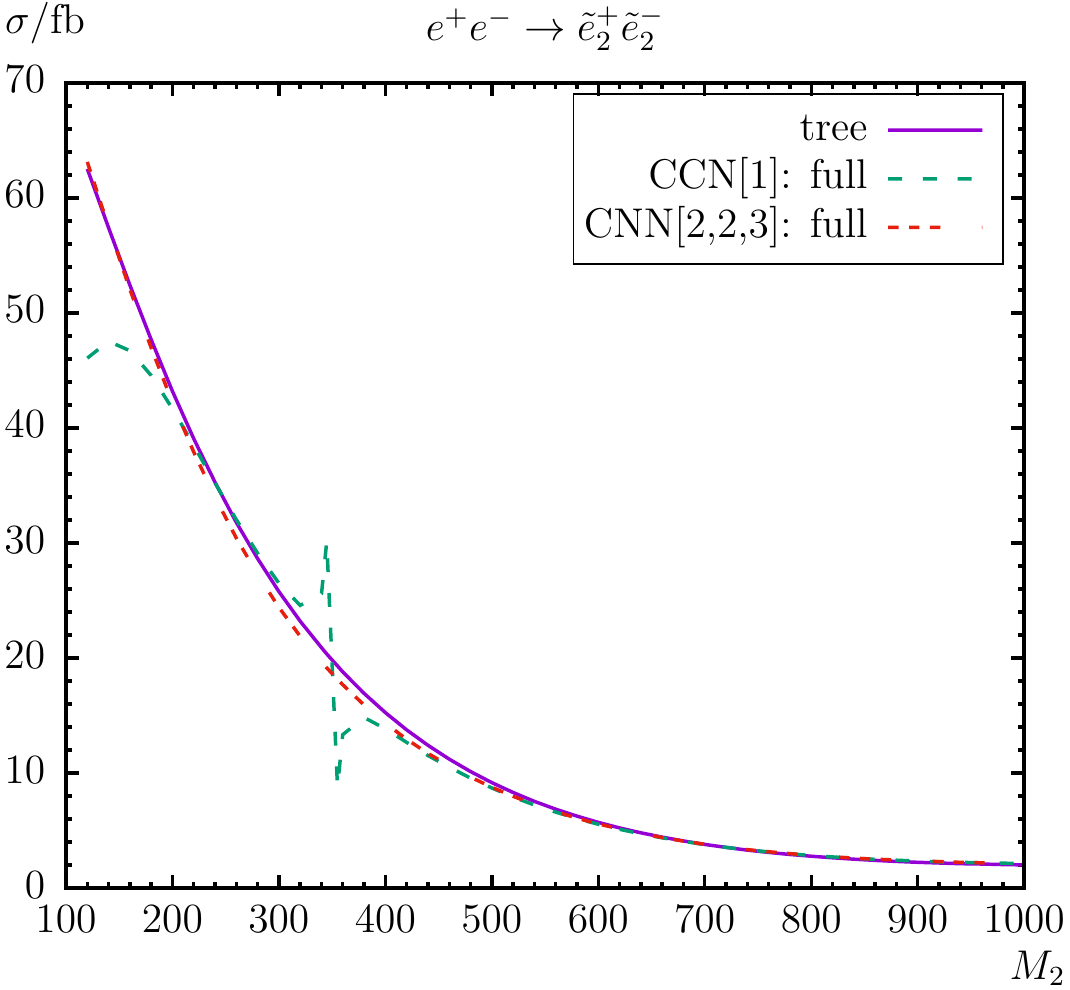}
\includegraphics[width=0.48\textwidth,height=6cm]{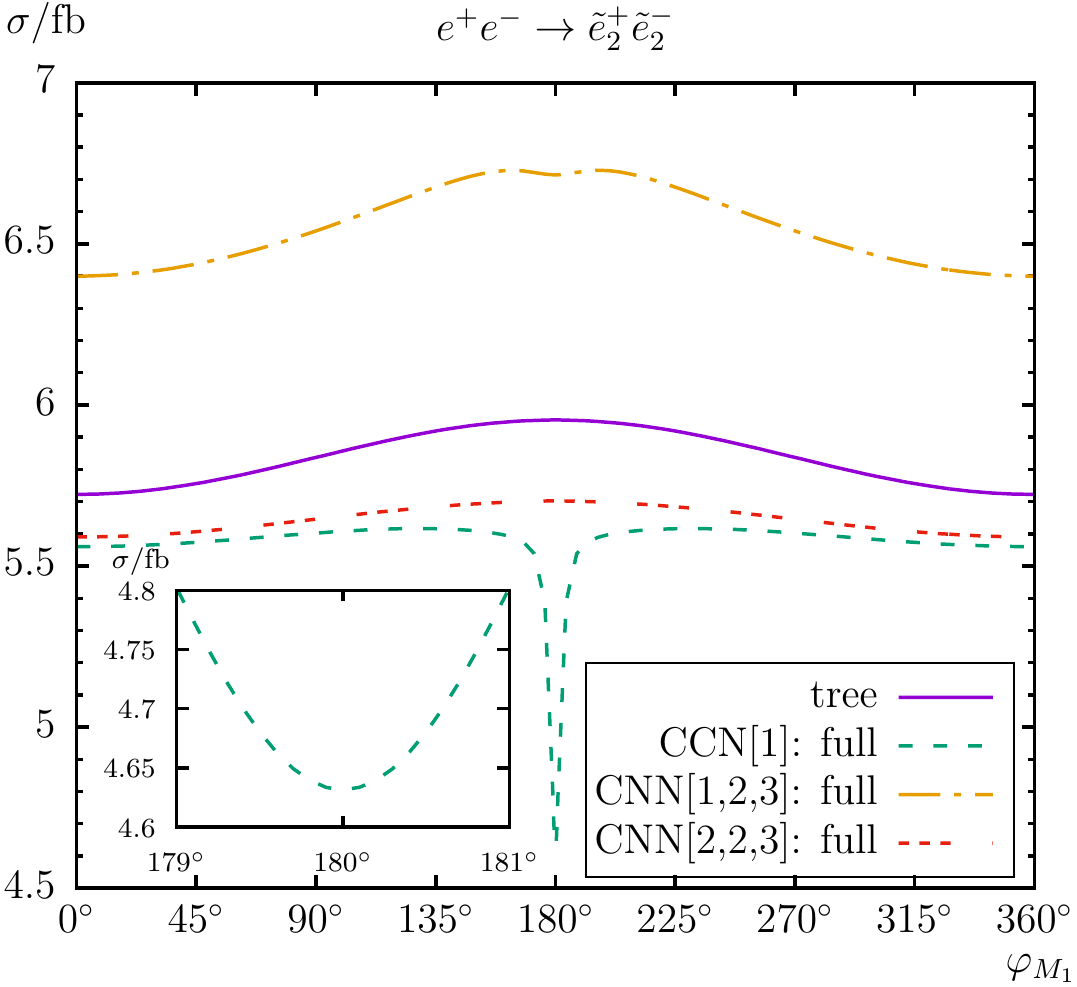}
\end{tabular}
\caption{\label{fig:eeSe2Se2}
  $\sig(\eeSezSez)$.
  Tree-level and full one-loop corrected cross sections are shown 
  with parameters chosen according to \Scs; see \refta{tab:para}.
  The upper plots show the cross sections with $\sqrt{s}$ (left) 
  and $\MSE$ (right) varied; the middle plots show $\mu$ (left) 
  and $|M_1|$ (right) varied; the lower plots show $M_2$ (left) 
  and $\phiMe$ (right) varied. All masses and energies are in GeV.
}
\end{center}
\end{figure}

\begin{figure}
\begin{center}
\begin{tabular}{c}
\includegraphics[width=0.48\textwidth,height=6cm]{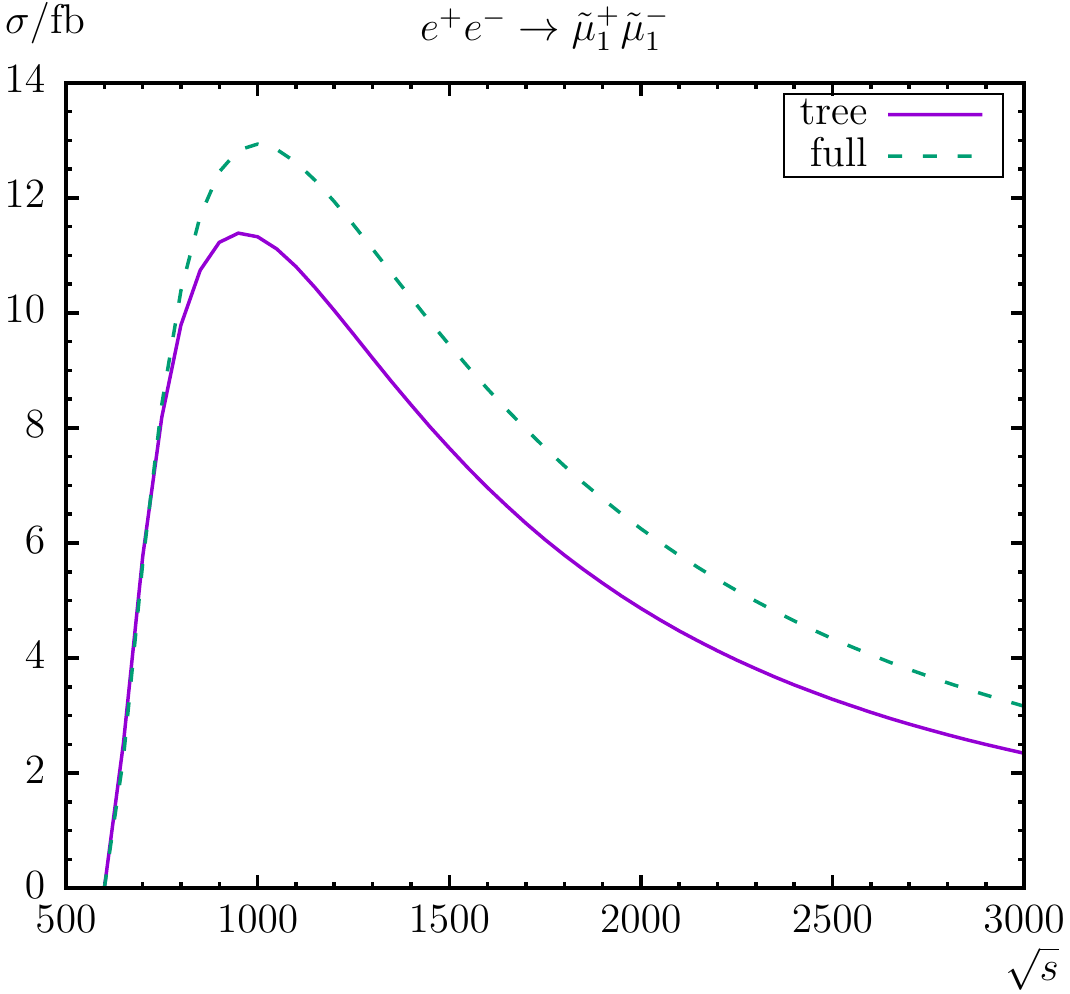}
\includegraphics[width=0.48\textwidth,height=6cm]{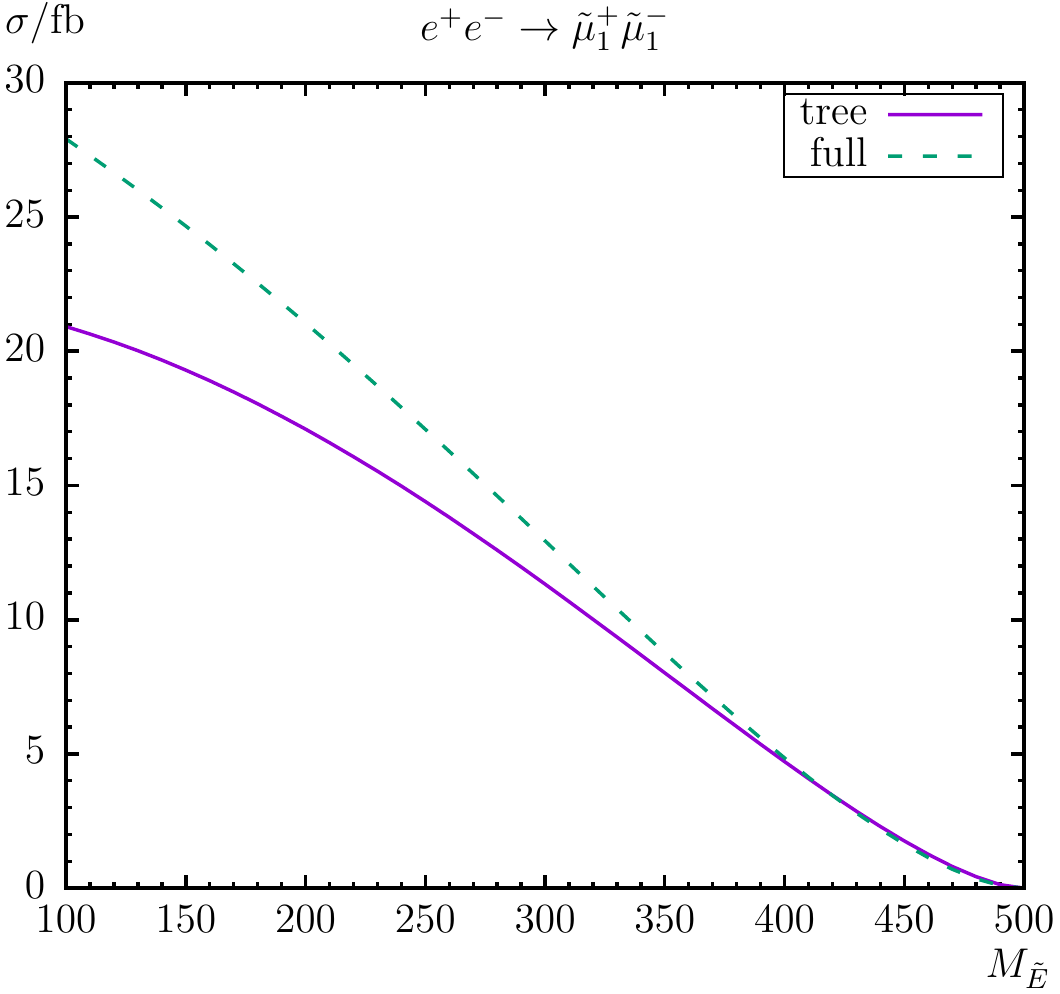}
\end{tabular}
\caption{\label{fig:eeSm1Sm1}
  $\sig(\eeSmeSme)$.
  Tree-level and full one-loop corrected cross sections are shown 
  with parameters chosen according to \Scs; see \refta{tab:para}.
  The plots show the cross sections with $\sqrt{s}$ (left) and 
  $\MSE$ (right) varied. All masses and energies are in GeV.
}
\end{center}
\end{figure}

\begin{figure}
\begin{center}
\begin{tabular}{c}
\includegraphics[width=0.48\textwidth,height=6cm]{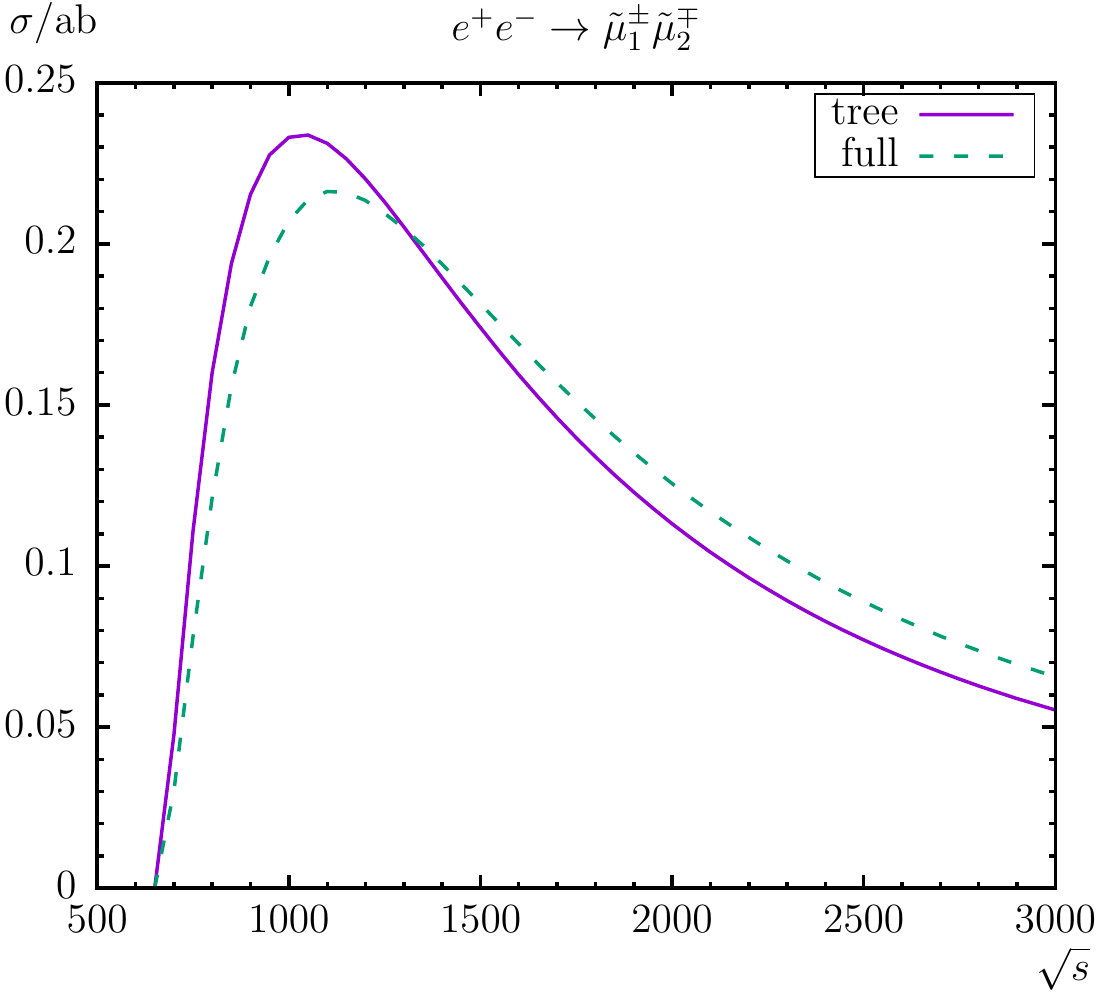}
\includegraphics[width=0.48\textwidth,height=6cm]{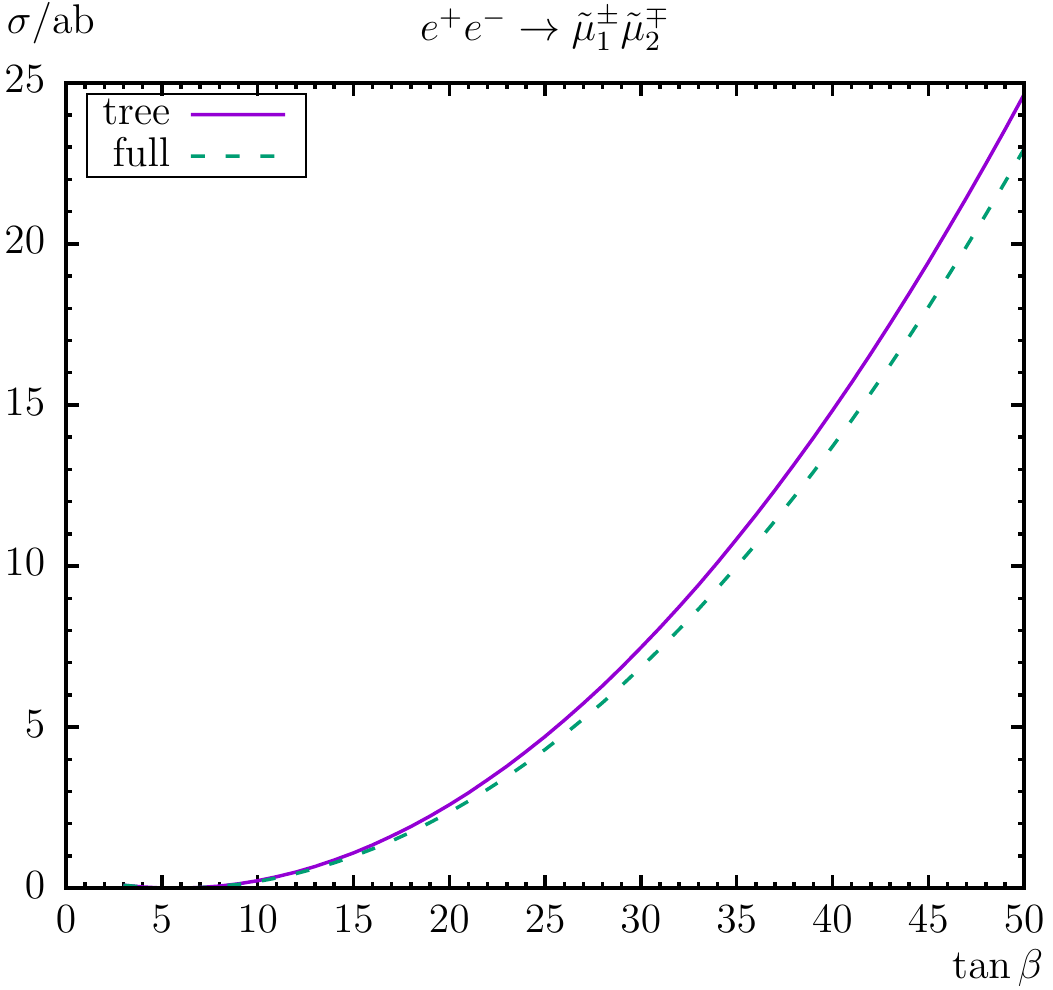}
\end{tabular}
\caption{\label{fig:eeSm1Sm2}
  $\sig(\eeSmeSmz)$.
  Tree-level and full one-loop corrected cross sections are shown 
  with parameters chosen according to \Scs; see \refta{tab:para}.
  The plots show the cross sections with $\sqrt{s}$ (left) and 
  $\MSE$ (right) varied.  All masses and energies are in GeV and the
  cross sections are in atobarn.}
\end{center}
\end{figure}

\begin{figure}
\begin{center}
\begin{tabular}{c}
\includegraphics[width=0.48\textwidth,height=6cm]{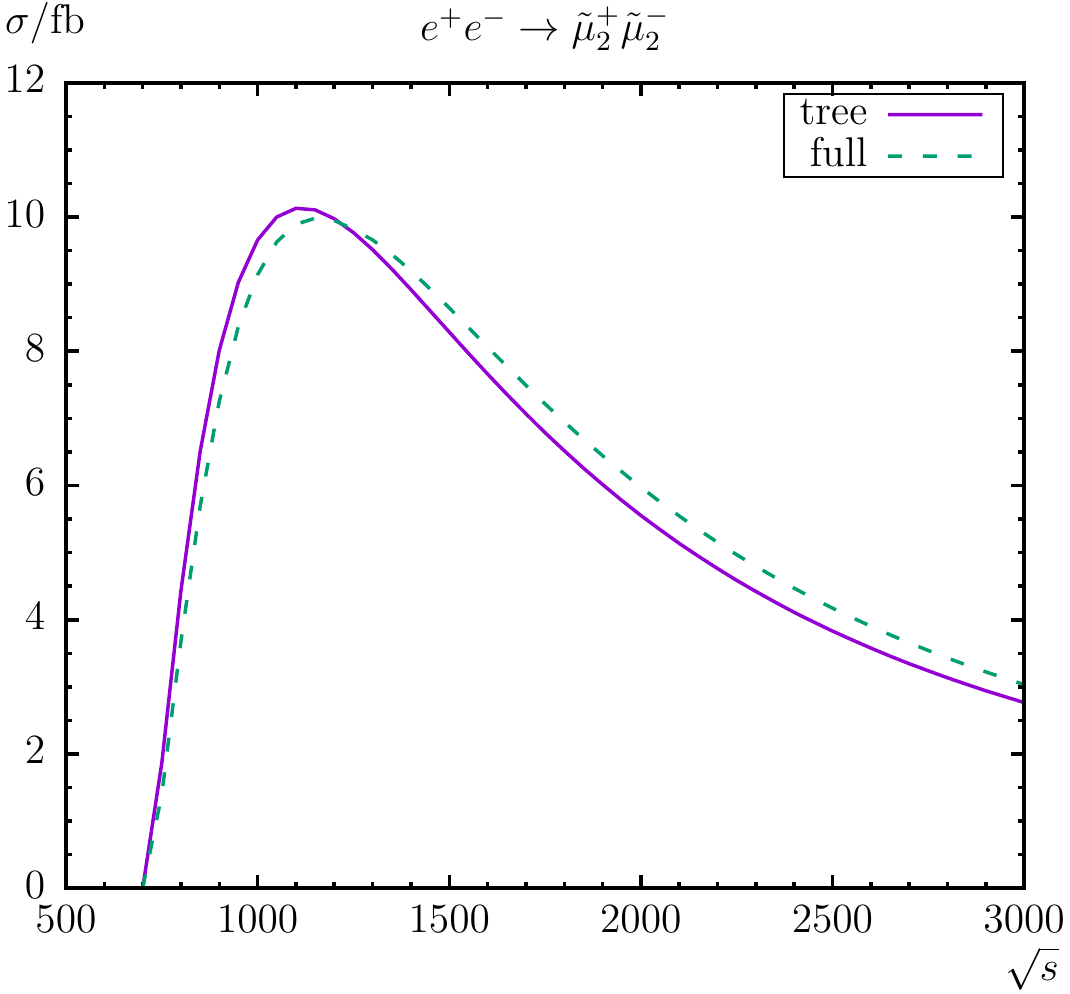}
\includegraphics[width=0.48\textwidth,height=6cm]{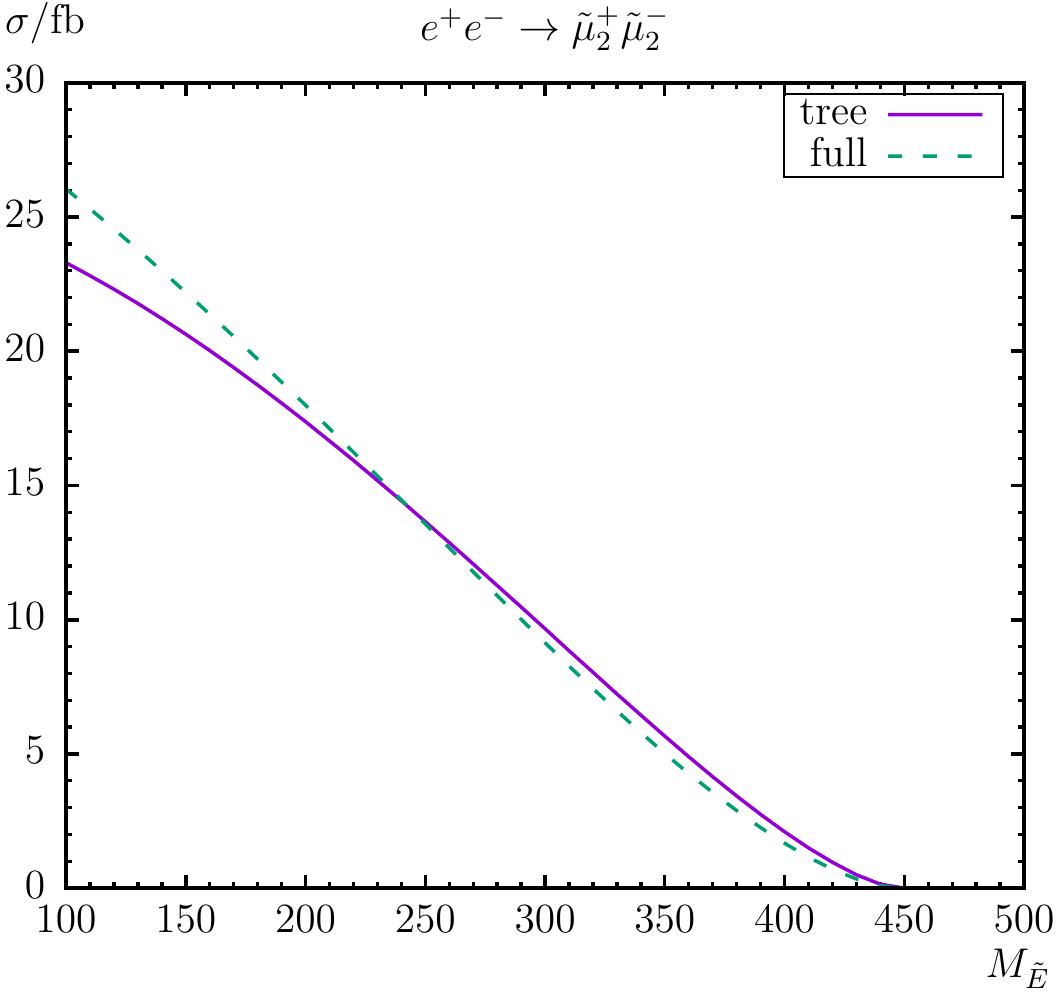}
\end{tabular}
\caption{\label{fig:eeSm2Sm2}
  $\sig(\eeSmzSmz)$.
  Tree-level and full one-loop corrected cross sections are shown 
  with parameters chosen according to \Scs; see \refta{tab:para}.
  The plots show the cross sections with $\sqrt{s}$ (left) and 
  $\MSE$ (right) varied. All masses and energies are in GeV.
}
\end{center}
\end{figure}

\begin{figure}[t]
\begin{center}
\begin{tabular}{c}
\includegraphics[width=0.48\textwidth,height=6cm]{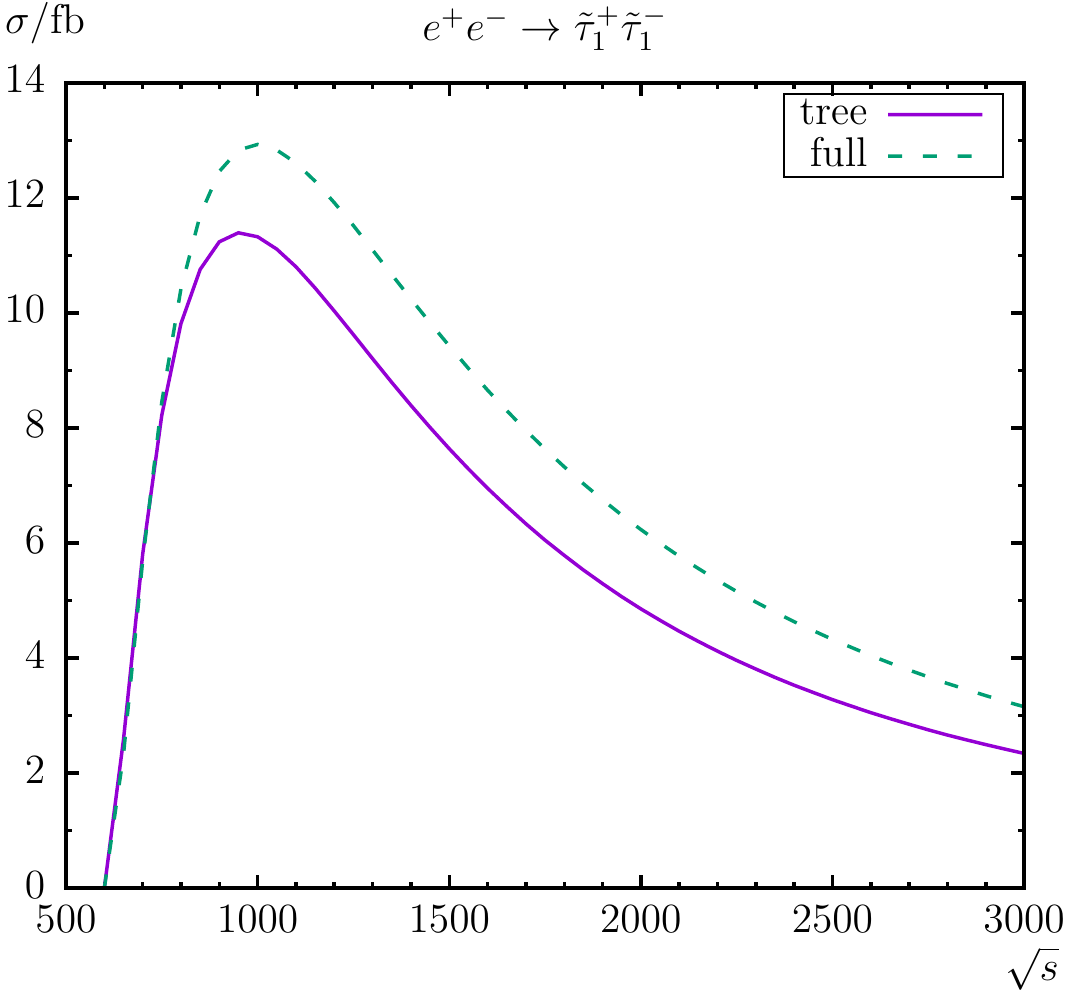}
\includegraphics[width=0.48\textwidth,height=6cm]{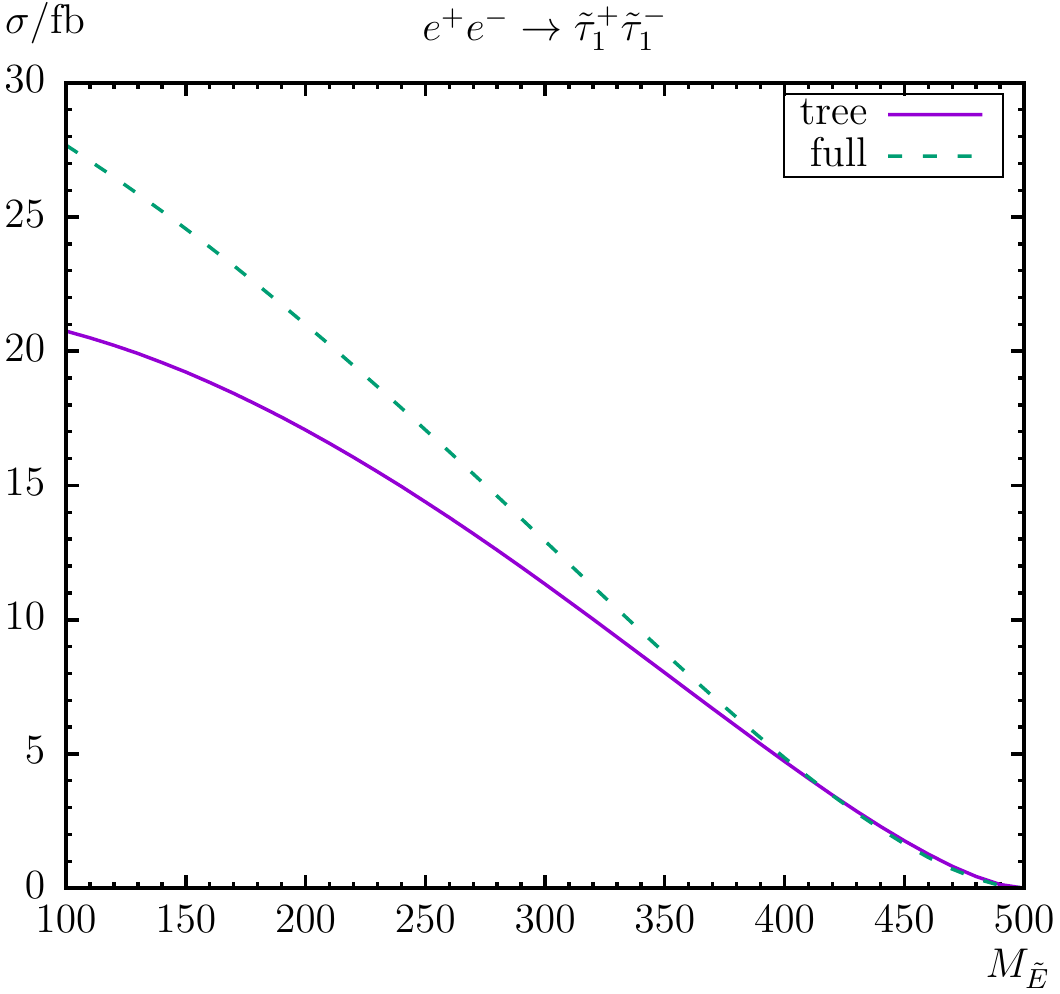}
\\[1em]
\includegraphics[width=0.48\textwidth,height=6cm]{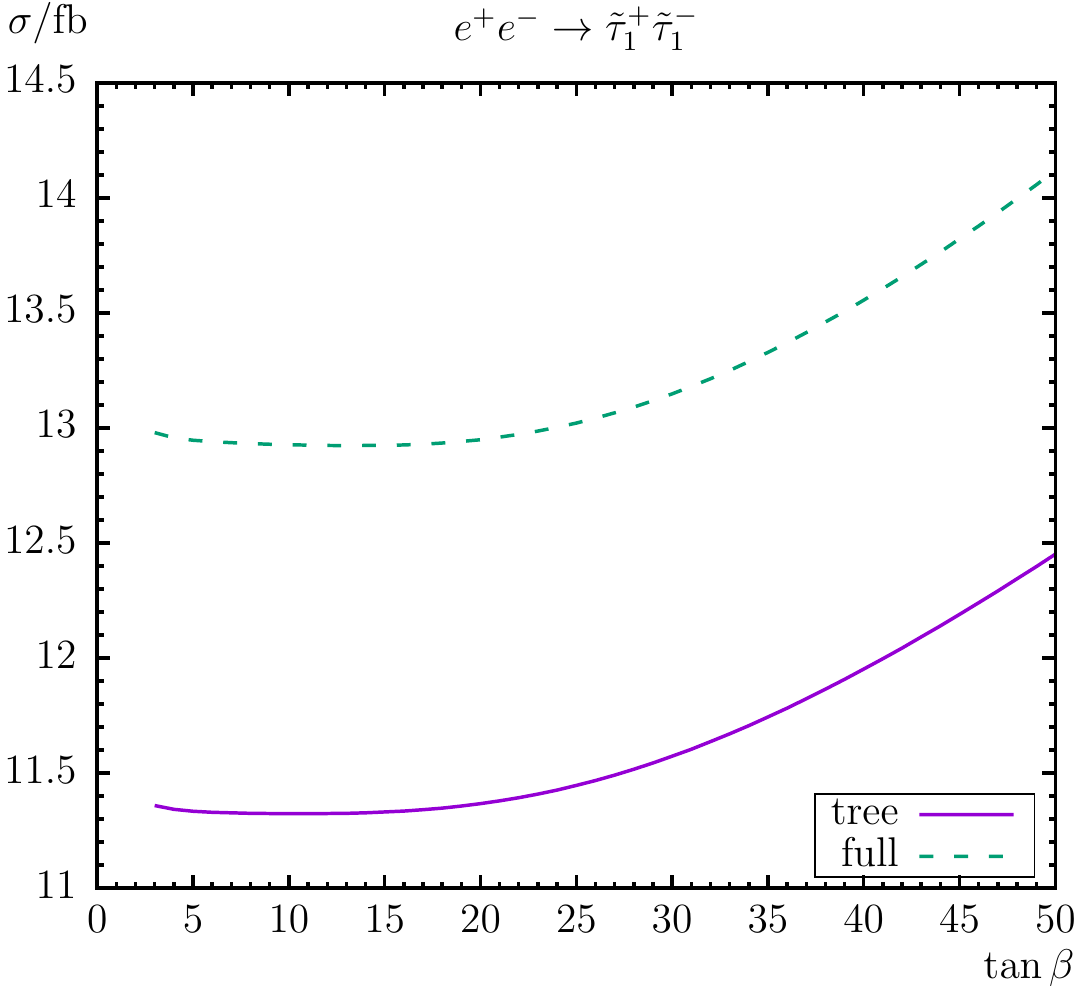}
\includegraphics[width=0.48\textwidth,height=6cm]{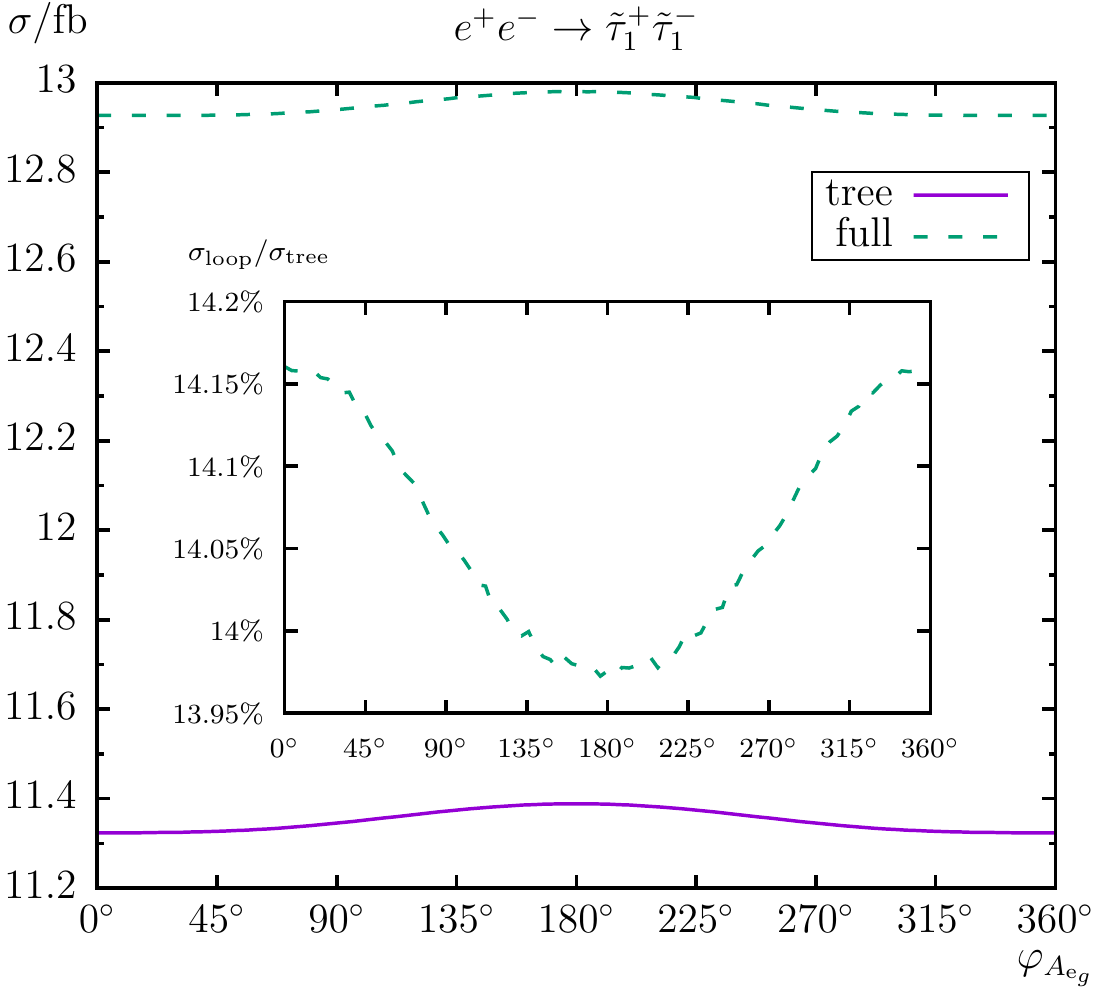}
\end{tabular}
\caption{\label{fig:eeSa1Sa1}
  $\sig(\eeSaeSae)$.
  Tree-level and full one-loop corrected cross sections are shown 
  with parameters chosen according to \Scs; see \refta{tab:para}.
  The upper plots show the cross sections with $\sqrt{s}$ (left) 
  and $\MSE$ (right) varied; the lower plots show $\TB$ (left) and 
  $\phiAeg$ (right) varied. All masses and energies are in GeV.
}
\end{center}
\end{figure}

\begin{figure}
\begin{center}
\begin{tabular}{c}
\includegraphics[width=0.48\textwidth,height=6cm]{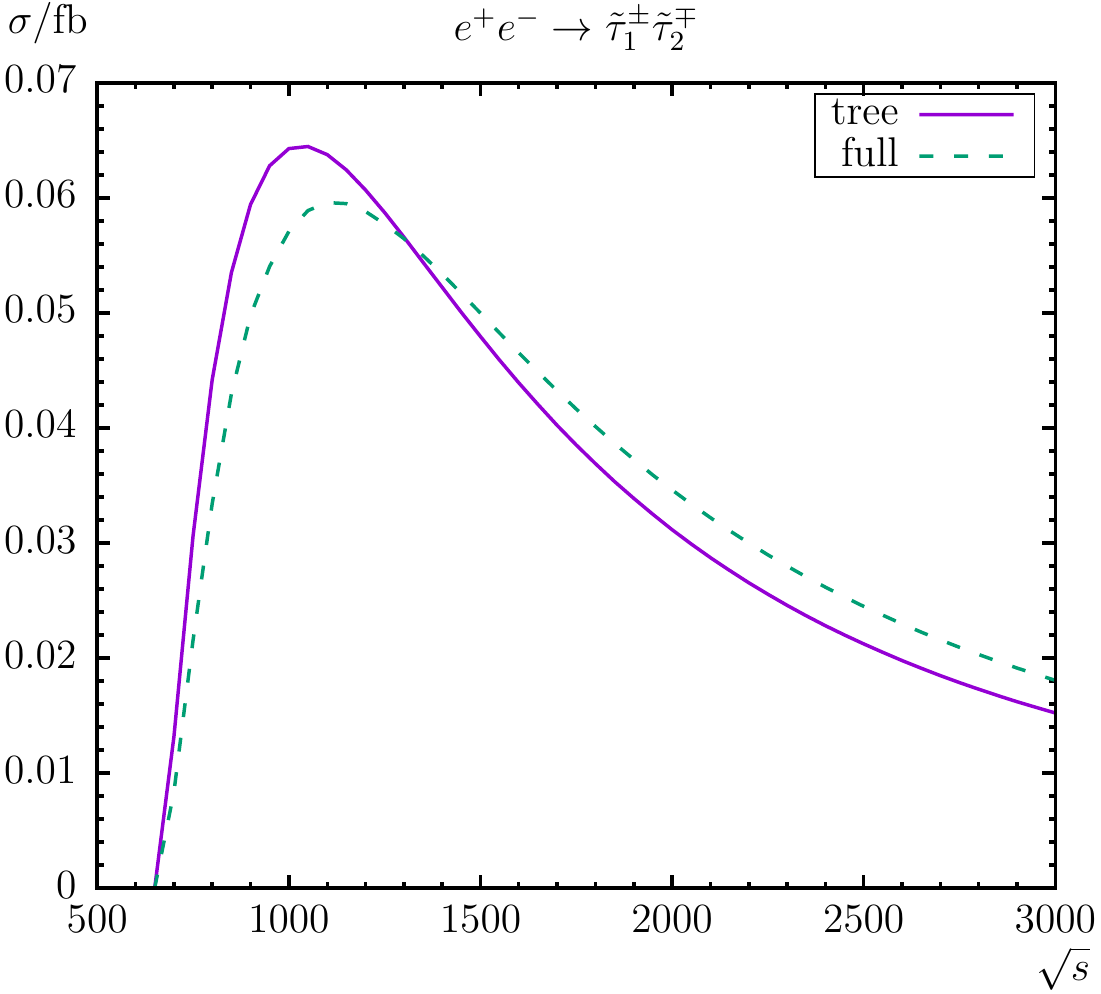}
\includegraphics[width=0.48\textwidth,height=6cm]{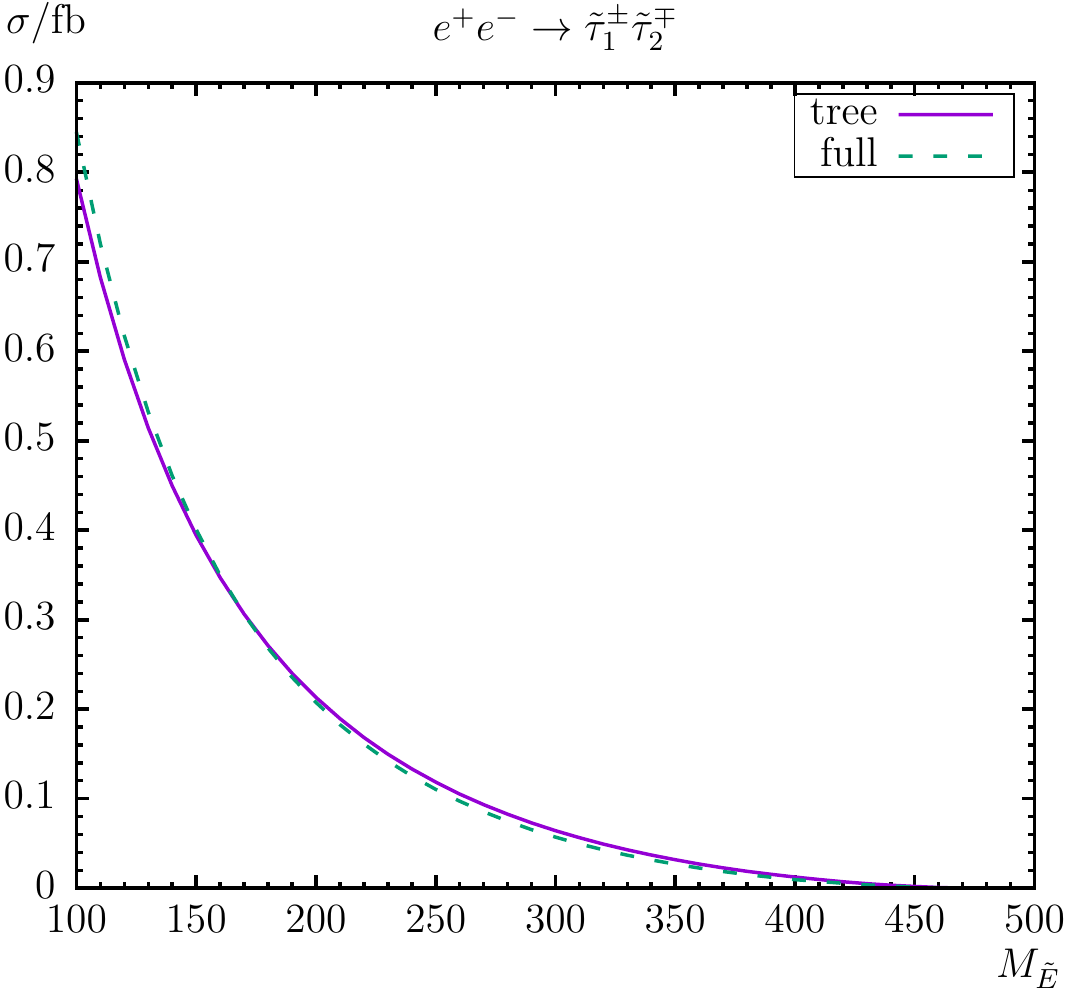}
\\[1em]
\includegraphics[width=0.48\textwidth,height=6cm]{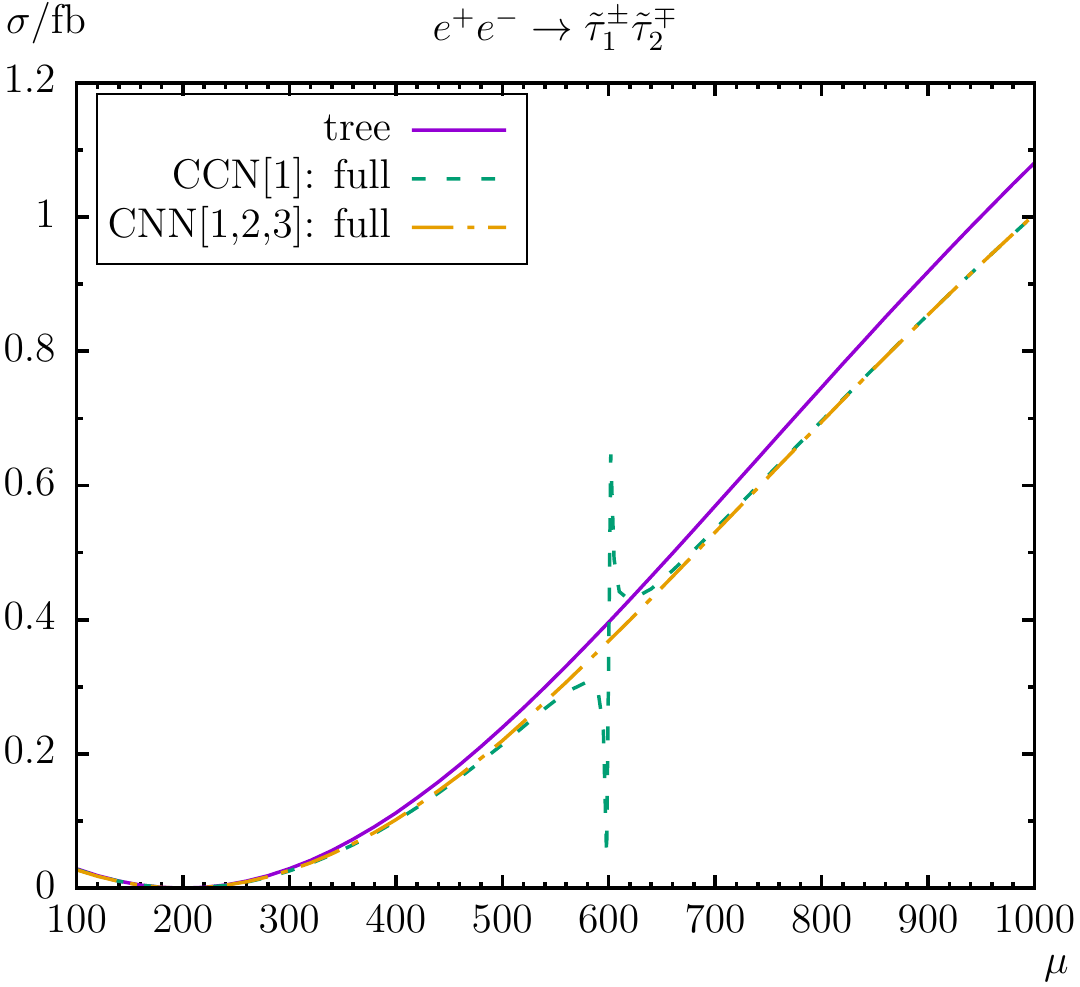}
\includegraphics[width=0.48\textwidth,height=6cm]{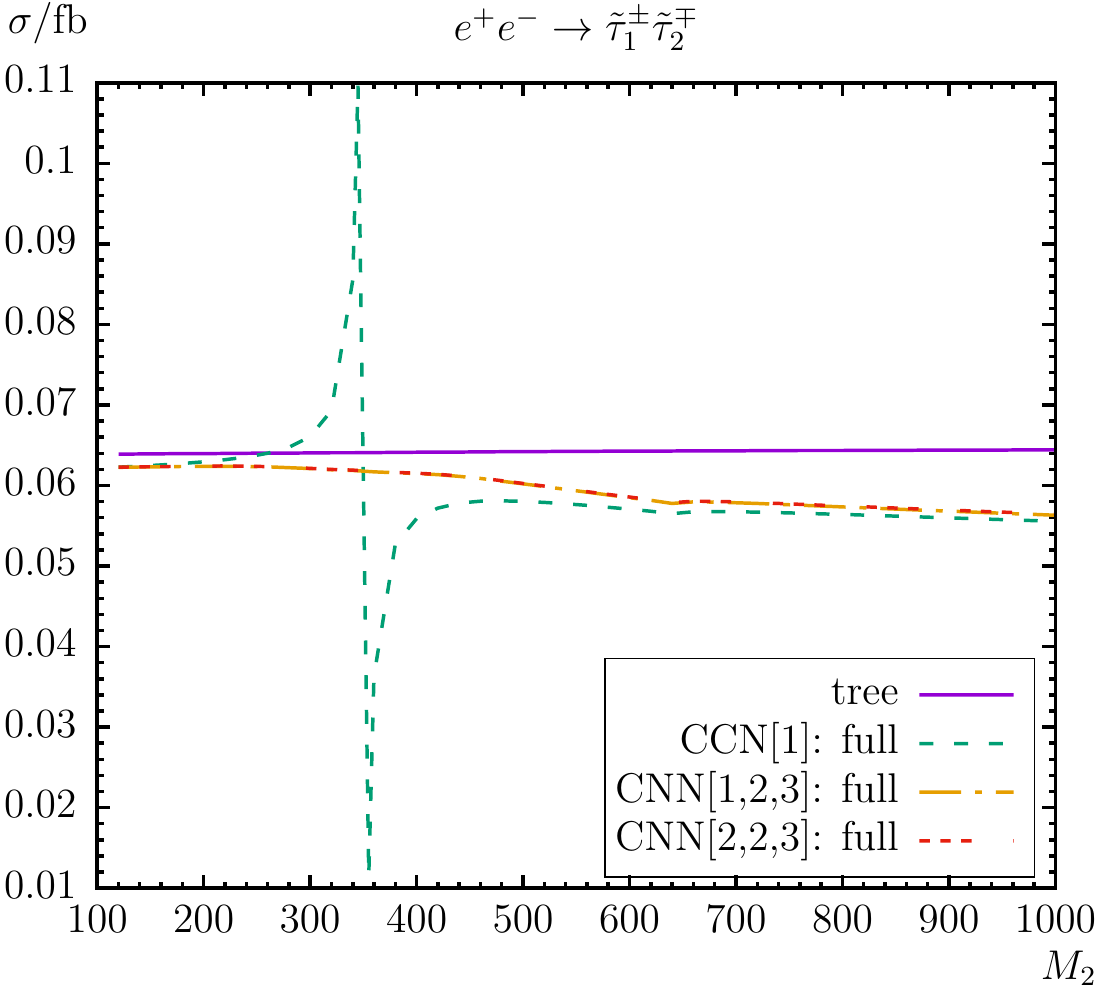}
\\[1em]
\includegraphics[width=0.48\textwidth,height=6cm]{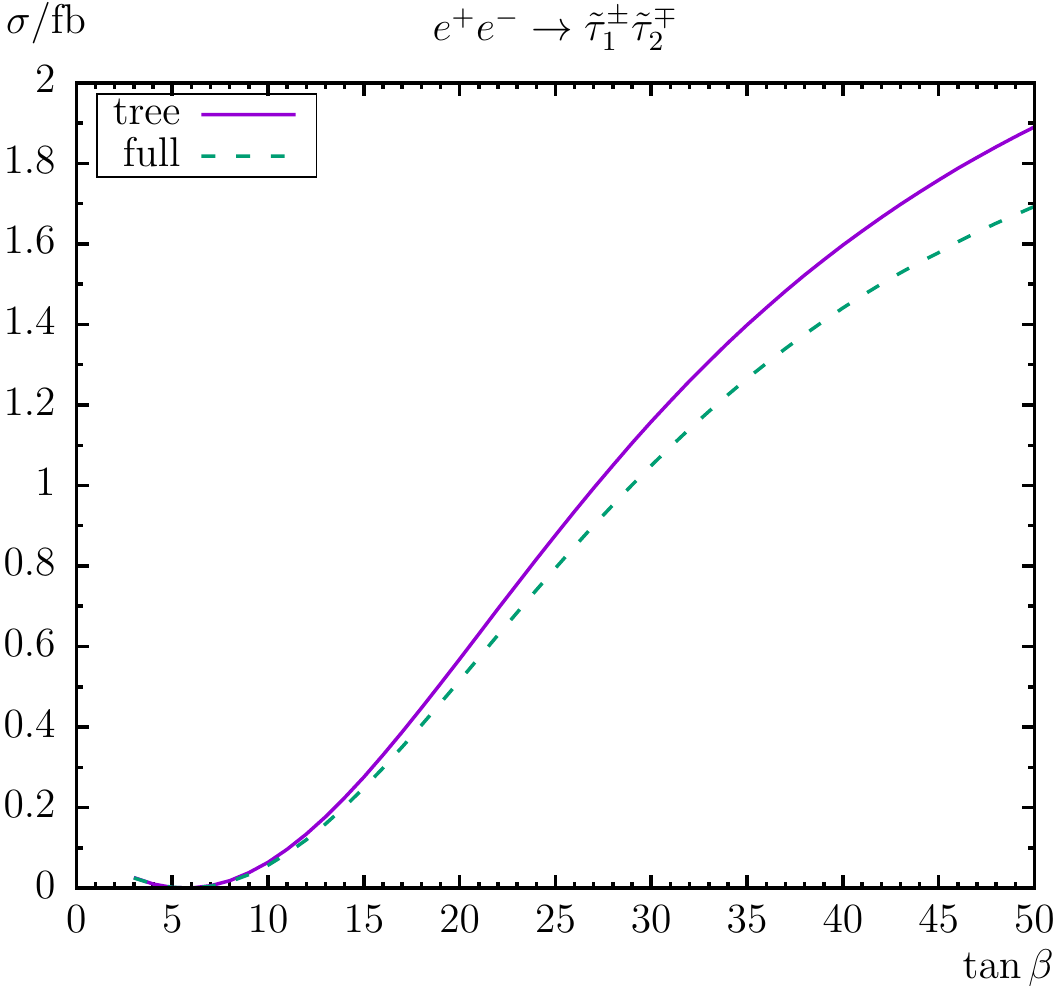}
\includegraphics[width=0.48\textwidth,height=6cm]{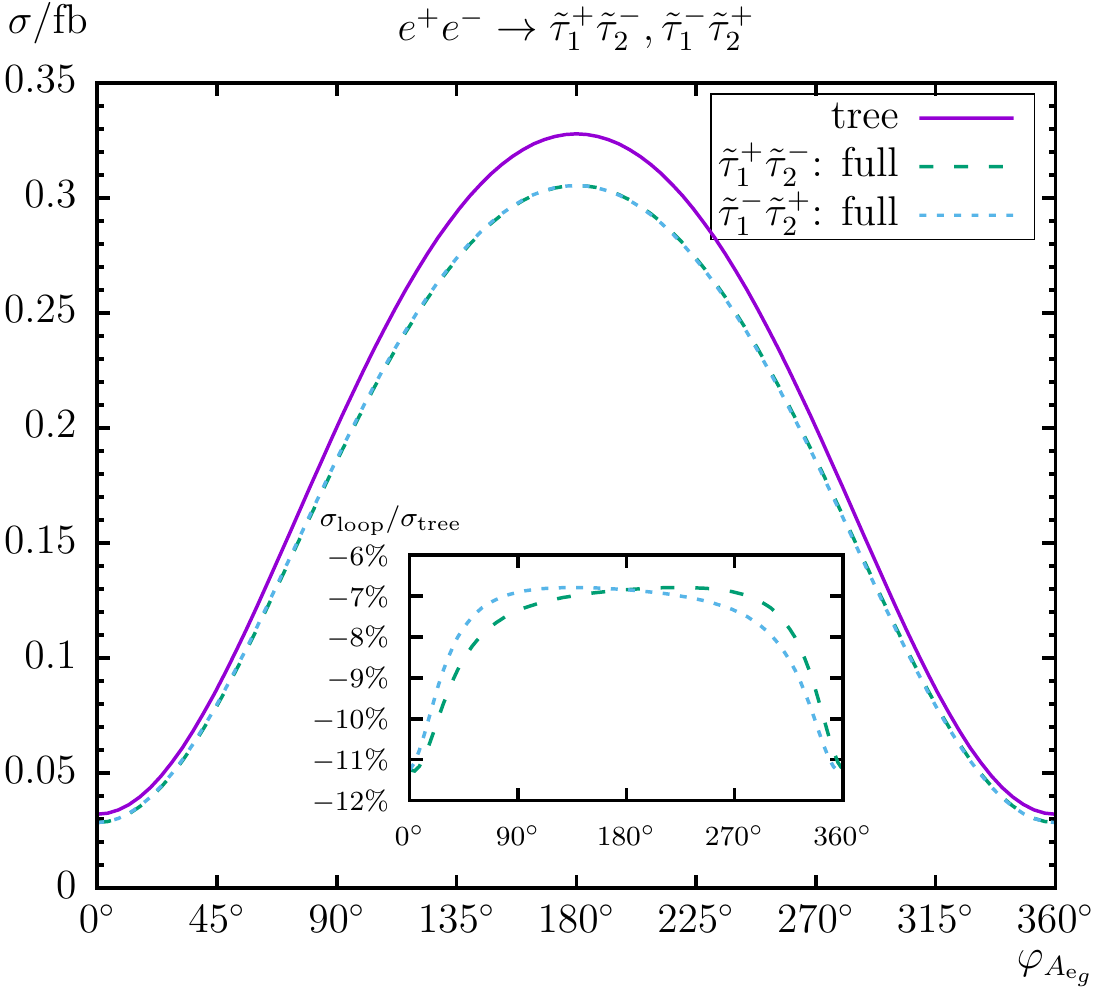}
\end{tabular}
\caption{\label{fig:eeSa1Sa2}
  $\sig(\eeSaeSaz)$.
  Tree-level and full one-loop corrected cross sections are shown 
  with parameters chosen according to \Scs; see \refta{tab:para}.
  The upper plots show the cross sections with $\sqrt{s}$ (left) 
  and $\MSE$ (right) varied; the middle plots show $\mu$ (left) 
  and $M_2$ (right) varied; the lower plots show $\TB$ (left) 
  and $\phiAeg$ (right) varied. All masses and energies are in GeV.
}
\end{center}
\end{figure}

\begin{figure}
\begin{center}
\begin{tabular}{c}
\includegraphics[width=0.48\textwidth,height=6cm]{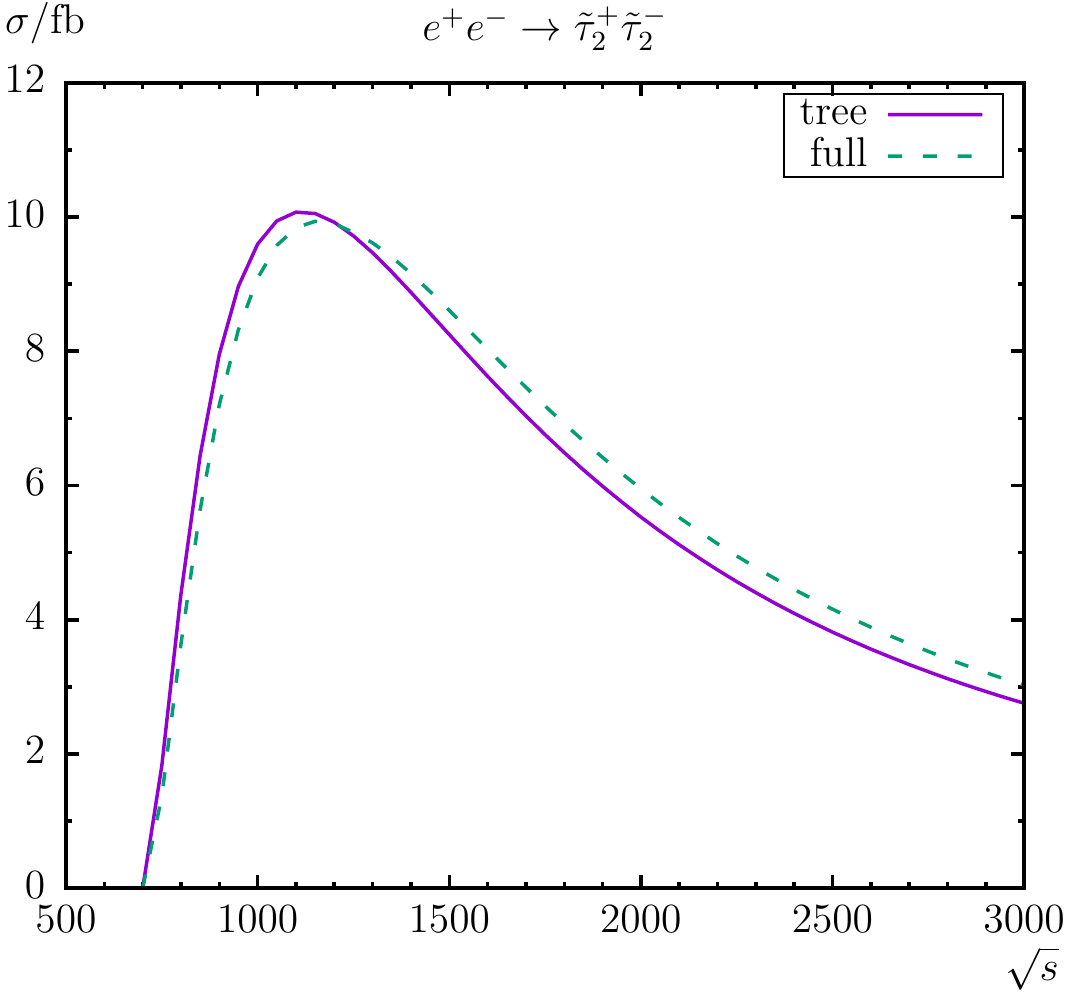}
\includegraphics[width=0.48\textwidth,height=6cm]{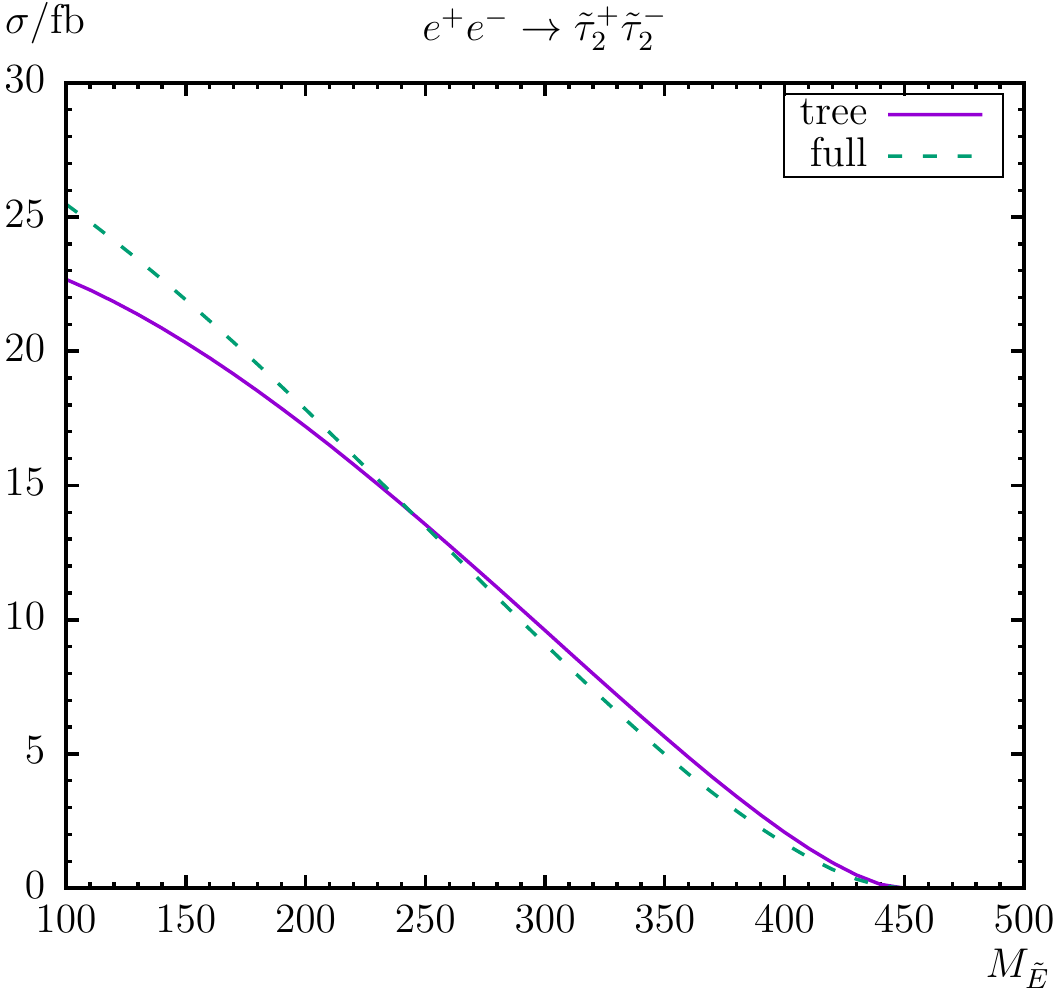}
\\[1em]
\includegraphics[width=0.48\textwidth,height=6cm]{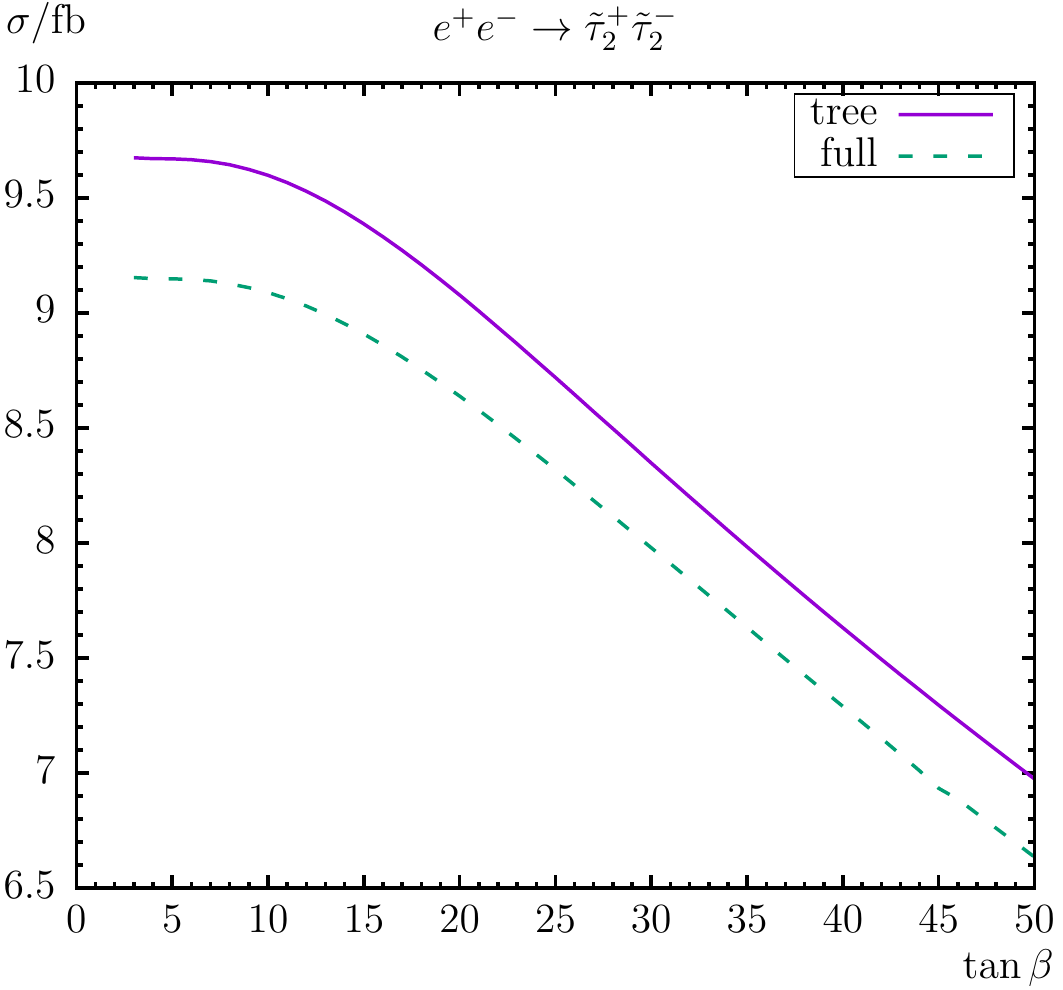}
\includegraphics[width=0.48\textwidth,height=6cm]{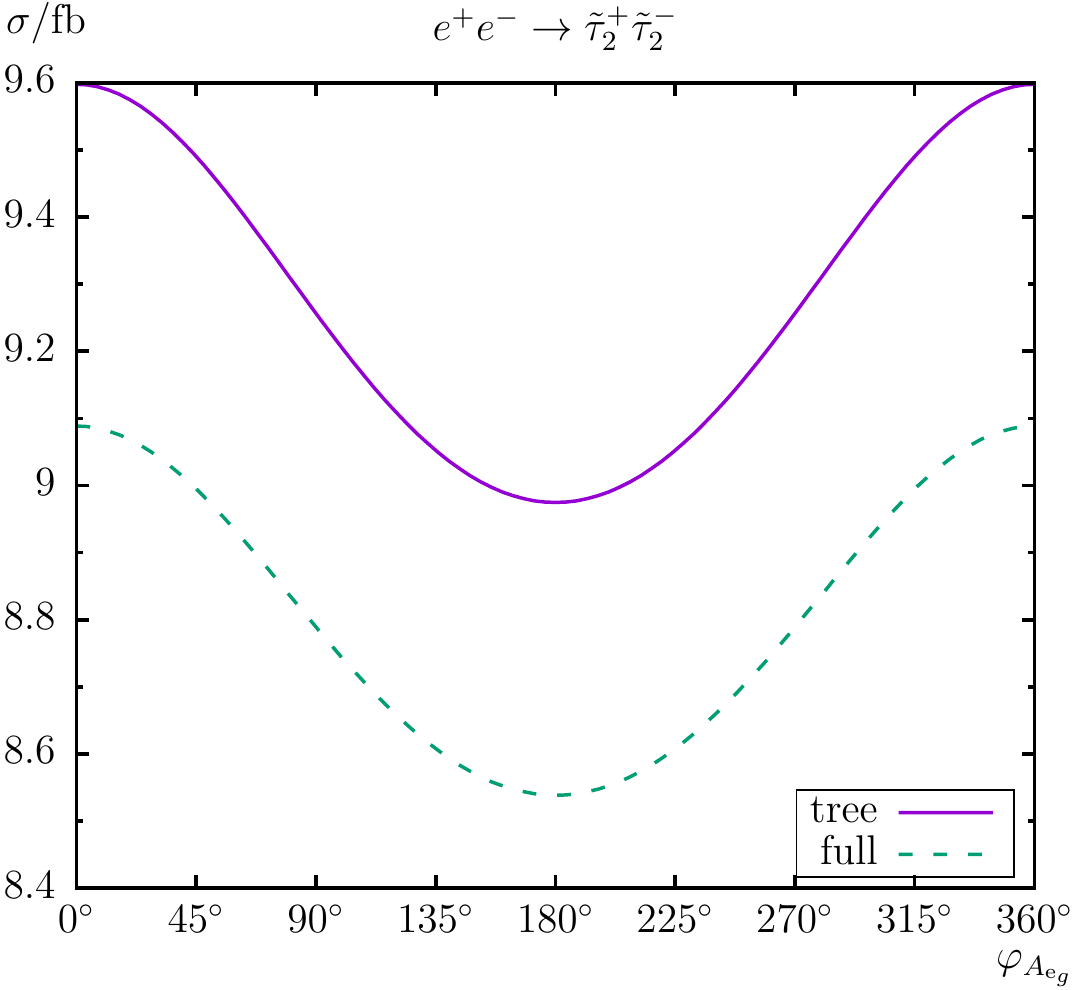}
\end{tabular}
\caption{\label{fig:eeSa2Sa2}
  $\sig(\eeSazSaz)$.
  Tree-level and full one-loop corrected cross sections are shown 
  with parameters chosen according to \Scs; see \refta{tab:para}.
  The upper plots show the cross sections with $\sqrt{s}$ (left) 
  and $\MSE$ (right) varied; the lower plots show $\TB$ (left) and 
  $\phiAeg$ (right) varied. All masses and energies are in GeV.
}
\end{center}
\end{figure}

In \reffis{fig:eeSe1Se1} -- \ref{fig:eeSa2Sa2} we show the results for the 
processes \eeSeSe\ ($g = 1,2,3;\, s,\spri = 1,2$) as a function of $\sqrt{s}$, 
$\MSE$, $\mu$, $|M_1|$, $M_2$, $\TB$, $\phiMe$, and $\phiAeg$.  It should be 
noted that for $s \to \infty$ decreasing cross sections $\propto \ln{(s)}/s$
for the first and $\propto 1/s$ for the second and third slepton generations 
are expected; see \citere{FrMaZe2004}.
We also remind the reader that $\sig(\eeSeSe)$ denotes the sum of the two 
charge conjugated processes $\forall\; s \neq s^{\prime}$; 
see \refeq{eq:eeSeSesum}.

\medskip

We start with the process $\eeSeeSee$ shown in \reffi{fig:eeSe1Se1}.
Away from the production threshold, loop corrections of $\sim +6\,\%$ at 
$\sqrt{s} = 1000\gev$ are found in scenario \Scs\ (see \refta{tab:para}), 
with a maximum of $27\,\fb$ at $\sqrt{s} \approx 1700\gev$. 
The relative size of the loop corrections increase with increasing 
$\sqrt{s}$ and reach $\sim +21\,\%$ at $\sqrt{s} = 3000\gev$.
A ``tree crossing'' (\ie where the loop corrections become 
zero and therefore cross the tree-level result) can be found at 
$\sqrt{s} \approx 825\gev$. 

The cross sections are decreasing with increasing $\MSE$ due to 
kinematics, and the full one-loop result has its maximum of $\sim 28\,\fb$ 
at $\MSE = 100\gev$. Analogously the relative corrections are decreasing 
from $\sim +13\,\%$ at $\MSE = 100\gev$ to $\sim -30\%$ at $\MSE = 490\gev$. 
The tree crossing takes place at $\MSE \approx 375\gev$.  For higher $\MSE$ 
values the loop corrections are negative, where the relative size becomes 
large due to the (relative) smallness of the tree-level results, which goes 
to zero for $\MSE \approx 500\gev$ due to kinematics.
For the other parameter variations one can conclude that a cross section 
roughly twice as large can be possible for very low $\MSE$ 
(which, however, are challenged by the current ATLAS/CMS exclusion bounds).

With increasing $\mu$ in \Scs\ (middle left plot) we find a decrease of 
the loop corrected production cross section within \Code{CCN[1]}
(green dashed line). 
The relative loop corrections reach $\sim +18\,\%$ at $\mu = 100\gev$ 
(at the border of the experimental exclusion bounds) and $\sim +7\,\%$ 
at $\mu = 600$-$1000\gev$.
One can see the expected breakdown of the \Code{CCN[1]} scheme for 
$\mu = M_2$, \ie in our case at $\mu \approx M_2 = 600\gev$ 
(see also \citeres{LHCxN,LHCxNprod}).  Therefore, in the middle left 
plot of \reffi{fig:eeSe1Se1} also the corresponding results are shown 
for the \Code{CNN[1,2,3]} (yellow dash-dotted line) and 
\Code{CNN[2,1,3]} (black dash-dotted line) schemes, which are smooth 
at $\mu = M_2$.
Outside the region of $\mu \sim M_2$ the scheme \Code{CCN[1]} is
expected to be reliable, since each of the three OS conditions is
strongly connected to one of the three input parameters, $|M_1|$, $M_2$ 
and $\mu$.  Similarly, \Code{CNN[1,2,3]} is expected to be reliable for 
$\mu$ smaller than $M_2$, as in this case, again each of the three OS 
renormalization conditions is strongly connected to the three input 
parameters.  This behavior can be observed in the plot: for 
$\mu \lsim M_2 = 600\gev$ \Code{CNN[1,2,3]} is reliable, while for
$\mu \gsim M_2 = 600\gev$ it becomes unreliable.
While $\delta \mu$ in the \Code{CNN[2,1,3]} scheme has a strong minimum  
at $\mu \approx 421\gev$, dominating the loop corrections, it approximates 
\Code{CCN[1]} very good for $\mu \gsim 500\gev$.  A rising deviation 
between the schemes can be observed for $\mu < 250\gev$, where the 
schemes \Code{CNN[1,2,3]} and \Code{CNN[2,1,3]} are nearly constant, 
\ie independent of $\mu$. 
Overall, it is possible to find for every value of $\mu$ a renormalization 
scheme that behaves stable and ``flat'' \wrt the tree-level cross section.

With increasing $|M_1|$ in \Scs\ (middle right plot) we find a strong 
decrease of the production cross section, due to the change in the 
interference of the $\neu1$ (dominant for $|M_1| \lsim 340\gev$),
$\neu3$ (dominant for $340\gev \lsim |M_1| \lsim 610\gev$), and
$\neu4$ (dominant for $|M_1| \gsim 610\gev$) in the $t$-channel.
It should be noted that there is \textit{no} tree crossing in this plot.
The loop corrections decrease from $\sim +11\%$ at $|M_1| = 100\gev$ to
$\sim +3.8\%$ at $|M_1| = 530\gev$ and then increase to $\sim +15\%$ at 
$|M_1| = 920\gev$.  However, for $|M_1| \gsim 700\gev$ the production
cross section becomes relatively small.

The dependence $M_2$ of the cross section in \Scs\ is shown in the 
lower left plot.  One can clearly see the expected breakdown of the 
\Code{CCN[1]} scheme for $M_2 = \mu$, \ie in our case at 
$M_2 \approx \mu = 350\gev$ (see also \citeres{LHCxN,LHCxNprod}) 
and the smooth behavior of \Code{CNN[2,2,3]} (red dashed line)
around $M_2 \sim \mu = 350\gev$.
Within \Code{CCN[1]} the cross section even turns out to be negative for 
$M_2 < 154\gev$ due to a maximum of $\delta M_1$ at $M_2 \approx 20\gev$, 
dominating the loop corrections.  This renders the \Code{CCN[1]} scheme 
to be unreliable for $M_2 \lsim \mu$.
For $M_2 \gsim \mu$, on the other hand, the scheme \Code{CCN[1]} is 
expected to be reliable, since each of the three OS conditions is strongly 
connected to one of the three input parameters, $|M_1|$, $M_2$ and $\mu$.
The loop corrections at the level of $\sim +6\,\%$ are found to be nearly 
independent of $M_2$ for $M_2 \gsim \mu$ within the \Code{CCN[1]} scheme. 
And corrections of $\sim +7\,\%$ are found to be (nearly) independent of 
$M_2$ within the \Code{CNN[2,2,3]} scheme.

Now we turn to the complex phase dependence.  We find that the phase 
dependence $\phiMe$ of the cross section in \Scs\ is large for 
\Code{CCN[1]} (lower right plot, green dashed line).  Loop corrections 
at the level of $\sim +6\,\%$ at $\phiMe = 0^{\circ},360^{\circ}$ and 
$\sim -48\,\%$ at $\phiMe = 180^{\circ}$ are found.  It should be noted 
here, that there is \textit{no} divergency or threshold at 
$\phiMe = 180^{\circ}$, the cross section is smooth/finite; 
see the inlay in the lower right plot.  This large structure is caused
by a strong (local) minimum of the renormalization constant 
$\delta M_1$ at $\phiMe = 180^{\circ}$, dominating the loop corrections.
Using another renormalization scheme (\eg\ \Code{CNN[2,2,3]}) the dip 
disappears, as can be seen in the plot.  The one-loop corrections then 
reach $\sim +7\,\%$ and are nearly independent from $\phiMe$.
The loop effects of $\phiAeg$, on the other hand, are tiny and therefore 
not shown explicitely.

\medskip

The relative corrections for the process $\eeSeeSez$, as shown in 
\reffi{fig:eeSe1Se2}, are rather large for the parameter set chosen; 
see \refta{tab:para}.  In the upper left plot of \reffi{fig:eeSe1Se2} 
the relative corrections grow from $\sim +16.6\,\%$ at 
$\sqrt{s} \approx 1000\gev$ (\ie \Scs) up to $\sim +35\,\%$ at 
$\sqrt{s} \approx 3000\gev$.  The tree crossing takes place at 
$\sqrt{s} \approx 730\gev$ (where the higher-order corrections are 
relatively small around the crossing) 
and the maximum cross section of $52\,\fb$ is reached at 
$\sqrt{s} \approx 850\gev$.

The dependence on $\MSE$ is shown in the upper right plot of
\reffi{fig:eeSe1Se2} and follows the same pattern as for $\eeSeeSee$,
\ie a strong decrease with increasing $\MSE$ as obvious from kinematics. 
The loop corrections decrease from $\sim +28\%$ at $\MSE = 100\gev$ to 
$\sim -31\%$ at $\MSE = 470\gev$ (the latter is due to the smallness of 
the tree-level cross section), with a tree crossing at $\MSE = 425\gev$.

The Higgs mixing parameter $\mu$ dependence is shown in the middle left
plot.  It is rather linear and decreasing from $\sim 48.6\,\fb$ at small 
$\mu$ down to $\sim 46.9\,\fb$ at $\mu = 1000\gev$.
The relative corrections are $\sim +17.5\,\%$ at $\mu = 100\gev$ and 
$\sim 16.6\,\%$ at $\mu = 1000\gev$.
In this case the (expected) breakdown of the \Code{CCN[1]} scheme for 
$\mu \approx M_2 = 600\gev$ is rather weakly pronounced.
The corresponding results are shown for the \Code{CNN[1,1,2]} and 
\Code{CNN[2,1,3]} schemes, which are smooth at $\mu = M_2$.  For small 
and large values of $\mu$ \Code{CNN[1,1,2]} differs from \Code{CCN[1]}, 
whereas the other scheme, \Code{CNN[2,1,3]}, is very close to \Code{CCN[1]}. 
While the \Code{CNN[2,1,3]} scheme has a strong (local) minimum 
of $\delta \mu$ at 
$\mu \approx 421\gev$, dominating the loop corrections, it approximates 
\Code{CCN[1]} rather good for all other values of $\mu$.

The dependence on $|M_1|$ is shown in the middle right plot of 
\reffi{fig:eeSe1Se2}.
A strong dependence of the tree-level cross section can be observed
(from the dominant neutralino $t$-channel exchange), which is amplified 
at the one-loop level.  The size of the loop corrections varies from 
$\sim +16.6\%$ at $|M_1| = 350\gev$ (\ie \Scs) to $\sim +15.2\%$ at 
$|M_1| = 750\gev$ and then increase again to $\sim +16.1\%$ at 
$|M_1| = 1000\gev$.

The cross section dependence of $M_2$ in \Scs\ is shown in the lower left 
plot.  Again, one can see the (expected) breakdown of the \Code{CCN[1]} 
scheme for $M_2 \approx \mu = 350\gev$ 
and the smooth behavior of \Code{CNN[2,2,3]} around $M_2 \approx \mu = 350\gev$.
For $M_2 \gsim \mu$ the scheme \Code{CCN[1]} is expected to be reliable, 
while \Code{CNN[2,2,3]} is reliable for all values of $M_2$; see above.
The combined (reliable) one-loop corrections of \Code{CCN[1]} and 
\Code{CNN[2,2,3]} at the level of $\sim +16\,\%$ are found to be only 
weakly dependent of $M_2$.

The phase dependence $\phiMe$ of the cross section in \Scs\ is shown in the 
lower right plot of \reffi{fig:eeSe1Se2}.
In this case it turns out to be substantial, already changing the tree-level 
cross section by up to $3.7\,\%$.
The relative loop corrections ($\sigloop/\sigtree$) vary with $\phiMe$ between 
$\sim +14.6\,\%$ at $\phiMe = 180^{\circ}$ and $\sim +17.3\,\%$ at 
$\phiMe = 158^{\circ},203^{\circ}$ for \Code{CCN[1]}.
Again, the cross section does not diverge at $\phiMe = 180^{\circ}$. 
This structure is caused by a minimum of $\delta M_1$ at $\phiMe = 180^{\circ}$, 
dominating the loop corrections.  Using the \Code{CNN[2,2,3]} scheme 
(red dashed line) the dip disappears, as can be seen in the plot. 
The relative loop corrections reach $\sim +17.7\,\%$ at $\phiMe = 180^{\circ}$.
For our parameter set \Scs, with the complex phase $\phiMe$,
the $\CP$ asymmetry turn out to be (numerically) zero, which can be seen 
from the \textit{identical} green dashed ($\tilde{e}^+_1 \tilde{e}^-_2$) 
and yellow dash-dotted ($\tilde{e}^-_1 \tilde{e}^+_2$) lines.
Finally the variation with $\phiAeg$ is again negligible and not shown 
here.

\medskip

We now turn to the process $\eeSezSez$ shown in \reffi{fig:eeSe2Se2},
which is found to be of \order{10\,\fb}. 
As a function of $\sqrt{s}$ (upper left plot) the loop corrections range 
from $\sim -3\,\%$ at $\sqrt{s} = 1000\gev$ (\ie \Scs) to $\sim -10\,\%$ 
at $\sqrt{s} = 3000\gev$.

The cross sections are decreasing with increasing $\MSE$ due to 
kinematics, and the full one-loop result has its maximum of $\sim 11.6\,\fb$ 
at $\MSE = 100\gev$. Analogously the relative corrections are decreasing 
from $\sim +4\,\%$ at $\MSE = 100\gev$ to $\sim -23\%$ at $\MSE = 440\gev$. 
The tree crossing takes place at $\MSE \approx 190\gev$.  For higher $\MSE$ 
values the loop corrections are negative, where the relative size becomes 
large due to the (relative) smallness of the tree-level results, which goes 
to zero for $\MSE \approx 450\gev$ due to kinematics.
For the other parameter variations one can conclude that a cross section  
twice as large can be possible for very low $\MSE$ 
(which however are challenged by the current ATLAS/CMS exclusion bounds).

The dependence on $\mu$ (middle left plot) is rather small.  
The one-loop corrections within the \Code{CCN[1]} scheme are $\sim +2\,\%$ 
at $\mu = 100\gev$, $\sim -2\,\%$ at $\mu = 1000\gev$ and have a tree 
crossing at $\mu \approx 150\gev$.  The corresponding smooth%
\footnote{
  The peak at $\mu \approx 437\gev$ in the \Code{CNN[1,1,2]} 
  scheme is the anomalous threshold generated by the 
  $C_0(s,\mse2^2,\mse2^2,\mneu1^2,\mneu4^2,m_e^2)$ function.
}
results \Code{CNN[1,1,2]} and \Code{CNN[1,2,3]} are shown, but both differ 
for small values (and \Code{CNN[1,2,3]} also for large values) of $\mu$ 
significantly from \Code{CCN[1]}.  
Therefore, in addition we also show the \Code{CNN[2,1,3]} scheme 
(black dash-dotted line) which approximates \Code{CCN[1]} very good for 
all values of $\mu$ except for $\mu \approx 421\gev$ where $\delta \mu$ 
has a strong (local) minimum, that dominates the loop corrections in this 
part of the parameter space.

With increasing $|M_1|$ in \Scs\ (middle right plot) we find again a strong 
decrease of the production cross section, due to the change in the 
interference of the (dominant) neutralinos in the $t$-channel.
It should be noted that there is \textit{no} tree crossing in this plot.
The loop corrections are rather small and decrease from $\sim -0.6\%$ at 
$|M_1| = 100\gev$ to $\sim -3\%$ at $|M_1| = 470\gev$ and then increase to 
$\sim -0.9\%$ at $|M_1| = 1000\gev$.  

The dependence $M_2$ of the cross section in \Scs\ is shown in the lower 
left plot of \reffi{fig:eeSe2Se2}. 
The tree cross section decreases strongly from $\sim 62\,\fb$ 
at $M_2 = 120\gev$ down to $\sim 2\,\fb$ at $M_2 = 1000\gev$,
because of the change in the interference of the $\neu1$ 
(dominant for $M_2 \lsim 300\gev$), $\neu3$ 
(dominant for $300\gev \lsim M_2 \lsim 480\gev$), and $\neu4$ 
(dominant for $M_2 \gsim 480\gev$) in the $t$-channel.
As before, one can see the breakdown of the \Code{CCN[1]} scheme for 
$M_2 \approx \mu = 350$, and the smooth behavior of \Code{CNN[2,2,3]} 
(red dashed line) around $M_2 \approx \mu = 350\gev$.  Again, \Code{CCN[1]} 
is reliable for $M_2 \gsim 400\gev$ while \Code{CNN[2,2,3]} is reliable for 
all values of $M_2$ and very close to \Code{CCN[1]}.  The (reliable) one-loop 
corrections of \Code{CCN[1]} and \Code{CNN[2,2,3]} are at the level of 
$\sim \pm 5\,\%$.  The \Code{CNN[1,2,3]} scheme (not shown) is also smooth 
for all values of $M_2$ and can reach $\sim +20\,\%$ at $M_2 \approx 800\gev$.

Now we turn to the complex phase dependence.  We find that the phase 
dependence $\phiMe$ of the cross section in \Scs\ is rather large 
(lower right plot) for the \Code{CCN[1]} scheme.  Loop corrections at the 
level of $\sim -3\,\%$ at $\phiMe = 0^{\circ},360^{\circ}$ and $\sim -22\,\%$ 
at $\phiMe = 180^{\circ}$ are found.  It should be noted again, that there 
is \textit{no} divergency or threshold at $\phiMe = 180^{\circ}$, the cross 
section does not diverge; see the inlay in the lower right plot. 
This large structure is caused by a strong minimum of $\delta M_1$ at 
$\phiMe = 180^{\circ}$, dominating the loop corrections.
The \Code{CNN[2,2,3]} scheme (red dashed line) is close to \Code{CCN[1]}
at the level of $\sim -3\,\%$ without this peculiar structure.
In contrast the full corrections of \Code{CNN[1,2,3]} reach $\sim +13\,\%$.
The loop effects of $\phiAeg$, on the other hand, are tiny and therefore 
not shown explicitely.

\medskip

We now turn to the second generation charged slepton production.
The process $\eeSmeSme$ is shown in \reffi{fig:eeSm1Sm1}, which is found 
in \Scs\ at the level of $10\,\fb$, but can be substantially
larger by roughly a factor of two for small $\MSE$; see below.
Away from the production threshold, loop corrections of $\sim +14\,\%$ at 
$\sqrt{s} = 1000\gev$ (\ie \Scs) are found.  They reach their maximum of 
$\sim +35\,\%$ at $\sqrt{s} = 3000\gev$. The tree crossing takes place
at $\sqrt{s} \approx 725\gev$.

The cross section depends strongly on $\MSE$, as can be seen in the right
plot.  It is decreasing with increasing $\MSE$ and the full correction has 
its maximum of $\sim 28\,\fb$ at $\MSE = 100\gev$. 
The variation of the relative corrections are rather large, $\sim +33\,\%$ 
at $\MSE = 100\gev$, with a tree crossing at $\MSE = 440\gev$, and 
$\sim -25\,\%$ at $\MSE = 490\gev$ where the cross section goes to zero
due to kinematics.

The dependence on the remaining parameters is (rather) negligible 
and therefore we have omitted showing the corresponding plots here.

\medskip

The process $\eeSmeSmz$ is shown in \reffi{fig:eeSm1Sm2}.  It should be 
noted that the smuon (and stau) tree-level process consist of only one 
$Z$ exchange diagram (see \reffi{fig:eeSlSl}) and is%
\footnote{
  With no slepton mixing 
  (\ie no off diagonal entries in the slepton mixing matrix)
  there is no tree-level cross section at all. On the other 
  hand, large off diagonal entries (\eg\ in our case large $\TB$) 
  should be able to enhance $\sigtree$; see below.
}
\begin{align}
\label{eq:SigTree}
\sigtree &\propto |U^{\Se_g}_{22} U^{\Se_g*}_{12}|^2 
               = |-U^{\Se_g}_{21} U^{\Se_g*}_{11}|^2
    = \frac{m_{\Fe_g}^2 |A_{\Fe_g} - \mu^*\TB|^2}{(\mse{g1}^2 - \mse{g2}^2)^2}\,,
\end{align}
with the generation index $g = 2,3$. (In the case of selectrons there 
is an additional tree-level diagram with neutralino exchange; 
see \reffi{fig:eeSlSl}.) 
Setting $\MSL = \MSE + m$ and neglecting off-diagonal contributions,
$D$-terms, and $m_{\Fe_g}^2$ contributions in the slepton mass matrix, 
yields
\begin{align}
(\mse{g1}^2 - \mse{g2}^2)^2 &= m^2\, (2 \MSE + m + \ldots)^2 + \ldots\,,
\end{align}
and consequently
\begin{align}
\sigtree &\propto \frac{m_{\Fe_g}^2 |A_{\Fe_g} - \mu^*\TB|^2}
                       {m^2\, (2 \MSE + m + \ldots)^2 + \ldots}
          \propto \frac{m_{\Fe_g}^2}{m^2}\,.
\label{prod12}
\end{align}
For vanishing $m$ the cross section can be relatively large (but remains
finite). 

In our scenario \Scs\ we have chosen a (more realistic) setting with 
$\MSL \ne \MSE$ (with an off-set of $m = 50\gev$).  Therefore within 
this scenario the cross section for $\eeSmeSmz$ turns out to be 
strongly suppressed by $m_{\mu}^2/m^2$, of $\order{0.1\,\ab}$; 
see the left plot in \reffi{fig:eeSm1Sm2}.  This is below the reach 
of a linear collider.  For this reason we refrain from a more detailed 
discussion here.
The only expected exception is the variation with $\TB$ (see footnote 10)
which we show in the right plot of \reffi{fig:eeSm1Sm2}. 
The loop corrected cross section increases from \order{0.01\,\ab} at small 
$\TB$ to $\sim 23\,\ab$ at $\TB = 50$, as expected. 
The relative corrections for the $\TB$ dependence are increasing from 
$\sim -12\,\%$ at $\TB = 10$ to $\sim -7\,\%$ at $\TB = 50$.

\medskip

Now we turn to the process $\eeSmzSmz$ shown in \reffi{fig:eeSm2Sm2}.
As a function of $\sqrt{s}$ (left plot) we find relative corrections of 
$\sim -5\,\%$ at $\sqrt{s} = 1000\gev$ (\ie \Scs), and $\sim +10\,\%$ at
$\sqrt{s} = 3000\gev$ with a tree crossing at $\sqrt{s} \approx 1200\gev$.

In the analysis as a function of $\MSE$ (right plot) the cross section is 
decreasing with increasing $\MSE$, but can vary roughly by a factor of two 
\wrt \Scs.  The full (relative) one-loop correction has its maximum of 
$\sim 26\,\fb$ ($\sim +12\,\%$) at $\MSE = 100\gev$, decreasing to 
$\sim 0.1\,\fb$ ($\sim -40\,\%$) at $\MSL = 440\gev$ with a tree crossing 
at $\MSE \approx 245\gev$.

The dependence on the other parameters is again (rather) negligible and 
therefore not shown here.

\medskip

Turning to the third generation charged slepton production, the
process $\eeSaeSae$ is shown in \reffi{fig:eeSa1Sa1}. 
As a function of $\sqrt{s}$ we find loop corrections of $\sim +14\,\%$ 
at $\sqrt{s} = 1000\gev$ (\ie \Scs), a tree crossing at 
$\sqrt{s} \approx 725\gev$ (where the one-loop corrections are
between $\pm 10\,\%$ for $\sqrt{s} \lsim 900 \gev$) and $\sim +35\,\%$ 
at $\sqrt{s} = 3000\gev$, very similar to $\eeSmeSme$.

In the analysis as a function of $\MSE$ (upper right plot) the cross 
sections are decreasing with increasing $\MSE$ as obvious from 
kinematics and the full corrections have their maximum of $\sim 28\,\fb$ 
at $\MSE = 100\gev$, more than two times larger than in \Scs.  The relative 
corrections are changing from $\sim +33\,\%$ at $\MSE = 100\gev$ to 
$\sim -25\,\%$ at $\MSE = 490\gev$ with a tree crossing at $\MSE = 415\gev$.

Here we show in the lower left plot of \reffi{fig:eeSa1Sa1} the dependence 
on $\TB$.  Contrary to other slepton production cross sections analyzed 
before%
\footnote{
  We have omitted showing these plots because 
  the dependence on $\TB$ was indeed negligible
  (with exception of $\eeSmeSmz$; see above).
}%
, $\sigfull(\eeSaeSae)$ increases with $\TB$. 
The relative corrections for the $\TB$ dependence vary between 
$\sim +14.2\,\%$ at $\TB = 5$ and $\sim +13.4\,\%$ at $\TB = 50$.

The phase dependence $\phiAeg$ of the cross section in \Scs\ is shown 
in the lower right plot of \reffi{fig:eeSa1Sa1}.  
The loop correction increases the tree-level result by $\sim +14\,\%$.
The phase dependence of the relative loop correction is very small and 
found to be below $0.2\,\%$.
The variation with $\phiMe$ is negligible and therefore not shown here.

\medskip

The process $\eeSaeSaz$ is shown in \reffi{fig:eeSa1Sa2}.
The overall size of this cross section turns out to be rather small,
including all analyzed parameter variations, but enhanced \wrt $\eeSmeSmz$ 
by about a factor of $m_\tau^2/m_\mu^2$; see \refeq{prod12}.
The loop corrections have a noticeable impact, as can be seen in all six 
panels of \reffi{fig:eeSa1Sa2}, but never lift the cross section above 
$2\,\fb$.

As a function of $\sqrt{s}$ (upper left plot) we find relative corrections 
of $\sim -11\,\%$ at $\sqrt{s} = 1000\gev$ (\ie \Scs), and $\sim +19\,\%$ at 
$\sqrt{s} = 3000\gev$ with a tree crossing at $\sqrt{s} \approx 1300\gev$.

As for other slepton production cross sections, $\sig(\eeSaeSaz)$ 
depends strongly on $\MSE$, where values one order of magnitude larger
than in \Scs\ (with $\MSE = 300\gev$) are possible for small $\MSE$.
One can see that the full corrections have their maximum of $\sim 0.8\,\fb$ 
at $\MSE = 100\gev$.  The relative corrections are decreasing from 
$\sim +6\,\%$ at $\MSE = 100\gev$ to $\sim -51\,\%$ at $\MSE = 470\gev$ with 
a tree crossing at $\MSE \approx 170\gev$.

With increasing $\mu$ in \Scs\ (middle left plot) we find a nearly linear 
increase of the production cross section.
While the \Code{CCN[1]} scheme has its expected singularity%
\footnote{
  It should be noted that, as discussed in the beginning of
  \refse{sec:numeval}, within the \Code{CCN[1]} scheme a small 
  divergence in the tree-level result (at $\mu = M_2 = 600\gev$) 
  is induced by $\delta \mu$ through the shifted scalar tau masses. 
  $\delta \mu$ in \refeq{eq:SlepAux} enters via \refeq{eq:SlepY} 
  into \refeq{eq:dSfShift} from which the slepton shifts are 
  calculated.  The divergence is suppressed with $\mfe{g}$ 
  (see \refeq{eq:SlepAux}) and therefore not visible in the tree-level 
  results for selectron production; see above.  In order to overcome 
  the problem with a divergent tree-level result, we used here the 
  tree-level result of the \Code{CNN[1,2,3]} scheme, which is free 
  of such a divergence.
}
at $\mu = 600\gev$, the \Code{CNN[1,2,3]} scheme is smooth around this 
point.  The relative loop corrections are nearly identical for both 
schemes and reach $\sim -10\,\%$ at $\mu = 350\gev$ (\ie \Scs) and go 
up to $\sim -7\,\%$ at $\mu = 1000\gev$.  It should be noted that the 
tree cross section is zero at $\mu = 200\gev$ where $U^{\Se_3}_{11,22} = 0$ 
($U^{\Se_3}_{12,21} = 1$); see \refeq{eq:SigTree}.

The $M_2$ dependence of the cross section in \Scs\ is shown in the middle 
right plot.  One can see again the (expected) breakdown of the 
\Code{CCN[1]} scheme for $M_2 = \mu = 350\gev$ and the smooth behavior 
of \Code{CNN[1,2,3]} and \Code{CNN[2,2,3]} around $M_2 \sim \mu = 350\gev$. 
The loop corrections at the level of $\sim -5\,\%$ at $M_2 = 120\gev$
and $\sim -7\,\%$ at $M_2 = 1000\gev$ are found to be rather independent 
of $M_2$ within all three schemes.  The tiny dip (hardly visible) at 
$M_2 \approx 644\gev$ in all three schemes is the chargino production 
threshold $\mcha1 + \mcha2 = \sqrt{s} = 1000\gev$.

Here we also show the variation with $\TB$ in the lower left plot of 
\reffi{fig:eeSa1Sa2}.  The loop corrected cross section increases from 
$\order{0.1\,\ab}$ at small $\TB$ to $\sim 1.7\,\fb$ at $\TB = 50$. 
The relative corrections for the $\TB$ dependence are changing from 
$\sim -12\,\%$ at $\TB = 10$ to $\sim -10\,\%$ at $\TB = 50$.

The phase dependence $\phiAeg$ of the cross section in \Scs\ is shown in 
the lower right plot of \reffi{fig:eeSa1Sa2}.  It is very pronounced
and can vary $\sigfull(\eeSaeSaz)$ from $0.03\,\fb$ to $0.3\,\fb$.
The (relative) loop corrections are at the level of $\sim 10\,\%$ and 
vary with $\phiAeg$ below $\pm 4.5\,\%$ \wrt the tree cross section.
For our parameter set \Scs, with the complex phase $\phiAeg$, the $\CP$ 
asymmetry turn out to be very small, as can be seen in the inlay.

\medskip

Finally we turn to the process $\eeSazSaz$ shown in \reffi{fig:eeSa2Sa2}, 
which turns out to be sizable at the level of $10\,\fb$.
As a function of $\sqrt{s}$ (upper left plot) we find loop corrections 
of $\sim -5\,\%$ at $\sqrt{s} = 1000\gev$ (\ie \Scs), and $\sim +10\,\%$ at 
$\sqrt{s} = 3000\gev$, with a tree crossing at $\sqrt{s} \approx 1200\gev$.

The dependence on $\MSE$ is shown in the upper right plot. 
The relative corrections are $\sim +12\,\%$ at $\MSE = 100\gev$, 
$\sim -5\,\%$ at $\MSE = 300\gev$ (\ie \Scs), and $\sim -39\,\%$ 
at $\MSE = 440\gev$, where the cross section goes to zero at 
$\mu = 450\gev$ (because of the choice $\MSL = \MSE + 50 \gev$). 
The tree crossing is found at $\mu \approx 245\gev$.

Again, here we show in the lower left plot of \reffi{fig:eeSa2Sa2} the 
dependence on $\TB$.  Contrary to other slepton production cross 
sections analyzed before, $\sigfull(\eeSazSaz)$ decreases with $\TB$. 
The relative corrections for the $\TB$ dependence vary between 
$\sim -5.4\,\%$ at $\TB = 5$ and $\sim -4.4\,\%$ at $\TB = 34$.
A dip (not visible) in the dotted line at $\TB \approx 46$ is the 
threshold $\mstau1 + \Mh = \mstau2$.

The phase dependence $\phiAeg$ of the cross section in \Scs\ is shown 
in the lower right plot of \reffi{fig:eeSa2Sa2}.  
The loop correction decreases the tree-level result by $\sim -5\,\%$.
The phase dependence of the relative loop correction is small and found 
to be below $0.5\,\%$.
The variation with $\phiMe$ is (again) negligible and therefore not 
shown here.

\medskip

Overall, for the charged slepton pair production we observed a decreasing 
cross section $\propto \ln{(s)}/s$ for the first and $\propto 1/s$ for 
the second and third slepton generations for $s \to \infty$; 
see \citere{FrMaZe2004}.
The (loop corrected) cross sections for the slepton pair production can 
reach a level of \order{10\,\fb}, depending on the SUSY parameters, but 
is very small for the production of two different smuons $\Sm_1\Sm_2$ at 
the \order{10\,\ab}.  This renders these processes difficult to observe 
at an $e^+e^-$ collider.%
\footnote{
  The limit of $10$~ab corresponds to ten events at 
  an integrated luminosity of $\cL = 1\, \iab$, which 
  constitutes a guideline for the observability of a 
  process at a linear collider.
}
The full one-loop corrections are very roughly $15\,\%$ of the tree-level 
results, but vary strongly on the size of $\MSE$ and in the case of
selectrons also on the size of $M_1$ and $M_2$.  Depending on the size of 
in particular these parameters the loop corrections can be either positive 
or negative.  This shows that the loop corrections, while being large, 
have to be included point-by-point in any precision analysis.
The dependence on $\phiMe$ ($\phiAeg$) was found at the level of 
$\sim 15\,\%$ ($\sim 8\,\%$), but can go up to $\sim 18\,\%$ 
($\sim 14\,\%$) for the extreme cases.  The relative loop corrections 
varied by up to $2\,\%$ ($4\,\%$) with $\phiMe$ ($\phiAeg$).
Consequently, the complex phase dependence must be taken into account 
as well.  Finally, for our parameter set \Scs\ the $\CP$ asymmetries 
turn out to be very small, well below $\pm 1\,\%$ (hardly measurable 
in future $e^+e^-$ collider experiments).

For all parameter choices it was possible to identify at least one
renormalization scheme that exhibited a ``smooth'' behavior.  It appears 
to be possible for any parameter variation to find a combination of 
schemes that yield a numerically stable (and nearly constant) one-loop 
level contribution.  A detailed analysis of which scheme yields this 
desired behavior as a function of the underlying SUSY parameters, 
however, is beyond the scope of this paper.


\subsection{\texorpdfstring{The process \boldmath{\eeSnSn}}
                           {The process e+e- -> Sneutrino Sneutrino}}
\label{sec:eeSnSn}

\begin{figure}
\begin{center}
\begin{tabular}{c}
\includegraphics[width=0.48\textwidth,height=6cm]{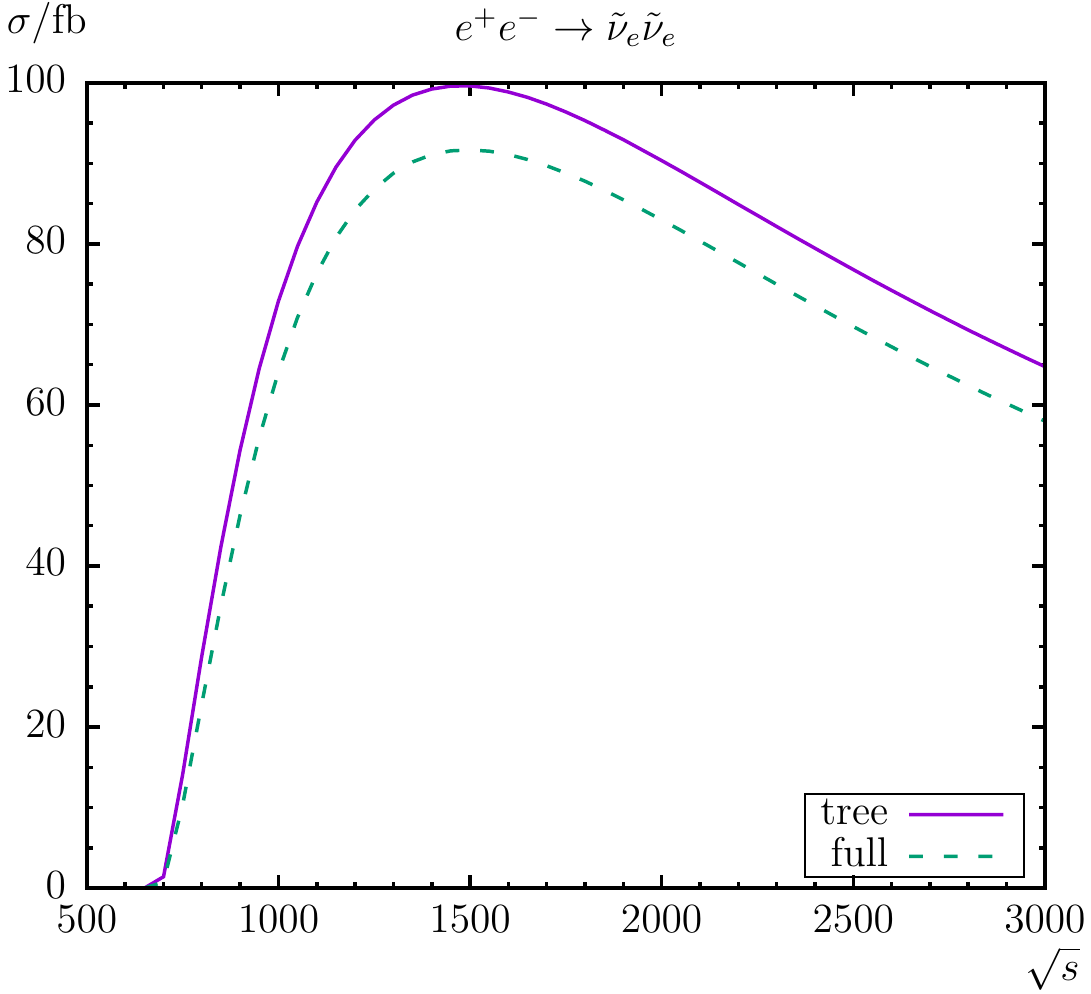}
\includegraphics[width=0.48\textwidth,height=6cm]{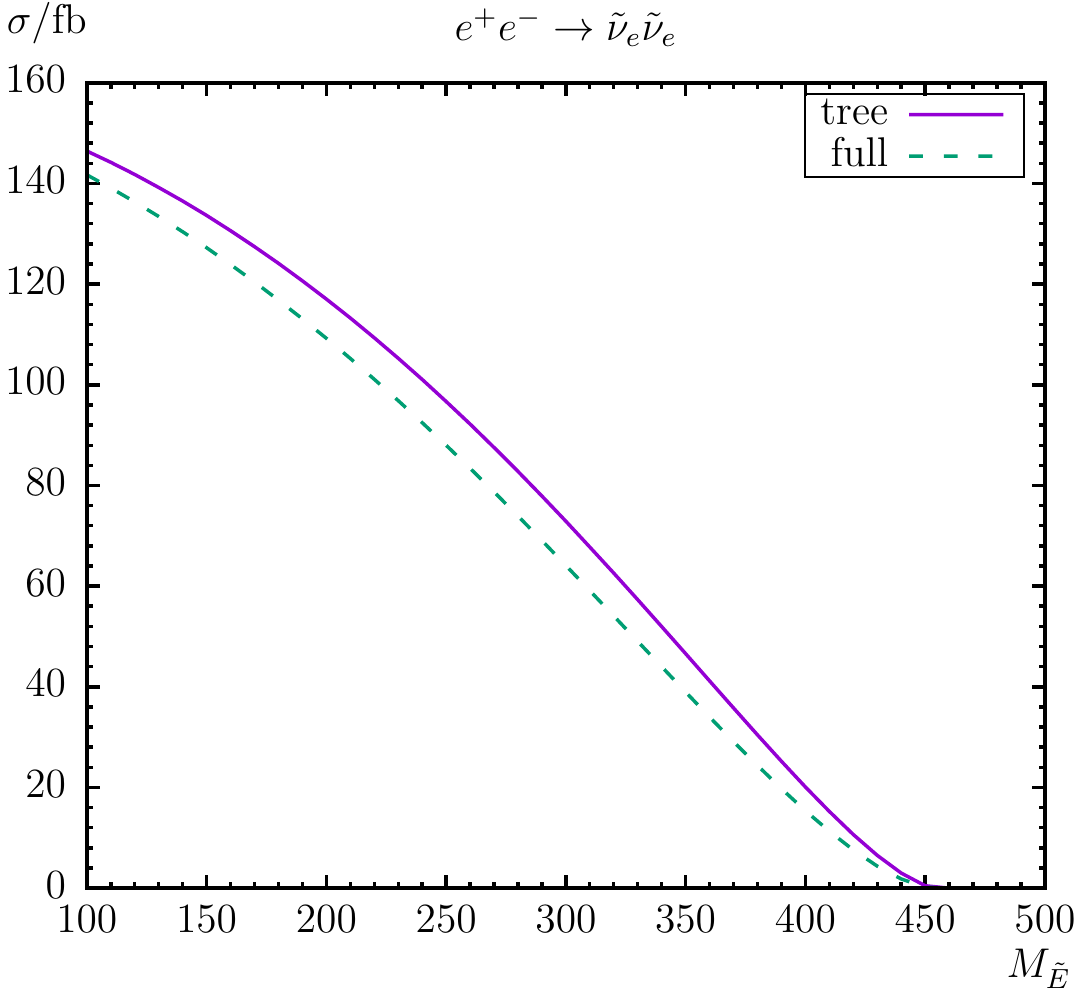}
\\[1em]
\includegraphics[width=0.48\textwidth,height=6cm]{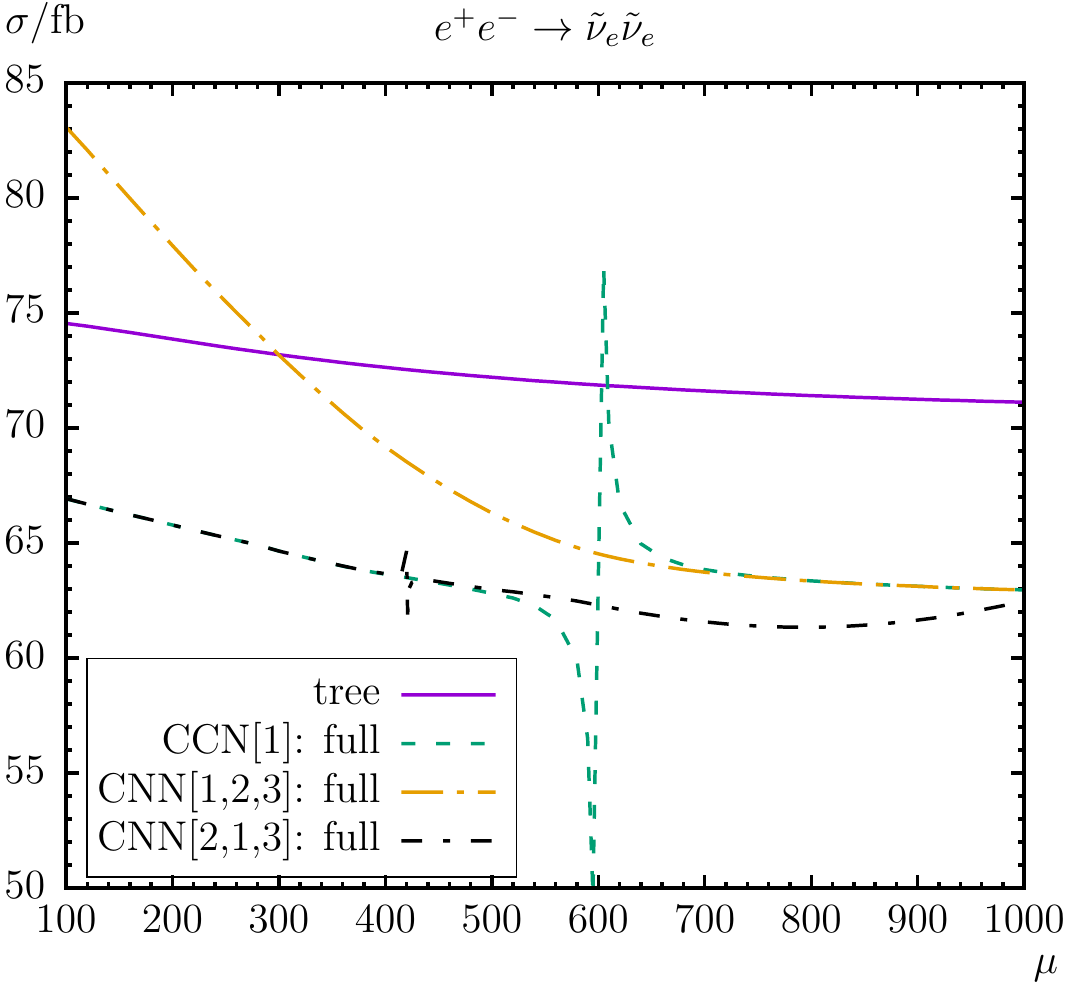}
\includegraphics[width=0.48\textwidth,height=6cm]{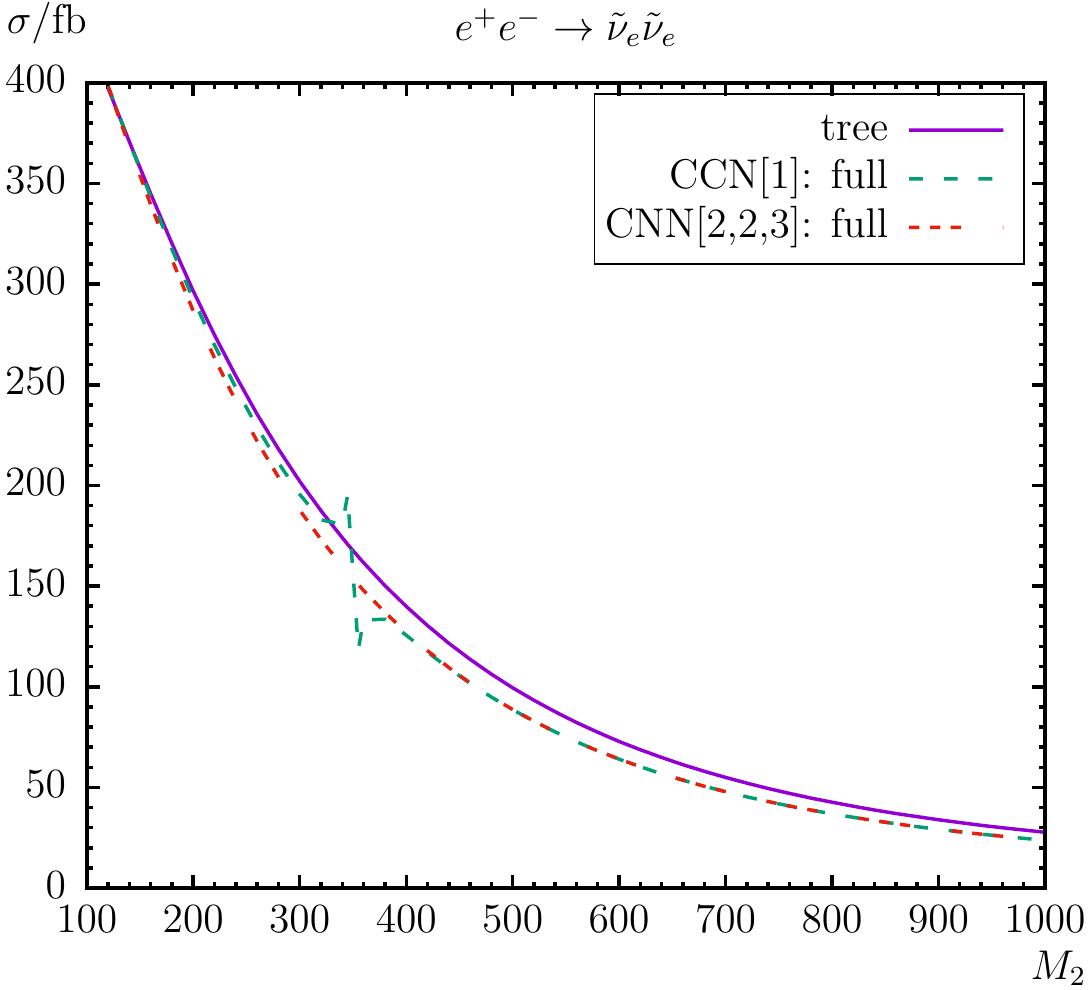}
\end{tabular}
\caption{\label{fig:eeSneSne}
  $\sig(\eeSneSne)$.
  Tree-level and full one-loop corrected cross sections are shown 
  with parameters chosen according to \Scs; see \refta{tab:para}.
  The upper plots show the cross sections with $\sqrt{s}$ (left) 
  and $\MSE$ (right) varied; the lower plots show $\mu$ (left) 
  and $M_2$ (right) varied. All masses and energies are in GeV.
}
\end{center}
\end{figure}

\begin{figure}
\begin{center}
\begin{tabular}{c}
\includegraphics[width=0.48\textwidth,height=6cm]{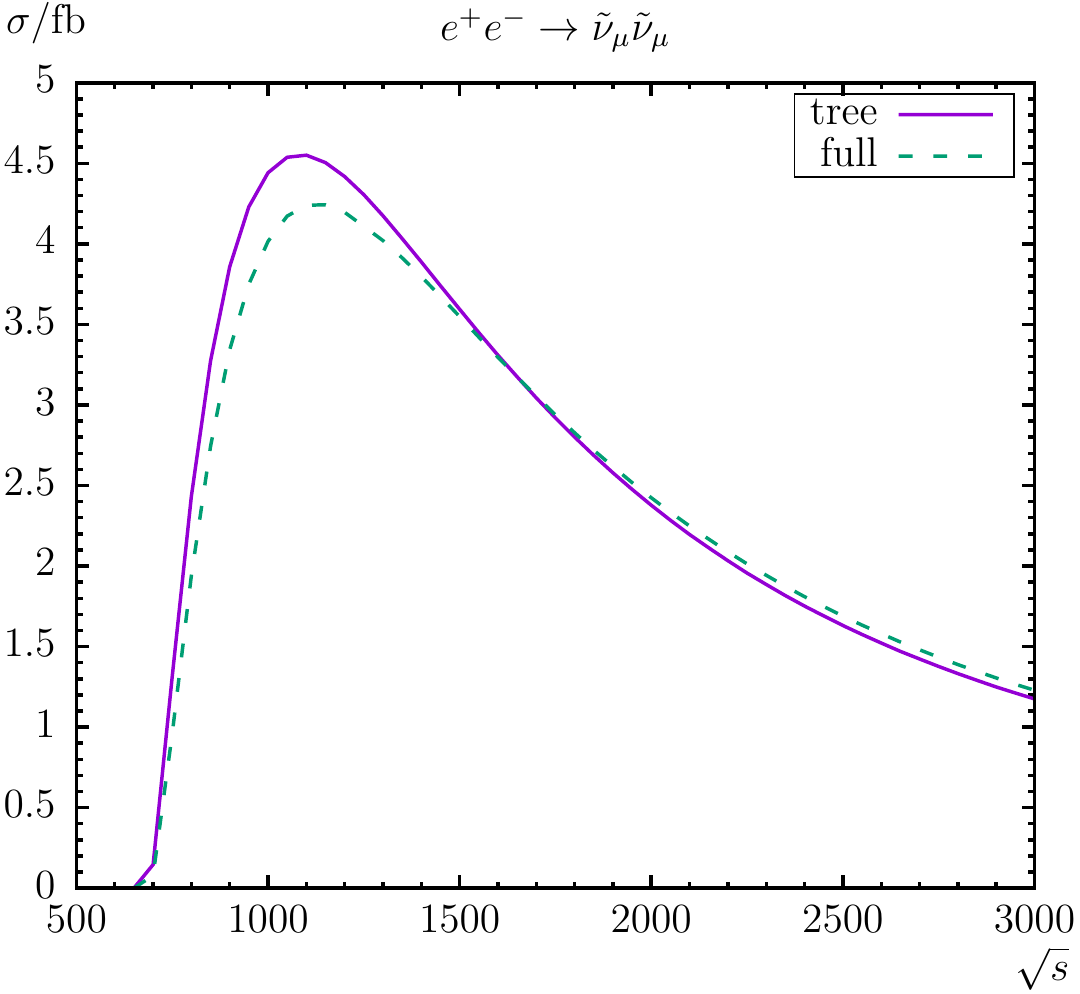}
\includegraphics[width=0.48\textwidth,height=6cm]{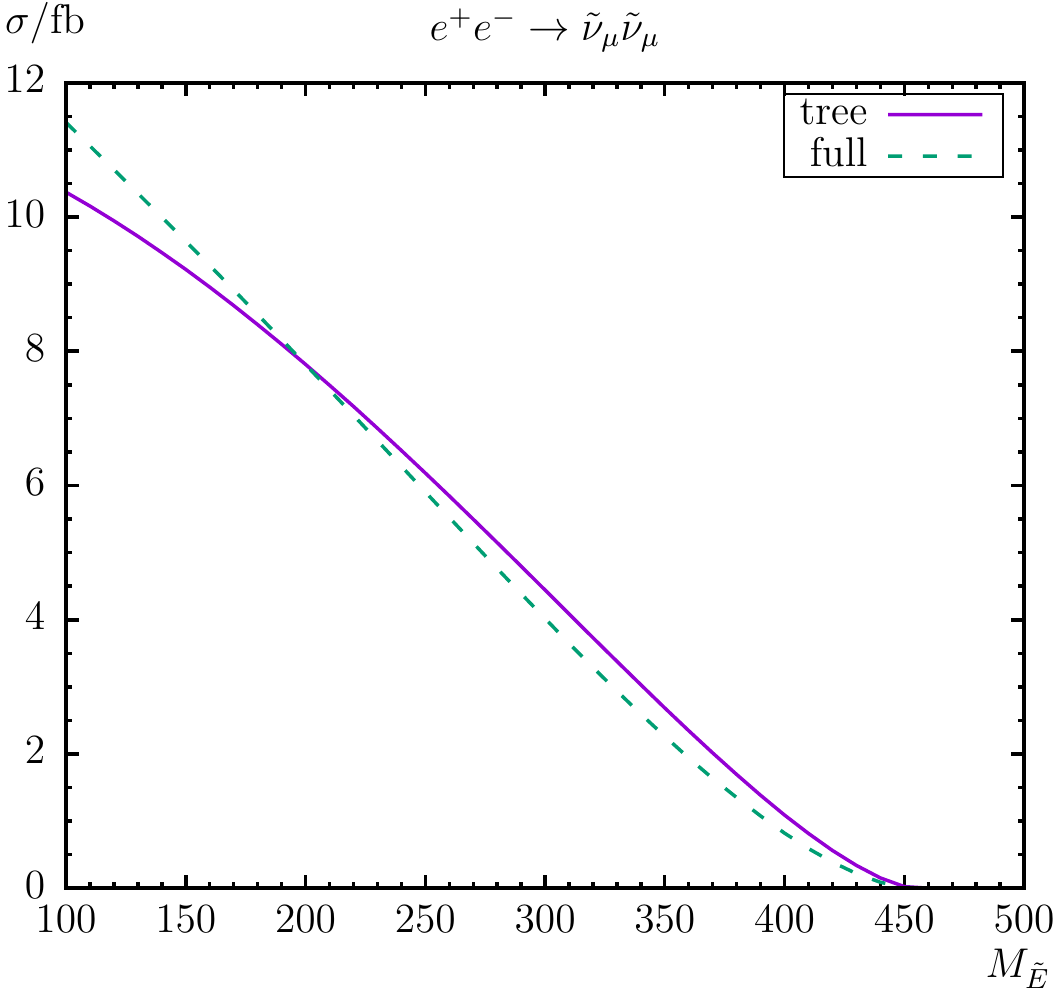}
\end{tabular}
\caption{\label{fig:eeSnmSnm}
  $\sig(\eeSnmSnm)$.
  Tree-level and full one-loop corrected cross sections are shown 
  with parameters chosen according to \Scs; see \refta{tab:para}.
  The plots show the cross sections with $\sqrt{s}$ (left) and 
  $\MSE$ (right) varied. All masses and energies are in GeV.
}
\end{center}
\end{figure}

\begin{figure}
\begin{center}
\begin{tabular}{c}
\includegraphics[width=0.48\textwidth,height=6cm]{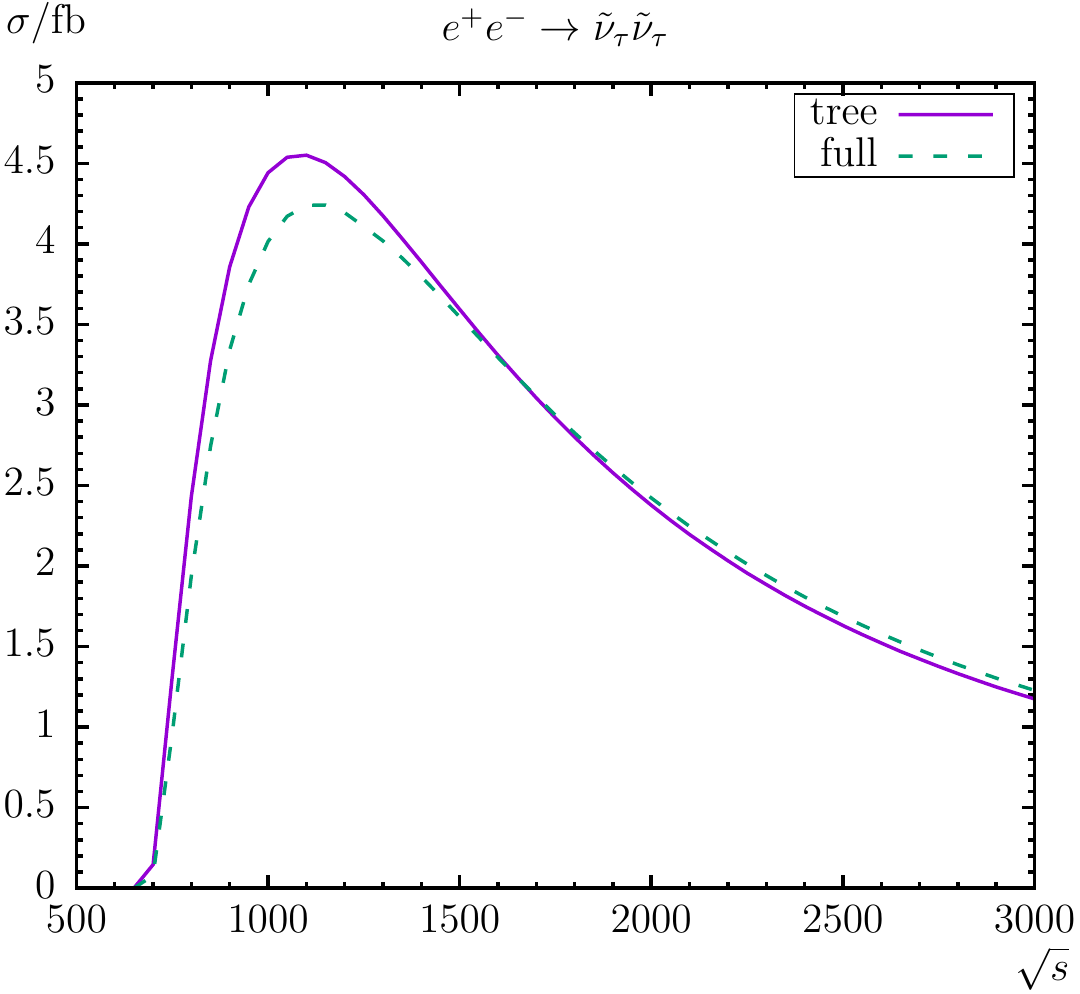}
\includegraphics[width=0.48\textwidth,height=6cm]{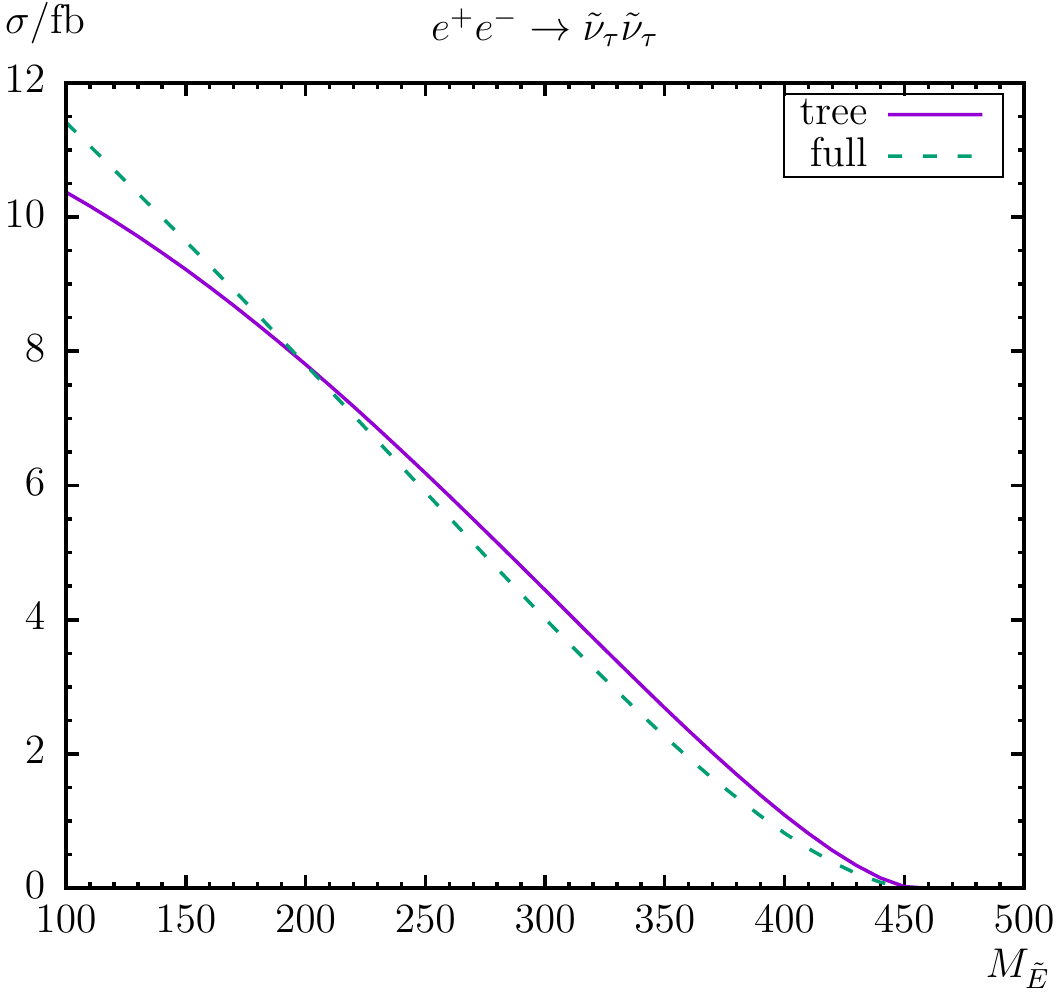}
\end{tabular}
\caption{\label{fig:eeSnaSna}
  $\sig(\eeSnaSna)$.
  Tree-level and full one-loop corrected cross sections are shown 
  with parameters chosen according to \Scs; see \refta{tab:para}.
  The plots show the cross sections with $\sqrt{s}$ (left) and 
  $\MSE$ (right) varied. All masses and energies are in GeV.
}
\end{center}
\end{figure}

In \reffis{fig:eeSneSne} -- (\ref{fig:eeSnaSna}) we present the results 
for scalar neutrino production at $e^+e^-$ colliders.  It should be 
noted that for $s \to \infty$ decreasing cross sections $\propto \ln{(s)}/s$
for the first and $\propto 1/s$ for the second and third slepton generations 
are expected; see \citere{FrMaZe2005}.
We start with the process $\eeSneSne$ that is shown in \reffi{fig:eeSneSne}. 

In the analysis of the production cross section as a function of $\sqrt{s}$ 
(upper left plot) we find the expected behavior: a strong rise close to the 
production threshold, followed by a decrease with increasing $\sqrt{s}$, 
where the $s$-channel dominates.
We find a very small shift \wrt $\sqrt{s}$ around the production threshold. 
Away from the production threshold, loop corrections of $\sim -12\,\%$ at 
$\sqrt{s} = 1000\gev$ are found in scenario \Scs\ (see \refta{tab:para}).
The relative size of the loop corrections increase up to $\sim -8\,\%$ at
$\sqrt{s} = 1650\gev$ and then reach $\sim -10\,\%$ at $\sqrt{s} = 3000\gev$.

The cross section as a function of $\MSE$ is shown in the upper right plot 
of \reffi{fig:eeSneSne}. The masses of the electron sneutrinos are governed 
by $\MSL = \MSE + 50 \gev$. 
Consequently, a strong, nearly linear decrease can be observed from 
$\sim 140\,\fb$ down to zero for $\MSE = 450\gev$ (\ie the sneutrino 
production threshold), as can be expected from kinematics.
In scenario \Scs\ we find a non-negligible decrease of the cross sections 
from the loop corrections.  They start at $\MSE = 100\gev$ with 
$\sim -3\,\%$ and reach $\sim -53\,\%$ at $\MSE \approx 450\gev$.
In the latter case these large loop corrections are due to the (relative) 
smallness of the tree-level results, which goes to zero at the sneutrino 
production threshold.

With increasing $\mu$ in \Scs\ (lower left plot) we find a small, rather
linear decrease of the production cross section within \Code{CCN[1]}, 
mainly induced by the change in the chargino $t$-channel contribution 
($\cha2$ is dominant for $\mu \lsim 570\gev$ and $\cha1$ dominant for 
$\mu \gsim 570\gev$). 
The relative loop corrections within \Code{CCN[1]} (green dashed line) reach 
$\sim -10\,\%$ at $\mu = 100\gev$ (at the border of the experimental limit), 
$\sim -12\,\%$ at $\mu = 350\gev$ (\ie \Scs) and $\sim -11.5\,\%$ at 
$\mu = 1000\gev$.  While \Code{CCN[1]} is unreliable for 
$500\gev < \mu < 700\gev$, \Code{CNN[2,1,3]} (black dash-dotted line) yields 
reliable higher-order corrections for $\mu < 600\gev$ and \Code{CNN[1,2,3]} 
(yellow dash-dotted line) for $\mu > 600\gev$; as can be seen in the lower 
left plot of \reffi{fig:eeSneSne}. 
Again, in the \Code{CNN[2,1,3]} scheme $\delta \mu$ has a strong minimum 
at $\mu \approx 421\gev$, dominating the loop corrections.

The dependence $M_2$ of the cross section in \Scs\ is shown in the lower 
right plot of \reffi{fig:eeSneSne}, where again the chargino $t$-channel 
exchange plays a dominant role ($\cha1$ is dominant for $M_2 \lsim 380\gev$ 
and $\cha2$ dominant for $M_2 \gsim 380\gev$). 
Again, one can see the (expected) breakdown of the \Code{CCN[1]} scheme 
for $M_2 = \mu = 350\gev$. 
Outside the region $M_2 \approx \mu$ the scheme \Code{CCN[1]} is expected 
to be reliable, since each of the three OS conditions is strongly connected 
to one of the three input parameters, $|M_1|$, $M_2$ and $\mu$.
The loop corrections are $\sim -12\,\%$ at $M_2 = 600\gev$ (\ie \Scs) and
$\sim -14\,\%$ at $M_2 = 1000\gev$, with a tree crossing at 
$M_2 \approx 130\gev$. The \Code{CNN[2,2,3]} scheme (red dashed line) 
is smooth for all values of $M_2$ shown and a perfect approximation for 
\Code{CCN[1]}.

Due to the absence of $\phiMe$ in the tree-level production cross section 
the effect of this complex phase is expected to be small.  Correspondingly
we find that the phase dependence $\phiMe$ of the cross section in our
scenario is tiny.  The same holds for the variation with $\phiAeg$, it 
also remains tiny and unobservable.  Therefore we omit showing these two 
complex phases explicitely.

\medskip

In \reffi{fig:eeSnmSnm} we present the cross sections $\sig(\eeSnmSnm)$.
In the analysis as a function of $\sqrt{s}$ (left plot) we find 
as before a tiny shift \wrt $\sqrt{s}$, where the position of
the maximum cross section shifts by about $+50\gev$.
The relative corrections are found to be of $\sim -9\,\%$ at 
$\sqrt{s} = 1000\gev$ (\ie \Scs), and $\sim +5\,\%$ at $\sqrt{s} = 3000\gev$. 

As a function of $\MSE$ (right plot) the cross section decreases rather 
linearly.  The relative corrections change from $\sim +10\,\%$ at 
$\MSE = 100\gev$ to $\sim -60\,\%$ at $\MSE = 450\gev$, with a tree 
crossing at $\MSE \approx 200\gev$.

The dependence on the other parameters is (again) negligible and therefore 
not shown here.

\medskip

We finish the \eeSnSn\ analysis with $\sig(\eeSnaSna)$ in 
\reffi{fig:eeSnaSna}.  The results are very similar as for 
$\sig(\eeSnmSnm)$ (see above), since the sneutrino masses are 
the same for all three generations, as are the contributing 
higher-order diagrams. 

\medskip

Overall, for the sneutrino pair production we observed a decreasing 
cross section $\propto \ln{(s)}/s$ for the first and $\propto 1/s$ for 
the second and third slepton generations for $s \to \infty$; 
see \citere{FrMaZe2005}.
The full one-loop corrections are very roughly $10\,\%$ of the
tree-level results, but depend strongly on the size of $\MSE$, where
larger values result even in negative loop corrections.
The cross sections are largest for $\eeSneSne$ and roughly smaller by 
one order of magnitude for $\eeSnmSnm$ and $\eeSnaSna$. 
This is caused by the \textit{absence} of the chargino $t$-channel diagram 
(\ie a $\champ{c}\, e^{\pm}\, \Sn_g$ coupling), 
which only contributes to $\eeSneSne$; see \reffi{fig:eeSlSl}.
The variation of the cross sections with $\phiMe$ and $\phiAeg$ is found 
extremely small and have not been shown explicitely.

As for charged slepton production, for all parameter choices it was
possible to identify at least one renormalization scheme that exhibited a 
``smooth'' behavior.  Also for sneutrino production it appears to be 
possible for any parameter variation to find a combination of schemes that 
yield numerically stable (and nearly constant) one-loop level contributions.


\section{Conclusions}
\label{sec:conclusions}

We have evaluated all slepton production modes at $e^+e^-$ colliders with 
a two-particle final state, \ie \eeSeSe\ and \eeSnSn\ allowing for complex 
parameters.  In the case of discovery of sleptons a subsequent precision 
measurement of their properties will be crucial to determine their nature 
and the underlying (SUSY) parameters. 
In order to yield sufficient accuracy, one-loop corrections to the various 
slepton production modes have to be considered.  This is particularly the 
case for the anticipated high accuracy of the slepton property determination 
at $e^+e^-$ colliders~\cite{LCreport}.

The evaluation of the processes (\ref{eq:eeSeSe}) and (\ref{eq:eeSnSn})
is based on a full one-loop calculation, also including hard, soft and
collinear QED radiation.  The renormalization is chosen to be identical 
as for the slepton decay calculations~\cite{Stau2decay}, or slepton 
production from heavy Higgs-boson decay~\cite{HiggsDecaySferm}, as well as 
from chargino and neutralino decays~\cite{LHCxC1,LHCxC2,LHCxN,LHCxNprod}.
Consequently, the predictions for the production and decay can be used 
together in a consistent manner (\eg, in a global phenomenological 
analysis of the slepton sector at the one-loop level).

We first briefly reviewed the relevant sectors including some details 
of the one-loop renormalization procedure of the cMSSM, which are 
relevant for our calculation.  In most cases we follow \citere{MSSMCT}. 
We have discussed the calculation of the one-loop diagrams, the
treatment of UV, IR, and collinear divergences that are canceled by the 
inclusion of (hard, soft, and collinear) QED radiation. 
As far as possible we have checked our result against the literature, 
and in most cases where a meaningful comparison could be performed
we found good agreement; parts of the differences can be attributed to 
problems with input parameters and/or different renormalization schemes 
(conversions).

For the analysis we have chosen a standard parameter set (see
\refta{tab:para}), which allows for the production of all combinations 
of sleptons at an $e^+e^-$ collider with a center-of-mass energy up to 
$\sqrt{s} = 1000\gev$.
In the analysis we investigated the variation of the various production
cross sections with the center-of-mass energy $\sqrt{s}$, the Higgs mixing
parameter $\mu$, the gaugino mass parameters $|M_1|$ and/or $M_2$, 
the slepton soft SUSY-breaking parameter $\MSE$ ($\MSL = \MSE + 50\gev$) 
and the complex phases $\phiAeg$ (of the trilinear Higgs-slepton coupling,
$A_{\Fe_g}$; $g=1,2,3$) and $\phiMe$ (of the gaugino mass parameter $M_1$), 
respectively.  Where relevant we also showed the variation with $\TB$. 

In our numerical scenarios we compared the tree-level production cross 
sections with the full one-loop corrected cross sections. 
The numerical results we have shown are, of course, dependent on the choice 
of the SUSY parameters.  Nevertheless, they give an idea of the relevance
of the full one-loop corrections.
For the slepton pair production, \eeSeSe\ and \eeSnSn\ 
we observed for $s \to \infty$ a decreasing cross section 
$\propto \ln{(s)}/s$ for the first and $\propto 1/s$ for the second and 
third slepton generations.
The (loop corrected) cross sections for the charged slepton pair production 
can reach a level of $10\,\fb$, depending on the SUSY parameters, but 
is very small for the production of two different smuons $\Sm_1\Sm_2$ at the 
\order{10\,\ab} (and with the cross section of $\Stau1\Stau2$ enhanced by a 
factor of $m_\tau^2/m_\mu^2$).
This renders these processes difficult to observe at an $e^+e^-$ collider.%
\footnote{
  The limit of $10$~ab corresponds to ten events at an 
  integrated luminosity of $\cL = 1\, \iab$, which 
  constitutes a guideline for the observability of a 
  process at a linear collider.
}
The full one-loop corrections are very roughly $15\,\%$ of the tree-level 
results, but vary strongly on the size of $\MSE$ and in the case of
selectrons also on the size of $M_1$ and $M_2$.  Depending on the size of 
in particular these parameters the loop corrections can be either positive 
or negative. 
The dependence on $\phiMe$ ($\phiAeg$) was found at the level of 
$\sim 15\,\%$ ($\sim 8\,\%$), but can go up to $\sim 18\,\%$ 
($\sim 14\,\%$) for the extreme cases.  The relative loop corrections 
varied by up to $2\,\%$ ($4\,\%$) with $\phiMe$ ($\phiAeg$).
This shows that the loop corrections, including the complex phase 
dependence, have to be included point-by-point in any precision analysis, 
or any precise determination of SUSY parameters from the production of 
cMSSM sleptons at $e^+e^-$ linear colliders.

Concerning scalar neutrino production, the full one-loop corrections 
are very roughly $10\,\%$ of the tree-level results, but depend strongly 
on the size of $\MSE$, where larger values result even in negative loop 
corrections.  The cross sections are largest for $\eeSneSne$ and roughly 
smaller by one order of magnitude for $\eeSnmSnm$ and $\eeSnaSna$. 
The variation of the cross sections with $\phiMe$ and $\phiAeg$ is found 
extremely small.  Also for scalar neutrino production the loop corrections
have to be included point-by-point in any precision analysis at $e^+e^-$ 
linear colliders.

For all cross section calculations and for all parameter choices it was 
possible to identify at least one renormalization scheme that exhibited a 
``smooth'' behavior.  It appears to be possible for any parameter variation
to find a combination of schemes that yield numerically stable (and nearly 
constant) one-loop level contributions. A detailed analysis of which scheme 
yields this desired behavior as a function of the underlying SUSY parameters, 
however, is beyond the scope of this paper.

We emphasize again that our full one-loop calculation can readily be used 
together with corresponding full one-loop corrections to slepton 
decays~\cite{Stau2decay} or other slepton production 
modes~\cite{HiggsDecaySferm,LHCxC1,LHCxC2,LHCxN,LHCxNprod}.


\subsection*{Acknowledgements}

\begingroup \small
We thank T.~Hahn and F.~von~der~Pahlen for helpful discussions.  
The work of S.H.\ is supported 
in part by the MEINCOP Spain under contract FPA2016-78022-P, 
in part by the ``Spanish Agencia Estatal de Investigaci?n'' (AEI) and the EU
``Fondo Europeo de Desarrollo Regional'' (FEDER) through the project
FPA2016-78022-P, and in part by the AEI through the grant IFT
Centro de Excelencia Severo Ochoa SEV-2016-0597. 
\endgroup


\newcommand\jnl[1]{\textit{\frenchspacing #1}}
\newcommand\vol[1]{\textbf{#1}}

\end{document}

%% file: paperdef.tex
\newcommand{\lsim}
{\;\raisebox{-.3em}{$\stackrel{\displaystyle <}{\sim}$}\;}
\newcommand{\gsim}
{\;\raisebox{-.3em}{$\stackrel{\displaystyle >}{\sim}$}\;}

\newcommand\Code[1]{\ensuremath{\texttt{#1}}}
\newcommand\Var[1]{\ensuremath{\mathit{#1}}}

\newcommand\Vg{\Var{g}}
\newcommand\Vs{\Var{s}}

\newcommand\al{\alpha}

\newcommand\tb{\tan\beta}
\newcommand\TB{t_\beta}

\newcommand\CBB{c_{2\beta}}

\newcommand\LP{\left(}
\newcommand\RP{\right)}
\newcommand\LB{\left[}
\newcommand\RB{\right]}
\newcommand\LV{\left\{}
\newcommand\RV{\right\}}

\newcommand\ReTilde{\mathop{\widetilde{\mathrm{Re}}}}
\newcommand\ReDiag{\mathop{%
  \raise .5pt\hbox{[}%
  \widetilde{\mathrm{Re}}%
  \raise .5pt\hbox{]}}}
\newcommand\ReOffDiag{\mathop{%
  \raise .5pt\hbox{$\llbracket$}%
  \widetilde{\mathrm{Re}}%
  \raise .5pt\hbox{$\rrbracket$}}}
\newcommand\SE[1]{\Sigma_{#1}}
\newcommand\OS{\mathrm{OS}}
\newcommand\DRbar{\ensuremath{\smash{\overline{\mathrm{DR}}}}}
\newcommand\MSbar{\ensuremath{\overline{\mathrm{MS}}}}
\newcommand\matr[1]{\mathbf{#1}}
\newcommand\mati[1]{\bigl(#1\bigr)}

\newcommand\cL{{\cal L}}

\newcommand\SW{s_\mathrm{w}}
\newcommand\CW{c_\mathrm{w}}
\newcommand\MW{M_W}
\newcommand\MZ{M_Z}
\newcommand\Mh{M_h}

\newcommand\MHp{M_{H^\pm}}
\newcommand\mf[1]{m_{f_{#1}}}
\newcommand\mb{m_b}

\newcommand\Sf{\tilde f}

\newcommand\msf[1]{m_{\Sf_{#1}}}

\newcommand\Sn{\tilde\nu}
\newcommand\Sl{\tilde l}
\newcommand\msn[1]{m_{\Sn_{#1}}}
\newcommand\Se{\mathrm{\tilde e}}
\newcommand\Sm{\mathrm{\tilde \mu}}
\newcommand\Fe{\mathrm{e}}
\newcommand\mfe[1]{m_{\Fe_{#1}}}
\newcommand\mse[1]{m_{\Se_{#1}}}

\newcommand\Su{\mathrm{\tilde u}}
\newcommand\Fu{\mathrm{u}}
\newcommand\mfu[1]{m_{\Fu_{#1}}}
\newcommand\msu[1]{m_{\Su_{#1}}}
\newcommand\Sd{\mathrm{\tilde d}}
\newcommand\Fd{\mathrm{d}}
\newcommand\mfd[1]{m_{\Fd_{#1}}}
\newcommand\msd[1]{m_{\Sd_{#1}}}

\newcommand\Sele[1]{{\tilde e_{#1}}}

\newcommand\Smue[1]{{\tilde\mu_{#1}}}

\newcommand\Stau[1]{{\tilde\tau_{#1}}}

\newcommand\spri{s^{\prime}}

\newcommand\dTB{\delta\TB}
\newcommand\ino[1]{\tilde\chi_{#1}}

\newcommand\chapm[1]{\ino{#1}^\pm}
\newcommand\champ[1]{\ino{#1}^\mp}

\newcommand\cha{\chapm}
\newcommand\mcha[1]{m_{\chapm{#1}}}

\newcommand\neu[1]{\ino{#1}^0}
\newcommand\mneu[1]{m_{\neu{#1}}}

\newcommand\refeq[1]{Eq.~(\ref{#1})}
\newcommand\refeqs[1]{Eqs.~(\ref{#1})}
\newcommand\refta[1]{Tab.~\ref{#1}}
\newcommand\refse[1]{Sect.~\ref{#1}}

\newcommand\citere[1]{Ref.~\cite{#1}}
\newcommand\citeres[1]{Refs.~\cite{#1}}

\newcommand\eg{e.g.}
\newcommand\ie{i.e.\ }
\newcommand\wrt{w.r.t.\ }

\newcommand{\CP}{{\cal CP}}
\newcommand{\cp}{{\CP}}

\newcommand{\onel}{one-loop}
\newcommand{\tev}{\,\, \mathrm{TeV}}
\newcommand{\gev}{\,\, \mathrm{GeV}}
\newcommand{\mev}{\,\, \mathrm{MeV}}

\newcommand{\He}{h_1}
\newcommand{\Hz}{h_2}
\newcommand{\Hd}{h_3}

\newcommand{\eeSlSl}{\ensuremath{e^+e^- \to \Sl_{gs} \Sl_{gs^{\prime}}}}
\newcommand{\eeSeSe}{\ensuremath{e^+e^- \to \Se^{\pm}_{gs} \Se^{\mp}_{gs^{\prime}}}}
\newcommand{\eeSeeSee}{\ensuremath{e^+e^- \to \tilde{e}^+_1 \tilde{e}^-_1}}
\newcommand{\eeSeeSez}{\ensuremath{e^+e^- \to \tilde{e}^{\pm}_1 \tilde{e}^{\mp}_2}}
\newcommand{\eeSezSez}{\ensuremath{e^+e^- \to \tilde{e}^+_2 \tilde{e}^-_2}}
\newcommand{\eeSmeSme}{\ensuremath{e^+e^- \to \tilde{\mu}^+_1 \tilde{\mu}^-_1}}
\newcommand{\eeSmeSmz}{\ensuremath{e^+e^- \to \tilde{\mu}^{\pm}_1 \tilde{\mu}^{\mp}_2}}
\newcommand{\eeSmzSmz}{\ensuremath{e^+e^- \to \tilde{\mu}^+_2 \tilde{\mu}^-_2}}
\newcommand{\eeSaeSae}{\ensuremath{e^+e^- \to \tilde{\tau}^+_1 \tilde{\tau}^-_1}}
\newcommand{\eeSaeSaz}{\ensuremath{e^+e^- \to \tilde{\tau}^{\pm}_1 \tilde{\tau}^{\mp}_2}}
\newcommand{\eeSazSaz}{\ensuremath{e^+e^- \to \tilde{\tau}^+_2 \tilde{\tau}^-_2}}
\newcommand{\eeSnSn}{\ensuremath{e^+e^- \to \Sn_{g} \Sn_{g}^*}}
\newcommand{\eeSneSne}{\ensuremath{e^+e^- \to \tilde{\nu}_e \tilde{\nu}_e^*}}
\newcommand{\eeSnmSnm}{\ensuremath{e^+e^- \to \tilde{\nu}_{\mu} \tilde{\nu}_{\mu}^*}}
\newcommand{\eeSnaSna}{\ensuremath{e^+e^- \to \tilde{\nu}_{\tau} \tilde{\nu}_{\tau}^*}}
\newcommand{\eeSfSf}{\ensuremath{e^+e^- \to \Sf_{s} \Sf_{s^{\prime}}^*}}
\newcommand{\eeSasSas}{\ensuremath{e^+e^- \to \tilde{\tau}^+_s \tilde{\tau}^-_{s^{\prime}}}}

\newcommand\FA{\texttt{FeynArts}}
\newcommand\FC{\texttt{FormCalc}}
\newcommand\LT{\texttt{LoopTools}}
\newcommand\FH{\texttt{FeynHiggs}}
\newcommand\FT{\texttt{FeynTools}}

\newcommand\fb{\ensuremath{\mbox{fb}}}
\newcommand\ab{\ensuremath{\mbox{ab}}}

\newcommand\iab{\ensuremath{\ab^{-1}}}

\newcommand\msele[1]{m_{\tilde{e}_{#1}}}
\newcommand\msmue[1]{m_{\tilde{\mu}_{#1}}}
\newcommand\mstau[1]{m_{\tilde{\tau}_{#1}}}
\newcommand\msneu{m_{\tilde{\nu}_{e,\mu,\tau}}}

\newcommand{\Scs}{$\mathcal S$}

\newcommand{\sig}{\sigma}
\newcommand{\sigfull}{\sigma_{\text{full}}}
\newcommand{\sigtree}{\sigma_{\text{tree}}}
\newcommand{\sigloop}{\sigma_{\text{loop}}}

\newcommand{\sigvirt}{\sigma_{\text{virt}}}
\newcommand{\sigsoft}{\sigma_{\text{soft}}}
\newcommand{\sighard}{\sigma_{\text{hard}}}
\newcommand{\sigcoll}{\sigma_{\text{coll}}}

\newcommand{\phiAeg}{\varphi_{A_{\Fe_g}}}

\def\order#1{\ensuremath{{\cal O}(#1)}}
\def\reffi#1{\mbox{Fig.~\ref{#1}}}
\def\reffis#1{\mbox{Figs.~\ref{#1}}}

\def\ga{\gamma}
\def\de{\delta}
\def\la{\lambda}

\def\phimu{\varphi_{\mu}}
\def\phiMe{\varphi_{M_1}}
\def\phiMz{\varphi_{M_2}}

\def\MSL{M_{\tilde L}}
\def\MSE{M_{\tilde E}}

\definecolor{Orange}{named}{orange}
\definecolor{Purple}{named}{purple}
\definecolor{Lightblue}{cmyk}{0.9,0.1,0.1,0.3}
\definecolor{dgelborange}{cmyk}{0.,0.3,0.5, 0.}
\definecolor{Lila}{rgb}{0.5,0.,1}

%% file: eeSlep.bbl
\begin{thebibliography}{99} 

\begingroup \raggedright \small

%
%

\bibitem{Ni1984}
H.~Nilles, 
\jnl{Phys. Rept.} \vol{110} (1984) 1.

\bibitem{Ba1988}
R.~Barbieri, 
\jnl{Riv. Nuovo Cim.} \vol{11} (1988) 1. 

\bibitem{HaK85}
H.~Haber, G.~Kane,
\jnl{Phys. Rept.} \vol{117} (1985) 75.

\bibitem{GuH86}
J.~Gunion, H.~Haber,
\jnl{Nucl. Phys.} \vol{B 272} (1986) 1.

\bibitem{ATLASdiscovery} 
G.~Aad et al.\ [ATLAS Collaboration],
\jnl{Phys. Lett.} \vol{B 716} (2012) 1
[\href{http://de.arxiv.org/abs/arXiv:1207.7214}{arXiv:1207.7214} [hep-ex]].

\bibitem{CMSdiscovery} 
S.~Chatrchyan et al.\ [CMS Collaboration],
\jnl{Phys. Lett.} \vol{B 716} (2012) 30 
[\href{http://de.arxiv.org/abs/arXiv:1207.7235}{arXiv:1207.7235} [hep-ex]].

\bibitem{Go1983} 
H.~Goldberg,
\jnl{Phys. Rev. Lett.} \vol{50} (1983) 1419.

\bibitem{ElHaNaOlSr1984}
J.~Ellis, J.~Hagelin, D.~Nanopoulos, K.~Olive, M.~Srednicki,
\jnl{Nucl. Phys.} \vol{B 238} (1984) 453.

\bibitem{mhiggsCPXgen} 
A.~Pilaftsis,
\jnl{Phys. Rev.} \vol{D 58} (1998) 096010
[\href{http://de.arxiv.org/abs/arXiv:hep-ph/9803297}{arXiv:hep-ph/9803297}].

\bibitem{Pi1998}
A.~Pilaftsis,
\jnl{Phys. Lett.} \vol{B 435} (1998) 88
[\href{http://de.arxiv.org/abs/arXiv:hep-ph/9805373}{arXiv:hep-ph/9805373}].

\bibitem{Demir}
D.~A.~Demir,
\jnl{Phys. Rev.} \vol{D 60} (1999) 055006
[\href{http://de.arxiv.org/abs/arXiv:hep-ph/9901389}{arXiv:hep-ph/9901389}].

\bibitem{mhiggsCPXRG1} 
A.~Pilaftsis, C.~E.~M.~Wagner, 
\jnl{Nucl. Phys.} \vol{B 553} (1999) 3
[\href{http://de.arxiv.org/abs/arXiv:hep-ph/9902371}{arXiv:hep-ph/9902371}].

\bibitem{mhiggsCPXFD1} 
S. Heinemeyer,
\jnl{Eur. Phys. J.} \vol{C 22} (2001) 521
[\href{http://de.arxiv.org/abs/arXiv:hep-ph/0108059}{arXiv:hep-ph/0108059}].

\bibitem{ATLAS-SUSY}
\url{https://twiki.cern.ch/twiki/bin/view/AtlasPublic/SupersymmetryPublicResults}\,. 

\bibitem{CMS-SUSY}
\url{https://twiki.cern.ch/twiki/bin/view/CMSPublic/PhysicsResultsSUS}\,.

\bibitem{ILC-TDR}
H.~Baer et al.,
\textit{The International Linear Collider Technical Design Report - Volume 2: Physics},
\href{http://de.arxiv.org/abs/arXiv:1306.6352}{arXiv:1306.6352} [hep-ph].

\bibitem{teslatdr} 
R.-D.~Heuer et al.\ [TESLA Collaboration], 
\textit{TESLA Technical Design Report, Part~III: Physics at an $e^+e^-$ Linear Collider},
\href{http://de.arxiv.org/abs/arXiv:hep-ph/0106315}{arXiv:hep-ph/0106315}, see:\\
\url{http://tesla.desy.de/new_pages/TDR_CD/start.html}\,.

\bibitem{Ac2004}
K.~Ackermann et al., 
\textit{Proceedings Summer Colloquium, Amsterdam, Netherlands, 4 April 2003}, 
DESY-PROC-2004-01.

\bibitem{ilc1}
J.~Brau et al.\ [ILC Collaboration],
\textit{ILC Reference Design Report Volume 1 - Executive Summary},
\href{http://de.arxiv.org/abs/arXiv:0712.1950}{arXiv:0712.1950} [physics.acc-ph].

\bibitem{ilc2}
A.~Djouadi et al.\ [ILC Collaboration],
\textit{International Linear Collider Reference Design Report Volume 2: Physics at the ILC},
\href{http://de.arxiv.org/abs/arXiv:0709.1893}{arXiv:0709.1893} [hep-ph].

\bibitem{LCreport}
G.~Moortgat-Pick et al.,
\jnl{Eur. Phys. J.} \vol{C 75} (2015) 8, 371
[\href{http://de.arxiv.org/abs/arXiv:1504.01726}{arXiv:1504.01726} [hep-ph]].

\bibitem{CLIC1} 
L.~Linssen, A.~Miyamoto, M.~Stanitzki, H.~Weerts,
\href{http://de.arxiv.org/abs/arXiv:1202.5940}{arXiv:1202.5940} [physics.ins-det].

\bibitem{CLIC2}
H.~Abramowicz et al.\ [CLIC Detector and Physics Study Collaboration],
\textit{Physics at the CLIC $e^+e^-$ Linear Collider -- Input to the Snowmass process 2013}, 
\href{http://de.arxiv.org/abs/arXiv:1307.5288}{arXiv:1307.5288} [hep-ex].

\bibitem{lhcilc1} 
G.~Weiglein et al.\ [LHC/ILC Study Group],
\jnl{Phys. Rept.} \vol{426} (2006) 47
[\href{http://de.arxiv.org/abs/arXiv:hep-ph/0410364}{arXiv:hep-ph/0410364}].

\bibitem{lhcilc2}
A.~De Roeck et al.,
\jnl{Eur. Phys. J.} \vol{C 66} (2010) 525
[\href{http://de.arxiv.org/abs/arXiv:0909.3240}{arXiv:0909.3240} [hep-ph]].

\bibitem{lhcilc3}
A.~De Roeck, J.~Ellis, S.~Heinemeyer,
\jnl{CERN Cour.} \vol{49N10} (2009) 27.

\bibitem{Stau2decay} 
S.~Heinemeyer, C.~Schappacher,
\jnl{Eur. Phys. J.} \vol{C 72} (2012) 2136
[\href{http://de.arxiv.org/abs/arXiv:1204.4001}{arXiv:1204.4001} [hep-ph]].

\bibitem{HiggsDecaySferm} 
S.~Heinemeyer, C.~Schappacher,
\jnl{Eur. Phys. J.} \vol{C 75} (2015) 5, 198
[\href{http://de.arxiv.org/abs/arXiv:1410.2787}{arXiv:1410.2787} [hep-ph]].

\bibitem{LHCxC1} 
S.~Heinemeyer, F.~von~der~Pahlen, C.~Schappacher,
\jnl{Eur. Phys. J.} \vol{C 72} (2012) 1892
[\href{http://de.arxiv.org/abs/arXiv:1112.0760}{arXiv:1112.0760} [hep-ph]].

\bibitem{LHCxC2}
S.~Heinemeyer, F.~von~der~Pahlen, C.~Schappacher,
\href{http://de.arxiv.org/abs/arXiv:1202.0488}{arXiv:1202.0488} [hep-ph].

\bibitem{LHCxN} 
A.~Bharucha, S.~Heinemeyer, F.~von~der~Pahlen, C.~Schappacher,
\jnl{Phys. Rev.} \vol{D 86} (2012) 075023
[\href{http://de.arxiv.org/abs/arXiv:1208.4106}{arXiv:1208.4106} [hep-ph]].

\bibitem{LHCxNprod}
A.~Bharucha, S.~Heinemeyer, F.~von~der~Pahlen,
\jnl{Eur. Phys. J.} \vol{C 73} (2013) 2629
[\href{http://de.arxiv.org/abs/arXiv:1307.4237}{arXiv:1307.4237} [hep-ph]].

\bibitem{SchWa1985}
D.~H.~Schiller, D.~W\"ahner,
\jnl{Nucl.Phys.} \vol{B 255} (1985) 505.

\bibitem{CaDi1995}
B.~de~Carlos, M.~A.~Diaz,
\jnl{Phys. Lett.} \vol{B 417} (1998) 72
[\href{http://de.arxiv.org/abs/arXiv:hep-ph/9511421}{arXiv:hep-ph/9511421}].

\bibitem{ChDrGaLe2001}
S.~Y.~Choi, M.~Drees, B.~Gaissmaier, J.~S.~Lee,
\jnl{Phys. Rev.} \vol{D 64} (2001) 095009
[\href{http://de.arxiv.org/abs/arXiv:hep-ph/0103284}{arXiv:hep-ph/0103284}].

\bibitem{FrMiZe2001}
A.~Freitas, D.~J.~Miller, P.~M.~Zerwas,
\jnl{Eur. Phys. J.} \vol{C 21} (2001) 361
[\href{http://de.arxiv.org/abs/arXiv:hep-ph/0106198}{arXiv:hep-ph/0106198}].

\bibitem{Fr2002}
A.~Freitas,
``Production of scalar leptons at linera colliders'',
PhD thesis, Hamburg, Germany, 2002.

\bibitem{BlFrMoPo2002}
C.~Bl\"ochinger, H.~Fraas, G.~Moortgat-Pick, W.~Porod,
\jnl{Eur. Phys. J.} \vol{C 24} (2002) 297 
[\href{http://de.arxiv.org/abs/arXiv:hep-ph/0201282}{arXiv:hep-ph/0201282}].

\bibitem{FrMaZe2004}
A.~Freitas, A.~von~Manteuffel, P.~M.~Zerwas,
\jnl{Eur. Phys. J.} \vol{C 34} (2004) 487
[\href{http://de.arxiv.org/abs/arXiv:hep-ph/0310182}{arXiv:hep-ph/0310182}].

\bibitem{FrMaZe2005}
A.~Freitas, A.~von~Manteuffel, P.~M.~Zerwas,
\jnl{Eur. Phys. J.} \vol{C 40} (2005) 435
[\href{http://de.arxiv.org/abs/arXiv:hep-ph/0408341}{arXiv:hep-ph/0408341}].

\bibitem{ArHo2003}
A.~Arhrib, W.~Hollik,
\jnl{JHEP} \vol{0404} (2004) 073
[\href{http://de.arxiv.org/abs/arXiv:hep-ph/0311149}{arXiv:hep-ph/0311149}].

\bibitem{KoWeEbMa2004}
K.~Kova\v{r}\'ik, C.~Weber, H.~Eberl, W.~Majerotto,
\jnl{Phys. Lett.} \vol{B 591} (2004) 242
[\href{http://de.arxiv.org/abs/arXiv:hep-ph/0401092}{arXiv:hep-ph/0401092}].

\bibitem{Ko2005}
K.~Kova\v{r}\'ik,
``Precise predictions for sfermion pair production at a linear collider'',
PhD thesis, Bratislava, Slovakia, 2005.

\bibitem{KoWeEbMa2005}
K.~Kova\v{r}\'ik, C.~Weber, H.~Eberl, W.~Majerotto,
\jnl{Phys. Rev.} \vol{D 72} (2005) 053010
[\href{http://de.arxiv.org/abs/arXiv:hep-ph/0506021}{arXiv:hep-ph/0506021}].

\bibitem{feynarts1}
J.~K\"ublbeck, M.~B\"ohm, A.~Denner, 
\jnl{Comput. Phys. Commun.} \vol{60} (1990) 165.

\bibitem{feynarts2}
T.~Hahn, 
\jnl{Comput. Phys. Commun.} \vol{140} (2001) 418
[\href{http://de.arxiv.org/abs/arXiv:hep-ph/0012260}{arXiv:hep-ph/0012260}].

\bibitem{feynarts3}
T.~Hahn, C.~Schappacher, 
\jnl{Comput. Phys. Commun.} \vol{143} (2002) 54
[\href{http://de.arxiv.org/abs/arXiv:hep-ph/0105349}{arXiv:hep-ph/0105349}].\\
Program, user's guide and model files are available via: 
\url{http://www.feynarts.de}\,.

\bibitem{MSSMCT} 
T.~Fritzsche, T.~Hahn, S.~Heinemeyer, F.~von~der~Pahlen, H.~Rzehak, C.~Schappacher,   
\jnl{Comput. Phys. Commun.} \vol{185} (2014) 1529
[\href{http://de.arxiv.org/abs/arXiv:1309.1692}{arXiv:1309.1692} [hep-ph]].

\bibitem{formcalc1}
T.~Hahn, M.~P\'erez-Victoria,
\jnl{Comput. Phys. Commun.} \vol{118} (1999) 153
[\href{http://de.arxiv.org/abs/arXiv:hep-ph/9807565}{arXiv:hep-ph/9807565}].\\
Program and user's guide are available via: 
\url{http://www.feynarts.de/formcalc/}\,.

\bibitem{formcalc2}	
T.~Hahn, S.~Pa{\ss}ehr, C.~Schappacher,
\jnl{PoS LL2016 (2016) 068, J. Phys. Conf. Ser.} \vol{762} (2016) 1, 012065
[\href{http://de.arxiv.org/abs/arXiv:1604.04611}{arXiv:1604.04611} [hep-ph]].

\bibitem{HiggsDecayIno} 
S.~Heinemeyer, C.~Schappacher,
\jnl{Eur. Phys. J.} \vol{C 75} (2015) 5, 230
[\href{http://de.arxiv.org/abs/arXiv:1503.02996}{arXiv:1503.02996} [hep-ph]].

\bibitem{SbotRen1} 
S.~Heinemeyer, H.~Rzehak, C.~Schappacher,
\jnl{Phys. Rev.} \vol{D 82} (2010) 075010
[\href{http://de.arxiv.org/abs/arXiv:1007.0689}{arXiv:1007.0689} [hep-ph]].

\bibitem{SbotRen2}
S.~Heinemeyer, H.~Rzehak, C.~Schappacher,
\jnl{PoSCHARGED} \vol{2010} (2010) 039
[\href{http://de.arxiv.org/abs/arXiv:1012.4572}{arXiv:1012.4572} [hep-ph]].

\bibitem{Stop2decay}
T.~Fritzsche, S.~Heinemeyer, H.~Rzehak, C.~Schappacher, 
\jnl{Phys. Rev.} \vol{D 86} (2012) 035014
[\href{http://de.arxiv.org/abs/arXiv:1111.7289}{arXiv:1111.7289} [hep-ph]].

\bibitem{Gluinodecay} 
S.~Heinemeyer, C.~Schappacher,
\jnl{Eur. Phys. J.} \vol{C 72} (2012) 1905
[\href{http://de.arxiv.org/abs/arXiv:1112.2830}{arXiv:1112.2830} [hep-ph]].

\bibitem{HiggsProd}
S.~Heinemeyer, C.~Schappacher,
\jnl{Eur. Phys. J.} \vol{C 76} (2016) 4, 220
[\href{http://de.arxiv.org/abs/arXiv:1511.06002}{arXiv:1511.06002} [hep-ph]].

\bibitem{HpProd}
S.~Heinemeyer, C.~Schappacher,
\jnl{Eur. Phys. J.} \vol{C 76} (2016) 10, 535
[\href{http://de.arxiv.org/abs/arXiv:1606.06981}{arXiv:1606.06981} [hep-ph]].

\bibitem{eeIno}
S.~Heinemeyer, C.~Schappacher,
\jnl{Eur. Phys. J.} \vol{C 77} (2017) 9, 649
[\href{http://de.arxiv.org/abs/arXiv:1704.07627}{arXiv:1704.07627} [hep-ph]].

\bibitem{mhcMSSMlong}
M.~Frank, T.~Hahn, S.~Heinemeyer, W.~Hollik, H.~Rzehak, G.~Weiglein,
\jnl{JHEP} \vol{0702} (2007) 047
[\href{http://de.arxiv.org/abs/arXiv:hep-ph/0611326}{arXiv:hep-ph/0611326}].

\bibitem{complexmassscheme}
A.~Denner, S.~Dittmaier, M.~Roth, D.~Wackeroth,
\jnl{Nucl. Phys. B} \vol{560} (1999) 33
[\href{http://de.arxiv.org/abs/arXiv:hep-ph/9904472}{arXiv:hep-ph/9904472}].

\bibitem{cdr}
F.~del Aguila, A.~Culatti, R.~Mu\~noz-Tapia, M.~P\'erez-Victoria,
\jnl{Nucl. Phys.} \vol{B 537} (1999) 561
[\href{http://de.arxiv.org/abs/arXiv:hep-ph/9806451}{arXiv:hep-ph/9806451}].

\bibitem{dred1}
W.~Siegel, 
\jnl{Phys. Lett.} \vol{B 84} (1979) 193.

\bibitem{dred2}
D.~Capper, D.~Jones, P.~van Nieuwenhuizen,
\jnl{Nucl. Phys.} \vol{B 167} (1980) 479. 

\bibitem{dredDS}
D.~St\"ockinger,
\jnl{JHEP} \vol{0503} (2005) 076
[\href{http://de.arxiv.org/abs/arXiv:hep-ph/0503129}{arXiv:hep-ph/0503129}].

\bibitem{dredDS2}
W.~Hollik, D.~St\"ockinger,
\jnl{Phys. Lett.} \vol{B 634} (2006) 63
[\href{http://de.arxiv.org/abs/arXiv:hep-ph/0509298}{arXiv:hep-ph/0509298}].

\bibitem{denner}
A.~Denner,
\jnl{Fortsch. Phys.} \vol{41} (1993) 307
[\href{http://de.arxiv.org/abs/arXiv:0709.1075}{arXiv:0709.1075} [hep-ph]].

\bibitem{slicing1}
K.~Fabricius, I.~Schmitt, G.~Kramer, G.~Schierholz, 
\jnl{Zeit. Phys.} \vol{C 11} (1981) 315.

\bibitem{slicing2}
G.~Kramer, B.~Lampe, 
\jnl{Fortschr. Phys.} \vol{37} (1989) 161.

\bibitem{slicing3}
H.~Baer, J.~Ohnemus, J.~Owens, 
\jnl{Phys. Rev.} \vol{D 40} (1989) 2844.

\bibitem{slicing4}
B.~Harris, J.~Owens,
\jnl{Phys. Rev.} \vol{D 65} (2002) 094032
[\href{http://de.arxiv.org/abs/arXiv:hep-ph/0102128}{arXiv:hep-ph/0102128}].

\bibitem{cuba1}
T.~Hahn, 
\jnl{Comput. Phys. Commun.} \vol{168} (2005) 78
[\href{http://de.arxiv.org/abs/arXiv:hep-ph/0404043}{arXiv:hep-ph/0404043}].

\bibitem{cuba2}
T.~Hahn,
\href{http://de.arxiv.org/abs/arXiv:1408.6373}{arXiv:1408.6373} [physics.comp-ph].\\
The program is available via: \url{http://www.feynarts.de/cuba/}\,.

\bibitem{FeynHiggs1}
S.~Heinemeyer, W.~Hollik and G.~Weiglein,
\jnl{Comput. Phys. Commun.} \vol{124} (2000) 76
[arXiv:hep-ph/9812320].

\bibitem{FeynHiggs2}  
S.~Heinemeyer, W.~Hollik and G.~Weiglein,
\jnl{Eur. Phys. J.} \vol{C 9} (1999) 343
[arXiv:hep-ph/9812472].

\bibitem{FeynHiggs3}
G.~Degrassi, S.~Heinemeyer, W.~Hollik, P.~Slavich and G.~Weiglein,
\jnl{Eur. Phys. J.} \vol{C 28} (2003) 133
[arXiv:hep-ph/0212020].

\bibitem{FeynHiggs5}
T.~Hahn, S.~Heinemeyer, W.~Hollik, H.~Rzehak and G.~Weiglein,
\jnl{Comput. Phys. Commun.} \vol{180} (2009) 1426.

\bibitem{FeynHiggs6}  
T.~Hahn, S.~Heinemeyer, W.~Hollik, H.~Rzehak and G.~Weiglein,
\jnl{Phys. Rev. Lett.} \vol{112} (2014) 14,  141801
[arXiv:1312.4937 [hep-ph]].

\bibitem{FeynHiggs7}  
H.~Bahl and W.~Hollik,
\jnl{Eur. Phys. J.} \vol{C 76} (2016) 499
[arXiv:1608.01880 [hep-ph]].

\bibitem{FeynHiggs8}
H.~Bahl, S.~Heinemeyer, W.~Hollik and G.~Weiglein,
\jnl{Eur. Phys. J.} \vol{C 78} (2018) no.1, 57
[arXiv:1706.00346 [hep-ph]].

\bibitem{feyncalc}
R.~Mertig, M.~B\"ohm and A.~Denner, 
\jnl{Comput. Phys. Commun.} \vol{64} (1991) 345.


\bibitem{SPS1a}
B.~C.~Allanach et al.,
\jnl{Eur. Phys. J.} \vol{C 25} (2002) 113
[\href{http://de.arxiv.org/abs/arXiv:hep-ph/0202233}{arXiv:hep-ph/0202233}].

\bibitem{pdg}
C.~Patrignani et al.\ (Particle Data Group), 
\jnl{Chin. Phys.} \vol{C 40} (2016 and 2017 update) 100001.   

\bibitem{ATLAS1}
ATLAS Collaboration, ATLAS-CONF-2017-039, see:\\
\url{https://atlas.web.cern.ch/Atlas/GROUPS/PHYSICS/CONFNOTES/ATLAS-CONF-2017-039}\,.

\bibitem{CMS1}
CMS Collaboration, CMS-PAS-SUS-17-004, see:\\
\url{http://cms-results.web.cern.ch/cms-results/public-results/preliminary-results/SUS-17-004/index.html}\,.

\bibitem{CMS2}
CMS Collaboration, CMS-PAS-SUS-16-039, see:\\
\url{http://cms-results.web.cern.ch/cms-results/public-results/preliminary-results/SUS-16-039/index.html}\,.

\bibitem{ccb1}
J.~Fr\`ere, D.~Jones, S.~Raby,
\jnl{Nucl. Phys.} \vol{B 222} (1983) 11.

\bibitem{ccb2}
M.~Claudson, L.~Hall, I.~Hinchliffe,
\jnl{Nucl. Phys.} \vol{B 228} (1983) 501.

\bibitem{ccb3}
C.~Kounnas, A.~Lahanas, D.~Nanopoulos, M.~Quiros,
\jnl{Nucl. Phys.} \vol{B 236} (1984) 438.

\bibitem{ccb4}
J.~Gunion, H.~Haber, M.~Sher,
\jnl{Nucl. Phys.} \vol{B 306} (1988) 1.

\bibitem{ccb5}
J.~Casas, A.~Lleyda, C.~Mu\~noz,
\jnl{Nucl. Phys.} \vol{B 471} (1996) 3
[\href{http://de.arxiv.org/abs/arXiv:hep-ph/9507294}{arXiv:hep-ph/9507294}].

\bibitem{ccb6}
P.~Langacker, N.~Polonsky,
\jnl{Phys. Rev.} \vol{D 50} (1994) 2199
[\href{http://de.arxiv.org/abs/arXiv:hep-ph/9403306}{arXiv:hep-ph/9403306}].

\bibitem{ccb7}
A.~Strumia,
\jnl{Nucl. Phys.} \vol{B 482} (1996) 24
[\href{http://de.arxiv.org/abs/arXiv:hep-ph/9604417}{arXiv:hep-ph/9604417}].

\bibitem{chargedmhiggs2L}
M.~Frank et al., 
\jnl{Phys. Rev.} \vol{D 88} (2013) 5, 055013
[\href{http://de.arxiv.org/abs/arXiv:1306.1156}{arXiv:1306.1156} [hep-ph]].

\bibitem{MSSMphases}
S.~Dimopoulos, S.~Thomas,
\jnl{Nucl. Phys.} \vol{B 465} (1996) 23
[\href{http://de.arxiv.org/abs/arXiv:hep-ph/9510220}{arXiv:hep-ph/9510220}].

\bibitem{SUSYphases}
M.~Dugan, B.~Grinstein, L.~Hall,
\jnl{Nucl. Phys.} \vol{B 255} (1985) 413.

\bibitem{EDMrev2}
D.~Demir, O.~Lebedev, K.~Olive, M.~Pospelov, A.~Ritz,
\jnl{Nucl. Phys.} \vol{B 680} (2004) 339
[\href{http://de.arxiv.org/abs/arXiv:hep-ph/0311314}{arXiv:hep-ph/0311314}].

\bibitem{EDMPilaftsis}
D.~Chang, W.~Keung, A.~Pilaftsis,
\jnl{Phys. Rev. Lett.} \vol{82} (1999) 900
[Erratum-ibid.\ \vol{83} (1999) 3972]
[\href{http://de.arxiv.org/abs/arXiv:hep-ph/9811202}{arXiv:hep-ph/9811202}].

\bibitem{Pi1999}
A.~Pilaftsis,
\jnl{Phys. Lett.} \vol{B 471} (1999) 174
[\href{http://de.arxiv.org/abs/arXiv:hep-ph/9909485}{arXiv:hep-ph/9909485}].

\bibitem{EDMRitz}
O.~Lebedev, K.~Olive, M.~Pospelov, A.~Ritz,
\jnl{Phys. Rev.} \vol{D 70} (2004) 016003
[\href{http://de.arxiv.org/abs/arXiv:hep-ph/0402023}{arXiv:hep-ph/0402023}].

\bibitem{EDMDoink1}
W.~Hollik, J.~Illana, S.~Rigolin, D.~St\"ockinger,
\jnl{Phys. Lett.} \vol{B 416} (1998) 345
[\href{http://de.arxiv.org/abs/arXiv:hep-ph/9707437}{arXiv:hep-ph/9707437}].

\bibitem{EDMDoink2}
W.~Hollik, J.~Illana, S.~Rigolin, D.~St\"ockinger,
\jnl{Phys. Lett.} \vol{B 425} (1998) 322
[\href{http://de.arxiv.org/abs/arXiv:hep-ph/9711322}{arXiv:hep-ph/9711322}].

\bibitem{EDMheavy1}
P.~Nath,
\jnl{Phys. Rev. Lett.} \vol{66} (1991) 2565.

\bibitem{EDMheavy2}
Y.~Kizukuri, N.~Oshimo,
\jnl{Phys. Rev.} \vol{D 46} (1992) 3025.

\bibitem{EDMmiracle1}
T.~Ibrahim, P.~Nath,
\jnl{Phys. Lett.} \vol{B 418} (1998) 98
[\href{http://de.arxiv.org/abs/arXiv:hep-ph/9707409}{arXiv:hep-ph/9707409}].

\bibitem{EDMmiracle2}
T.~Ibrahim, P.~Nath,
\jnl{Phys. Rev.} \vol{D 57} (1998) 478 
[Erratum-ibid.\ \vol{D 58} (1998) 019901] 
[Erratum-ibid.\ \vol{D 60} (1998) 079903] 
[Erratum-ibid.\ \vol{D 60} (1999) 119901]
[\href{http://de.arxiv.org/abs/arXiv:hep-ph/9708456}{arXiv:hep-ph/9708456}].

\bibitem{EDMmiracle3}
M.~Brhlik, G.~Good, G.~Kane,
\jnl{Phys. Rev.} \vol{D 59} (1999) 115004
[\href{http://de.arxiv.org/abs/arXiv:hep-ph/9810457}{arXiv:hep-ph/9810457}].

\bibitem{EDMrev1}
S.~Abel, S.~Khalil, O.~Lebedev,
\jnl{Nucl. Phys.} \vol{B 606} (2001) 151
[\href{http://de.arxiv.org/abs/arXiv:hep-ph/0103320}{arXiv:hep-ph/0103320}].

\bibitem{EDMrev3}
Y.~Li, S.~Profumo, M.~Ramsey-Musolf,
\jnl{JHEP} \vol{1008} (2010) 062
[\href{http://de.arxiv.org/abs/arXiv:1006.1440}{arXiv:1006.1440} [hep-ph]].

\bibitem{Ya2013}
N.~Yamanaka,
\jnl{Phys. Rev.} \vol{D 87} (2013) 011701
[\href{http://de.arxiv.org/abs/arXiv:1211.1808}{arXiv:1211.1808} [hep-ph]].

\bibitem{plehnix}
V.~Barger, T.~Falk, T.~Han, J.~Jiang, T.~Li, T.~Plehn,
\jnl{Phys. Rev.} \vol{D 64} (2001) 056007
[\href{http://de.arxiv.org/abs/arXiv:hep-ph/0101106}{arXiv:hep-ph/0101106}].


\endgroup

\end{thebibliography}
